\documentclass[12pt, openany]{ubthesis}
\usepackage{amsmath, amssymb}
\usepackage[numbers]{natbib}
\usepackage{graphicx}
\usepackage{setspace}
\usepackage{caption}
\usepackage{bm}
\newcommand{\repeatcaption}[2]{%
  \renewcommand{\thefigure}{\ref{#1}}%
  \captionsetup{list=no}%
  \caption{#2 (repeated from page \pageref{#1})}%
}

\title{Non-equilibrium Effects in Dissipative Strongly Correlated Systems}
\author{Jiajun Li}
\degree{phd}
\dept{Physics}
\conferral{June 1st, 2017}

\begin{document}
\begin{titlepage}
\maketitle
\end{titlepage}

\begin{ubfrontmatter}
\makecopyright
\cleardoublepage
\begin{abstract}
Non-equilibrium phenomena in strongly correlated lattice systems coupling to dissipative environment are studied. Novel physics arises when strongly correlated system is driven out of equilibrium by external fields. Dramatic changes in physical properties, such as conductivity, are empirically observed in strongly correlated materials under high electric field. In particular, electric-field driven metal-insulator transitions are well-known as resistive switching effect in a variety of materials, such as VO$_2$, V$_2$O$_3$ and other transition metal oxides.  To satisfactorily explain both the phenomenology and its underlying mechanism, it is required to model microscopically the out-of-equilibrium dissipative lattice system of interacting electrons. In this thesis, we developed a systematic method of modeling non-equilibrium steady state of dissipative lattice system by means of Non-equilibrium Green's function and Dynamical Mean Field Theory. We firstly establish a ``minimum model" to formulate the strong-field transport in non-interacting dissipative electron lattice. This model is exactly soluble and convenient for discussing energy dissipation and steady-state properties. Non-equilibrium electron distribution and effective temperature naturally emerge as a result of competing electric power and Joule dissipation. Building on this model, we explore the non-equilibrium phase transition in dissipative Hubbard model. Our result verifies the importance of thermal effect in the non-equilibrium interacting system. Correlated metallic systems undergo metal-insulator transition at fields much lower than the quasiparticle energy scale. And the hysteretic $I-V$ relation shows the possibility of spatially inhomogeneous state during non-equilibrium phase transition. In addition, formation of filamentary structures have been widely reported by many experimental groups. In order to further examine the spatial inhomogeneity, we conduct finite-sample simulation in the dissipative Hubbard model with Hartree-Fock approximation. The calculation successfully explains the main experimental features of the non-equilibrium phase transitions, like formation of conductive filament and negative differential resistance, and reveals the underlying electronic mechanism. It also justifies the thermal description that non-equilibrium effective temperature approaches equilibrium transition temperature. 

Finally, we apply the formulation to strong-field transport of Dirac electrons in graphene, concentrating on current saturation due to electron-phonon interactions. We show the novel momentum distribution of Dirac electrons under strong electric field, which has its origin in Landau-Zener physics. We discuss in detail its relation to the experimentally observed phenomena. The arXiv version has been updated with minor modifications and corrections.
\end{abstract}
\clearpage
\begin{acknowledgements}
At the very very beginning, I acknowledge my family, my parents Xingwen and Yanjun, my uncle Shaojun and aunt Xiaoling, my brother Junjie, my grandparents Cunliang, Gailing, Dongping and all others who have raised me, helped me throughout my life and ignited my passions in science. Without them, I cannot imagine how I can reach this point of life. I particularly acknowledge my wife Meng who has a husband desperately trying to express his love from the other side of this planet through the two video calls everyday for the long six years \footnote{I guess I should also acknowledge Tencent and their WeChat team}. 

I acknowledge my advisor Jong Han, who has led and assisted me to explore the world of physics and scientific research. As a knowledgeable and accessible advisor, Jong has taught me a great lot, particularly the spirit of sticking to the high standard of \emph{good} scientific research. I highly appreciate his guidance throughout my PhD research. Beginning as a first-year PhD student without even knowing much about many-body physics, I could have never understood so much physics without the kind assistance of Jong.

I acknowledge my collaborators, particularly Dr. Camille Aron and Dr. Gabriel Kotliar. The collaboration has always been very enjoyable. The thesis would be impossible without their valuable comments. Throughout the many years, discussions with Gabi and Camille have greatly improved my understanding of condensed matter physics and influenced my taste of conducting scientific research.

I appreciate the kind assistance and advice from Dr. Xuedong Hu and Dr. Sambandamurthy Ganapathy, not only for physics and science but also for the graduate life. I acknowledge Dr. Igor \v{Z}uti\'{c} for many helpful discussions and advice. I acknowledge Dr. Jonathan Bird for the insightful discussions on strong-field transport in graphene. The experimental work done by Dr. Bird's group is the motivation of my theoretical studies in graphene. The great experimental work by Sujay Singh has also inspired my research in filament formation. 

Last but not the least, I thank all my best friends in Buffalo, particuarly my colleagues Sun Fan and Ding Han, who have been great roommates for my entire PhD life. The hotpot nights hosted on Chestnut Ridge, featuring Liu Zeming the best runner of Chestnut Ridge, Xu Mengyang the Fresh, Gao Weiwei the Tall, Rich and Handsome, Zhang Meng, Deng Guo, Jin Weixiang the greatest football star of Kunz, Zhu Xuechen the Teacher, Zhao Xinyu the Grand Master, Xu Gaofeng the Paragon, Xia Weiyi (a.k.a. Hawaii) the Emperor of Europe, Wen Han the Ronaldo of Kunz, Wu Yabei, Deng Yanting, Dong Ruifeng, Shen Chenghao, Liu Xiaobin, Qiu Yizhi, Wang Jiawei, Hui Haolei, Xiao Jiayang, Wu Qiong, Zhao Chuan, Wang Zongye, Yoichi Takato and all the wonderful people (please forgive me for the impossibility of listing all my very best friends in Buffalo) have been indispensable excitation that has constantly pumped me out of the lowly-lying ground states and makes possible a \emph{driven steady-state} of my PhD research which becomes inevitably an uphill fight from time to time. 
\end{acknowledgements}
\tableofcontents
\clearpage
\listoffigures
\clearpage
\end{ubfrontmatter}

\chapter{Introduction}
\label{intro}
Describing non-equilibrium state has been one of the central goals of statistical physics for decades. Due to the fast development of nanolithography and strong-field techniques in these days, real systems can now be driven far from the equilibrium state, resulting in novel physics essentially different from those in equilibrium. In terms of theory, description of non-equilibrium state can be traced back to the beginning of statistical physics. However, people are still in the middle of finding a complete theoretical framework of non-equilibrium state that rivals equilibrium statistical physics. An equilibrium system embedded in an open environment is described successfully with a (grand) canonical ensemble. However, a non-equilibrium state is usually much more complex than this. A general formalism of non-equilibrium thermodynamics is still lacking, and a system in non-equilibrium state cannot generally be characterized by thermodynamic functions and their differential relations. Frequently, the time-evolution and dynamics is crucial to describe even a steady state in non-equilibrium. In addition, dissipation is another mechanism that complicates the non-equilibrium state. For example, in a system driven by an external field, a steady-state can only be realized in the presence of dissipative mechanisms so that energy injected by the driving field is subsequently dissipated into the environment. Otherwise the injected energy will accumulate and the infinitely increasing (non-equilibrium) temperature would overwhelm any interesting physics.

The task of describing the non-equilibrium state becomes even more challenging when interaction enters the picture. Dynamical mean field theory is one of the most powerful tools to study strong correlation physics in higher-dimensional systems. However, it is non-trivial to implement it in arbitrary non-equilibrium systems. In this thesis, we will establish a formulation to examine the non-equilibrium steady sate (NESS) of strongly correlated systems. These stationary non-equilibrium phenomena featuring time-independent physical observables are closely related to a variety of interesting experimental observations. They are also of industrial interests in many cases. Some examples are given below.

\section{Resistive Switching in Strongly Correlated Materials}
Strongly correlated materials undergo sudden resistive change under strong electric field of $10^4\sim10^6$ V/m. This phenomenon is called Resistive Switching (RS), and is frequently studied in transition metal oxides and chalcogenides. Experiments unveiled a large family of materials where the RS phenomenon is observed, covering a range from transition metal band Insulators, chalcogenides to Mott insulators/correlated metals. Canonical Mott insulators, such as chromium-doped vanadium sesquioxides, NiS$_{2-x}$Se$_x$ and narrow-gap Mott insulators AM$_4$Q$_8$ (A = Ca, Ge; M = V, Nb, Ta, Mo; Q = S, Se, Te).  

\begin{figure}
\centering
\includegraphics[scale=0.5]{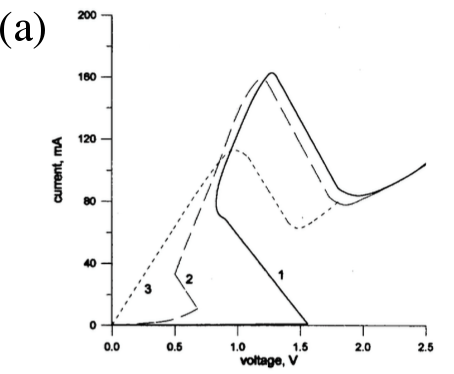}
\includegraphics[scale=0.3]{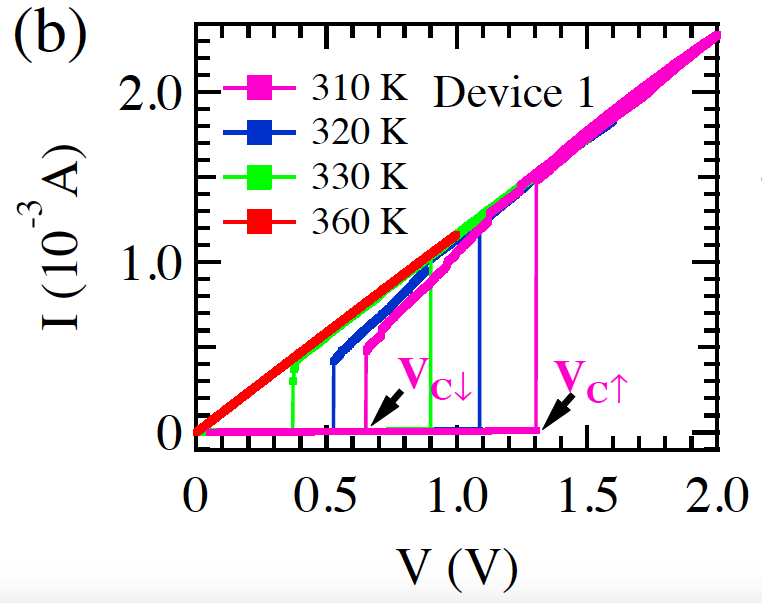}
\caption[$I-V$ curve in the resistive switching]{Current in the RS versus (a) sample voltage $V_s$ in V$_2$O$_3$ and (b) total voltage $V_t$ in VO$_2$. The sample is connected to the electric-generator and an external resistor in series, so that $V_t=IR+V_s$. Note that in the (a), the system transits from an antiferromagnetic insulator (AFI) to paramagnetic metal (PM) and then undergoes a transition to a Mott insulator. Panel (a) is adapted from Ref. \citenum{chudnovskii98} and the panel (b) is from Ref. \citenum{sujay15}.}
\label{RSsujay}
\end{figure}

 The RS effect has attracted attentions from the industry of electronics. Resistive random access memory (reRAM) has been proposed to be a strong candidate of the next generation storage technology. Despite plenty of experimental studies on RS phenomena, a microscopic description is still premature and its underlying mechanism is still on debate. In band insulators such as TiO$_2$, SrTiO$_3$, SrZrO$_3$ as well as some Ag/Cu based chalcogenides, it is proposed that electrochemical migration of ions is responsible for the RS phenomena\cite{pan14, waser07, waser09, kumai99, jeong13}. In Mott insulators, different mechanisms are proposed. Landau-Zener type of mechanisms are discussed in the literature\cite{oka03,oka10,oka12,sugimoto08,eckstein10}, where non-equilibrium excitations are created due to strong driving field and finally trigger the transition. On the other hand, avalanche mechanism is discussed in a family of narrow-gap Mott insulators, i.e. AM$_4$Q$_8$ (A = Ga, Ge; M = V, Nb, Ta, Mo; Q = S, Se)\cite{janod15,guiot13}. This mechanism is supported by the experimentally observed scaling law that threshold $E_\text{th}\sim E^{2.5}_\text{gap}$, and the phenomenology can be reproduced by calculations on a classical resistive network\cite{stoliar13}. Finally, it has been revealed that thermal mechanism due to Joule heating occurs in some oxides such as NiO\cite{sblee}, VO$_2$\cite{driscoll09,duchene71,zimmers13} and V$_2$O$_3$\cite{chudnovskii98}, etc. Despite the large family of materials showing RS phenomena, many features, such as filament formation and negative differential resistance, are shared by various materials. 
 
\begin{figure}
\centering
\includegraphics[scale=0.6]{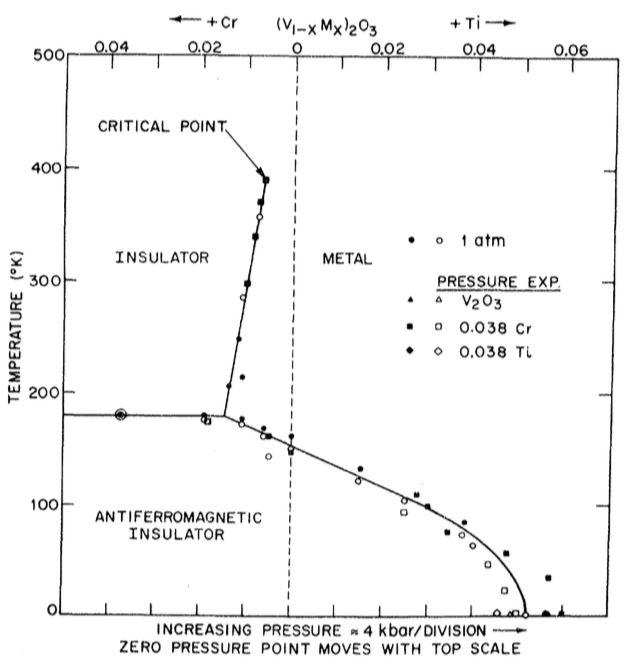}
\caption[Equilibrium phase diagram in V$_2$O$_3$]{Equilibrium phase diagram in V$_2$O$_3$. Left to the critical point, there is a small parameter window in which the material undergoes two metal-insulator transitions as temperature increases. The system firstly transits from AFI state to PM state, and then undergoes a transition to a Mott insulator at higher temperature. This process corresponds to the two non-equilibrium transitions shown in Fig. \ref{RSsujay}(a). The plot is adapted from Ref. \citenum{mcwhan73}.}
\label{RSmcwhan}
\end{figure}

Fig. \ref{RSsujay} shows typical $I-V$ relations of the RS in transition metal oxides. The samples of panel (a), (b) are corresonpondingly V$_2$O$_3$ and VO$_2$. Note that the device sample is connected to an external resistor $R$ to avoid overheating, so that sample voltage is related to total voltage by $V_s=V_t-IR$ provided that current is $I$. Later we will see that the external resistor plays a critical role in the theoretical calculations to reproduce the experimental results. The panel (a) shows two electric-field-driven transitions, switching the system from high-resistance state in equilibrium to a low-resistance state in the forward (increasing $V_t$) direction, and then again to a high-resistance state under higher electric fields. Specifically, the first insulator-to-metal transition (IMT) is from an antiferromagnetic insulator (AFI) to a paramagnetic metal (PM), where current sharply increases as the voltage bias reaches the threshold. The second metal-to-insulator transition (MIT) is from the PM to a Mott insulator in which no long-range order is present. These phase transitions correspond to the temperature-controlled metal-insulator transitions of V$_2$O$_3$ in equilibrium, as shown in Fig. \ref{RSmcwhan}. This resemblance between Electric-field-driven transitions and temperature-controlled transitions in equilibrium suggests a thermal scenario of resistive switching. The panel (b) shows both forward insulator-to-metal transition and backward (decreasing $V_t$) metal-to-insulator transitions. 

\begin{figure}
\centering
\includegraphics[scale=0.8]{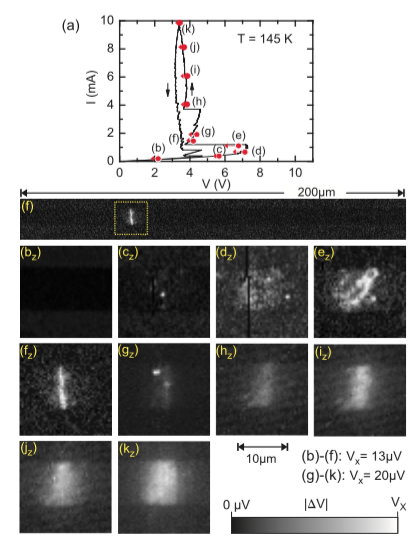}
\caption[Filament formation and $I-V$ characteristics]{Filament formation and $I-V$ characteristics in VO$_2$. The system undergoes dramatic drop in resistivity under strong voltage bias, which is accompanied by the formation of a conductive filament. The filament then gradually expands as total current increases. The graph is adapted from Ref. \citenum{guenon13}.}
\label{guenon}
\end{figure}

During the RS phenomena, it is widely observed that a filament suddenly forms out of the insulating oxide sample under strong voltage bias, and gradually expands to conduct increasing current. The process is shown in Fig. \ref{guenon}. 

\begin{figure}
\centering
\includegraphics[scale=0.5]{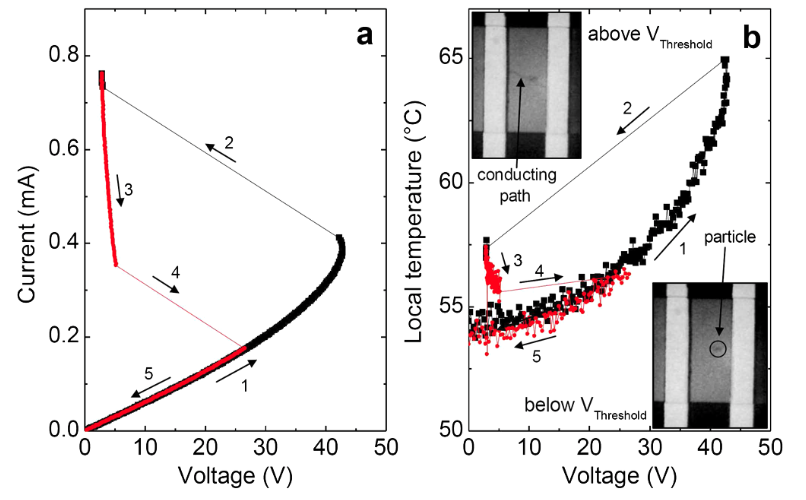}
\caption[Thermal heating in the RS in VO$_2$]{Current and temperature of the VO$_2$ sample under external voltage bias. An external resistor $R$ is connected in series to the sample, so sample voltage drops after the RS occurs. The sample starts to conduct current as a conductive filament forms, connecting the source/drain leads. Temperature rises before the RS and drops after it occurs. In the opposite direction, the temperature decreases and falls back to the initial level after the inverse RS. The graph is adapted from Ref. \citenum{zimmers13}.}
\label{zimmers}
\end{figure}

Further experiments also measured the temperature of a sample during the RS, and shows the thermal heating plays a critical role during the RS in VO$_2$. It is shown in Fig. \ref{zimmers}. A fluorescent particle is put in the sample to measure the temperature at its position. The temperature rises under external voltage bias and drops a few degrees as the RS occurs and the conducive filament forms. The filament is clearly shown in the inset of Fig. \ref{zimmers}(b).  After the RS, the system jumps to the NDR branch of the $I-V$ curve.  On the other hand, decreasing total voltage will induce inverse resistive switching. The current reduces on the NDR branch until the system jumps to the initial high-resistance state as the temperature drops back to the initial level.

These experimental studies have inspired phenomenological models based on resistor networks\cite{janod15}. However, it requires a microscopic theory to address the underlying mechanisms of the RS effect. In this thesis, we will construct a dissipative lattice model to characterize the non-equilibrium strongly-correlated quantum state of solids, and explore the microscopic mechanisms of the driven metal-insulator transition.

\section{Current saturation in graphene}
Graphene is one of the most studied 2D material. It is a semimetal with linear dispersion relation. It has high carrier mobility and critical current density, thus is a promising candidate for many applications in nanoscale devices. The research efforts of fabricating graphene field-effect transistors leads to observation of current saturation under strong electric field\cite{meric08}. The phenomenon limits the current that a graphene sample can conduct and quickly becomes a subject of intense research\cite{barreiro09,shishir09,fang11,ramamoorthy15}.  Optical phonon scattering of electrons at high-field regime is identified as the reason of current saturation\cite{perebeinos10}. Electrons are accelerated by the external field and rapidly lose energy by emitting optical phonons, causing the drift velocity to saturate. Semiclassical theories succeed to discuss the current saturation of samples with high carrier density, while it is necessary to establish a microscopic model to discuss the novel non-equilibrium physics occurring right at the Dirac point. Due to the rich prospective applications of graphene, understanding the current saturation phenomenon at different parameter regimes draws strong theoretical and practical interests. 

\begin{figure}
\centering
\includegraphics[scale=0.8]{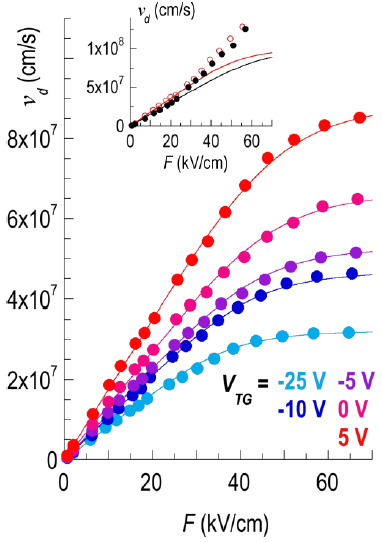}
\caption[Saturation of drift velocity in graphene]{Saturation of drift velocity in graphene. Drift velocity $v_d$ is plotted against external electric field at various gate voltages. The $v_d$ is defined as current density divided by carrier density, which is calculated using a capacitance model. The charge-neutrality point (Dirac point) is close to $V_\text{TG}=16.5V$, which is shown in the inset. The graph is adapted from Ref. \citenum{ramamoorthy15}}
\label{saturation}
\end{figure}

Fig. \ref{saturation} shows how current saturates under strong voltage bias in graphene. Although higher density of current carriers usually implies higher mobilities, the saturated velocity is inversely related to the carrier density. A simple field-effect model is introduced to explain the saturation effect, predicting the drift velocity saturates due to emission of optical phonons by electrons. This model assumes a finite Fermi sea around the Dirac point and an electron is immediately scattered when it is accelerated to reach the optical phonon energy $\hbar\omega_\text{ph}$. This model successfully predicts a scaling law of saturated velocity,
\begin{align}
v_d=\frac{2}{\pi}\frac{\omega_\text{OP}}{\sqrt{\pi n}},
\end{align}
where $n$ is the carrier density. This picture is obviously invalid in the vicinity of the Dirac point. In the case of Dirac electrons, it requires a quantum mechanical model to address the interacting non-equilibrium steady state. However, despite plenty of experimental studies in this subject, a microscopic theory is still lacking. In the last chapter of this thesis, we will discuss the non-equilibrium steady state of graphene under strong electric field. We will discuss the saturation of current and electronic drift velocity.

In the rest of the thesis, we will firstly discuss a non-interacting dissipative lattice model, which is the starting point of describing the non-equilibrium steady state of solids under dc-electric field. Then we examine the RS effect in a uniform strongly correlated system, and then turn to a finite-size sample to study the spatial inhomogeneities during the transitions. In the last chapter, we will examine the strong-field transport of graphene, in particular in the vicinity of the Dirac point.

\chapter{Formulation of Non-equilibrium Dissipative Lattice System}
\label{prb}

\section{Time-dependent theory in temporal gauge}
\subsection{Dissipation in quantum mechanics}
As its name indicates, dissipation causes a system to lose energy and/or information into the surrounding environment. Dissipative effect exists ubiquitously in realistic physical systems. It leads to line-broadening in spectral function as well as decoherence of quantum states, which is critical to achieving central goals of many research and technological fields\cite{QDS}. In non-equilibrium, the dissipative effect has been revealed as a critical mechanism necessary for understanding experimental observations and establishing well-defined non-equilibrium steady state\cite{Tsuji09, RMP-NEQDMFT, eckstein10}.

System-plus-reservoir method has been widely used to describe dissipative systems, where the complete model is divided to relevant part called system and irrelevant environment which is then ``integrated out". Caldeira-Leggett model has been a prototypical system-plus-reservoir model on which many theoretical studies are built\cite{Caldeira-Leggett83}. Here we will discuss a simpler model than Caldeira-Leggett model to demonstrate how dissipation effect can be included in a minimal formulation and its significance in physics.

Consider a free particle $d_k$ of dispersion relation $\epsilon(\boldsymbol{k}) = \epsilon_{\boldsymbol{k}}$ coupling to non-interacting fermion reservoir of orbitals $c_{k\alpha}$. Suppose the reservoir has energy levels of $\epsilon_\alpha$, the hamiltonian is
\begin{align}
H&=\sum_k\epsilon_k d^\dag_k d_k \nonumber\\
&+ \sum_{k\alpha}\epsilon_{\alpha}c^\dag_{k\alpha} c_{k\alpha} - g\sum_{k\alpha}\left(d^\dag_kc_{k\alpha}+H.c.\right)
\label{sys-res}
\end{align}
The coupling constant between particle and reservoir is $g$. The reservoirs are maintained at equilibrium state. Temperature is fixed at $T_\text{bath}$ and chemical potential is $\mu=0$.

This model is block-diagonal in $k$ and for a fixed $k$ it is simply a resonant level model connected to a fermion bath\cite{jauho94}. Dividing hamiltonian \eqref{sys-res} to system(first line) and reservoir parts(second line) allows us to treat system-reservoir coupling as ``interacting" hamiltonian and to apply method of Dyson equation. The ``non-interacting" hamiltonian gives retarded Green functions for individual system and reservoir particle:
\begin{align}
G^{r,0}_{d,\boldsymbol{k}}(\omega)=\frac{1}{\omega-\epsilon_k+i\eta}\nonumber\\
G^{r,0}_{c,\boldsymbol{k}\alpha}(\omega)=\frac{1}{\omega-\epsilon_\alpha+i\eta}\nonumber
\end{align}
Then the full retarded Green function is given by Dyson equation:
\begin{align}
&G^{r}_{d,\boldsymbol{k}}(\omega)^{-1}=G^{r,0}_{d,\boldsymbol{k}}(\omega)^{-1}-\Sigma^r(\omega)\nonumber\\
&\Sigma^r(\omega)=-g^2\sum_{\alpha}\frac{1}{\omega-\epsilon_\alpha+i\eta}
\end{align}
Using $\frac{1}{x+i\eta}=\mathcal{P}\frac{1}{x}-i\pi\delta(x)$, self energy has imaginary part
\begin{align}
\text{Im}\Sigma^r(\omega)=\pi g^2\sum_\alpha\delta(\omega-\epsilon_\alpha)=\pi g^2N(\omega),
\label{bathsret}
\end{align}
with reservoir density of states(DoS) $N(\omega)$. Assuming reservoir has flat energy band where $N(\omega)\approx N(0)$ for relevant energy scale, we define $\Gamma=\pi g^2 N(0)$ and 
\begin{align}
G^r_{d,\boldsymbol{k}}(\omega)=\frac{1}{\omega-\epsilon_k+i\Gamma}
\label{mingr}
\end{align}
A finite spectral width is obtained through mixing with reservoir levels. Fourier-transforming the Green's function to time domain, we get
\begin{align}
G^r_{d,\boldsymbol{k}}(t-t')=-i\theta(t-t')\exp\left(-i\epsilon_kt-\Gamma|t-t'|\right)
\label{gr-minmodel}
\end{align}
This expression is the same as that of free particle besides decaying factor $\exp\left(-\Gamma|t-t'|\right)$. Physically, it indicates the system perturbed at time $t'$ will lose the memory of perturbation in time scale $\Gamma^{-1}$. This is due to the dephasing and energy dissipation effects of fermion reservoir. Note the fermion reservoir resembles a bosonic reservoir, such as phonon bath, as long as ohmic dissipation is assumed, e.g. $\Gamma(\omega)\sim\text{const.}$ or $J(\omega)\sim \omega$ in low energy regime\cite{QDS}. 

As we shall see, excitations created by external field will increase indefinitely in non-dissipative systems, whereas a steady state may be established when energy dissipation effect is included\cite{Tsuji09,eckstein10}. As a result, this effect is critical to reproduce correct long-time behavior of driven non-equilibrium systems. 
\begin{figure}
\centering
\includegraphics{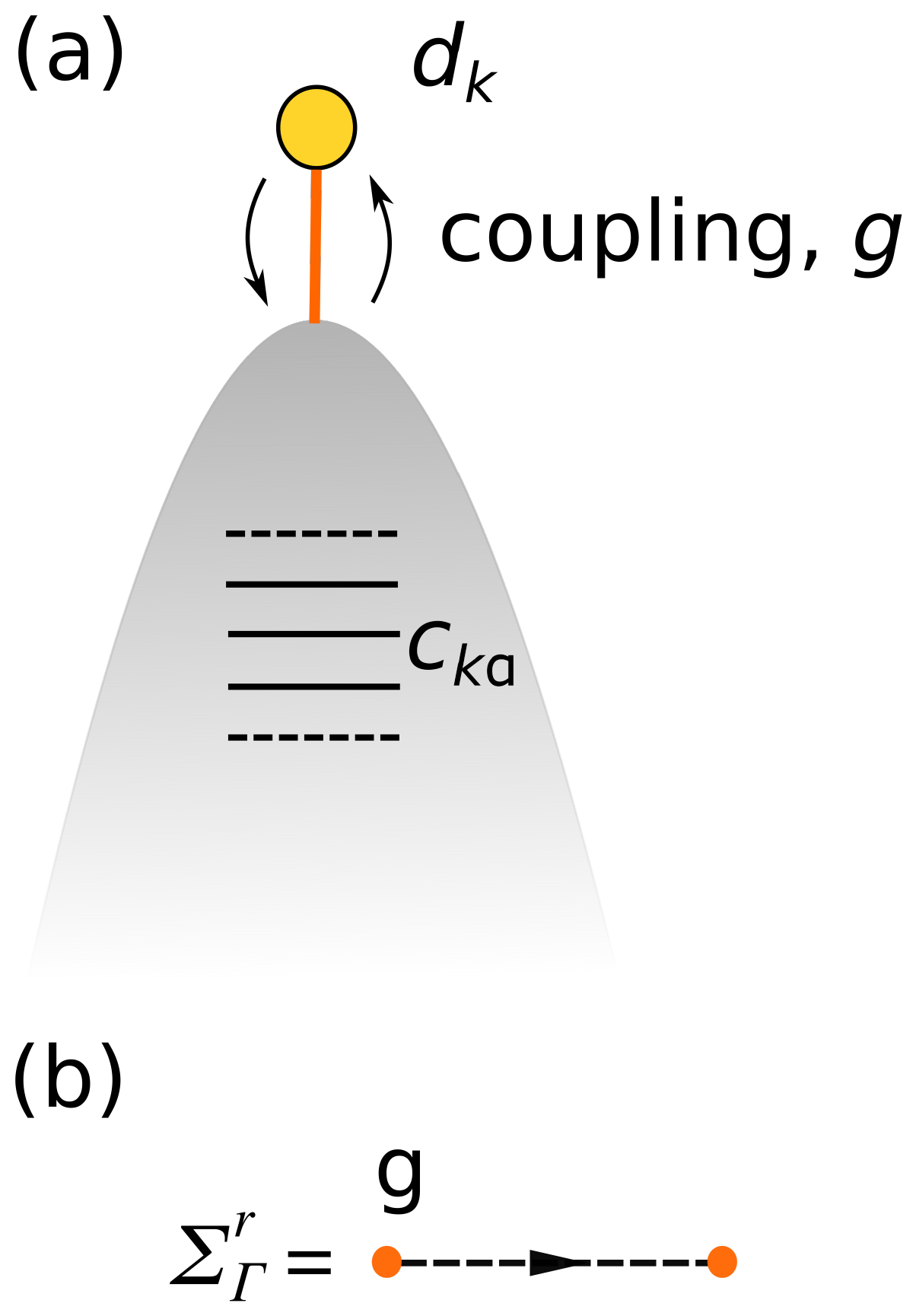}
\caption[System-plus-reservoir model]{(a) minimal system-plus-reservoir model, where system represented by a yellow dot is coupled to the fermion reservoir with coupling constant $g$, and (b) Feynman diagrams used to compute Green's functions. A dashed line represents an electron in the reservoir. The orange dot is the coupling vertex $g$. Since the electronic state in the system uniformly couples to all states in the reservoir, the self energy $\Sigma^r_\Gamma$ is summed over all reservoir states.}
\label{min-model}
\end{figure}

\subsection{Dissipative tight-binding model under electromagnetic fields}

We consider a general tight-binding model where each lattice site $d_\ell$ is connected to a fermion reservoir with orbitals $c_{\ell\alpha}$. Electrons hop between neighboring lattice sites as well as between lattice site and reservoirs.  In general, we may consider an arbitrary electromagnetic field applied in the lattice. The hamiltonian describing this system can be written as follow:
\begin{align}
H&=\gamma\sum_{\langle \ell,\ell'\rangle}\text{e}^{i\varphi(t)}\left(d^\dag_{\ell} d_{\ell'} + H.c.\right) +\nonumber\\
&+\sum_{\ell\alpha}\epsilon_\alpha c^\dag_{\ell\alpha}c_{\ell\alpha}-g\sum_{\ell\alpha} \left(c^\dag_{\ell\alpha}d_{\ell}+H.c.\right)+\nonumber\\
&+\sum_{\ell}\varepsilon_\ell d^\dag_\ell d_\ell,
\label{0inth}
\end{align}
where reservoirs are non-interacting and have energy levels of $\epsilon_\alpha$. The parameter $\gamma$ is wave function overlapping between electrons on neighboring sites, and $g$ is the coupling constant with reservoirs. 

\begin{figure}
\centering
\includegraphics[scale=0.6]{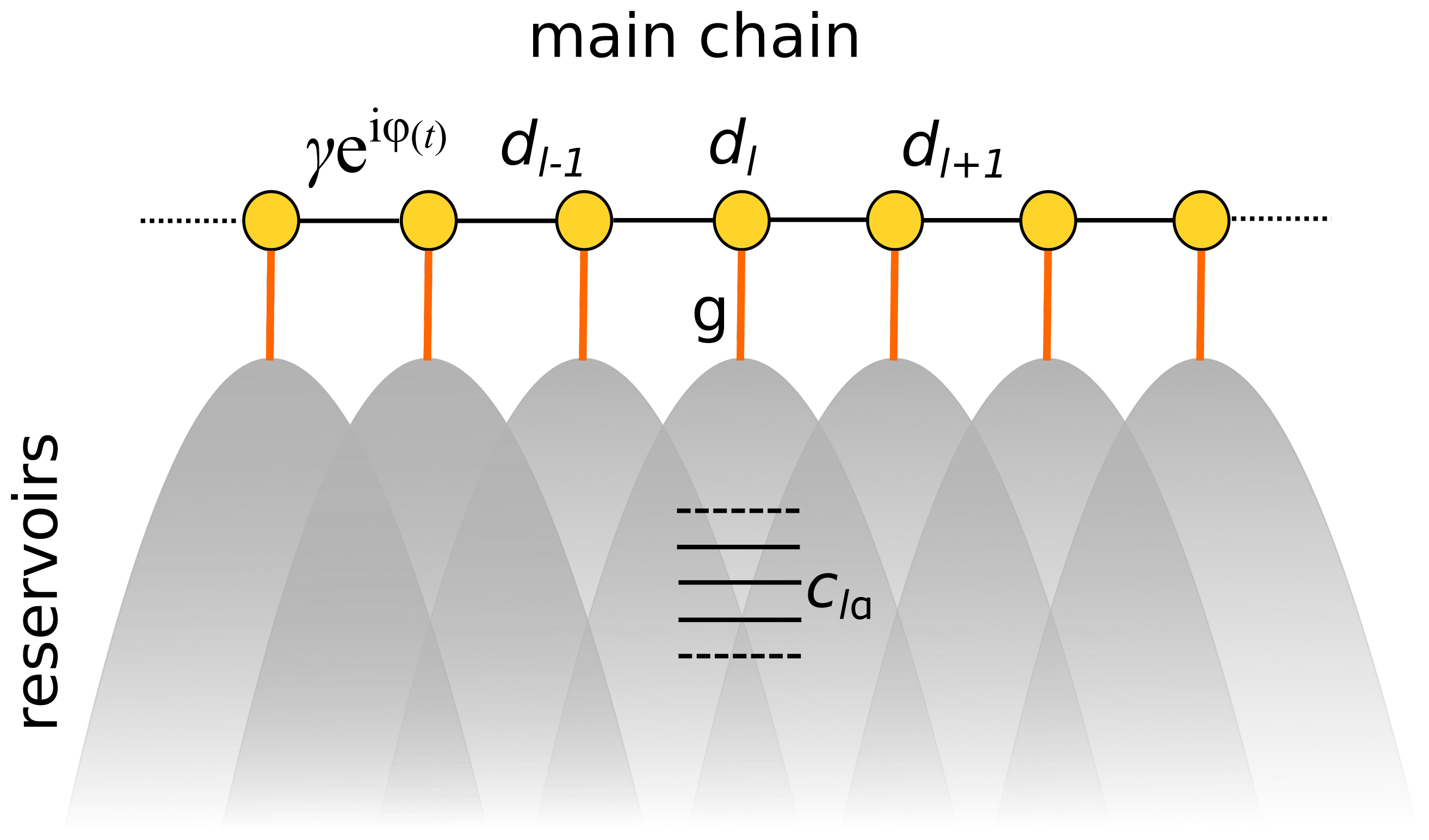}
\caption[Dissipative tight-binding chain]{Tight-binding lattice under external field. Each atom is assumed to have one orbital $d_\ell$ and connected fermion reservoirs with orbital $c_{\ell\alpha}$. Peierls factor $\exp(i\varphi(t))$ is multiplied with hopping parameter $\gamma$. The arbitrary potential $\varepsilon_\ell$ is not shown.}
\label{nointmodel}
\end{figure}

The hamiltonian is gauge-invariant under electromagnetic fields, where $\varepsilon_\ell = \phi_\ell$ is scalar potential and Peierls phase $\varphi(t)=\int_{\boldsymbol{r}_\ell}^{\boldsymbol{r}_{\ell'}} \boldsymbol{A}\cdot d\boldsymbol{s}$ is the line integral of vector potential\cite{Turkowski-Freericks, Jauho-Wilkins, graf-vogl}. This form of hamiltonian is gauge-invariant so that one is free to fix a convenient gauge for certain problem. In particular, people usually adopted temporal gauge, in which $\phi=0$, when dealing with homogeneous electric fields. However, Coulomb gauge, where $\boldsymbol{A}=0$, also has advantages in some circumstances. We will explicitly verify the gauge-invariance and articulate more details in later sections.

\subsection{Reaching non-equilibrium steady state}
We consider the case that a homogeneous electric field is applied in a one-dimensional tight-binding chain\cite{jong-prb}. Suppose homogeneous electric field is applied at $t=-T$, where $T$ is a large positive number, which is treated as infinity practically. For $t<-T$, the electron lattice is in the equilibrium state in contact with fermion reservoirs. After the external field is applied, the system is driven out of equilibrium and evolves to Non-equilibrium Steady State(NESS) through sufficient time of evolution. The reservoirs are maintained equilibrium with temperature $T_\text{b}=0$ and chemical potential $\mu=0$. 

To address the evolution after turning on the bias, it is natural to choose temporal gauge, where $\phi=0$ and $\varphi(t)=eEat\hat{\boldsymbol{x}}$. In the following discussion we will always assume $k_B=\hbar=e=a=1$. We firstly Fourier transform the hamiltonian by defining
\begin{align}
d^\dag_k=\frac{1}{\sqrt{N}}\sum_\ell \text{e}^{ik\ell}d^\dag_\ell,
\label{ftransform}
\end{align}
under which hamiltonian \eqref{0inth} is transformed to momentum representation:
\begin{align}
H&=\sum_k -2\gamma\cos\left(k+\varphi(t)\right)d^\dag_kd_k+\nonumber\\
&+\sum_{k\alpha}\epsilon_\alpha c^\dag_{k\alpha}c_{k\alpha}-\frac{g}{\sqrt{V}}\sum_{k\alpha} \left(c^\dag_{k\alpha}d_{k}+H.c.\right)
\label{0inthk}
\end{align}
we divide the hamiltonian \eqref{0inthk} into time-independent part $H_0=H(t=0)$ and time-dependent part $H_1=H(t)-H_0$:
\begin{align}
H_1=-2\gamma\sum_k \left(\cos\left(k+\varphi(t)\right)-\cos(k)\right)d^\dag_k d_k\equiv v(t)d^\dag_k d_k
\end{align}

This block-diagonal hamiltonian is nothing but above-mentioned resonant level model, with oscillating level energy. The corresponding Dyson equation becomes:
\begin{align}
\mathbf{G}^r_k&=\mathbf{G}^{r,0}_k+\mathbf{G}^{r,0}_k\mathbf{V}\mathbf{G}^{r}_k\label{dyson1:1}\\
\mathbf{G}^<_k&=[I+\boldsymbol{G}^{r}_k\mathbf{V}]\mathbf{G}^{<,0}_{k}[I+\mathbf{V}\mathbf{G}^{a}_k],
\label{dyson1:2}
\end{align}
in which $\mathbf{V}_{t,t'}=v(t)\delta(t-t')$ in time representation. The matrix multiplication in the above equations should be understood as convolutional integration in time variables. When steady state is considered, the oscillating term $\mathbf{V}$ is turned on at $t=-\infty$ therefore time integration is from $-\infty$ to $\infty$.

Following Ref. \citenum{jong-prb}, the retarded Green function is computed as:
\begin{align}
G^r_k(t,t')=-i\theta(t-t')\text{e}^{-\Gamma|t-t'|+2\gamma i\int_{t'}^t \cos(k+Es)ds}
\end{align}
Flat band is assumed and $\Gamma=\pi g^2N(0)$ is defined as in \eqref{bathsret}. This form is almost the same as \eqref{gr-minmodel}, besides a dynamical phase due to oscillating energy $v(t)$. As a result, excitations created by external field are constantly dissipated through fermion reservoirs. Physically, the system will have ``memory" limited to time scale $\Gamma^{-1}$. This mimics electron-impurity scattering in realistic system where electrons are described to become thermalized after scattering time $\tau^{-1}$  by semiclassical transport theories. It helps as well maintain a steady-state in which electric power is balanced with energy flux into reservoirs. Although no scattering really happens in this model, we can then identify the effective scattering time $\tau_\Gamma\sim \Gamma^{-1}$. In the later sections we will discuss in detail the electronic transport in this model, but now we would like to concentrate on clarifying the structure of Green's functions.

To obtain a gauge-invariant Green's function, we sum over all momenta resulting in local Green's function:
\begin{align}
G^r_\text{loc}(t-t')&=\frac{1}{2\pi}\int_{-\pi}^\pi dkG^r_k(t,t')\nonumber\\
&=-i\theta(t-t')\text{e}^{-\Gamma|t-t'|}J_0\left(\frac{4\gamma}{E}\sin\frac{E(t-t')}{2}\right),
\label{gloc0}
\end{align}
by noticing 0th Bessel function $J_0(x)=\int_{-\pi}^\pi dt\exp\left(-ix\sin(t)\right)$. This is a concise expression which is independent of gauge choices, thus could be compared with results obtained in other gauges later. As it is now a function of $(t-t')$ , we can Fourier transform it to frequency domain:
\begin{align}
G^r_\text{loc}(\omega)=\sum_{\ell=-\infty}^{\infty}\frac{J_\ell\left(\frac{\gamma}{2E}\right)^2}{\omega+\ell E+i\Gamma}
\label{gloc}
\end{align}
$G^<_k(t,t')$ can be computed with Dyson equation \eqref{dyson1:2}. Noticing \eqref{dyson1:1} amounts to $(\mathbf{G}^r_k)^{-1}=(\mathbf{G}^{r,0}_k)^{-1}-\mathbf{V}$, we have
\begin{align}
\mathbf{G}^<_k&=[I+\mathbf{G}^{r}_k\mathbf{V}]\mathbf{G}^<_{k,0}[I+\mathbf{V}\mathbf{G}^{a}_k]\nonumber\\
&=\mathbf{G}^{r}_k[(\mathbf{G}^{r}_k)^{-1}+\boldsymbol{V}]\mathbf{G}^{<,0}_{k}[(\mathbf{G}^{a}_k)^{-1}+\mathbf{V}]\mathbf{G}^{a}_k\nonumber\\
&=\mathbf{G}^{r}_k\left(\left(\mathbf{G}^{r,0}_k\right)^{-1}\mathbf{G}^{<,0}_{k}\left(\mathbf{G}^{a,0}_k\right)^{-1}\right)\mathbf{G}^{a}_k\nonumber\\
&=\mathbf{G}^{r}_k\mathbf{\Sigma}^<_\Gamma\mathbf{G}^{a}_k,
\end{align}
where $\Sigma_\Gamma^<(t-t')=\int d\omega\frac{i\Gamma}{\pi}f_\text{FD}(\omega)\text{e}^{-i\omega(t-t')}$ is nothing but equilibrium lesser self energy. Note $f_\text{FD}(\omega)=1/(\text{e}^{\beta\omega}+1)$ is Fermi-Dirac distribution. The same result has been worked out explicitly in Ref. \citenum{jong-prb}.
Local lesser Green's function is similarly defined and can be found as 
\begin{align}
G^<_\text{loc}(\omega)&=\frac{i\Gamma}{\pi}\int^\infty_{-\infty}dt\int^\infty_{-\infty}d\omega'f_\text{FD}(\omega')\int^0_{-\infty}ds\int^0_{-\infty}ds'\nonumber\\
&\times \text{e}^{i(\omega-\omega')t-i\omega(s-s')+\Gamma(s+s')}J_0\left(\frac{4\gamma}{E}\sqrt{R}\right)
\end{align}
with
\begin{align}
R=\sin^2\frac{Es}{2}\sin^2\frac{Es'}{2}-2\cos\left[E\left(t+\frac{s-s'}{2}\right)\right]\sin\frac{Es}{2}\sin\frac{Es'}{2}
\end{align}
This is a common form in trigonometry $R=a^2+b^2-2ab\cos(\alpha)$, and it is known that $J_0(\sqrt{R})=\sum_\ell J_\ell(a)J_\ell(b) \text{e}^{i\ell\alpha}$. The Fourier transformed $G^<$ is obtained as
\begin{align}
G^<_\text{loc}(\omega)=2i\Gamma\sum_\ell f_\text{FD}(\omega+\ell E)\left|\sum_m\frac{J_m\left(\frac{2\gamma}{E}\right){J_{m-\ell}\left(\frac{2\gamma}{E}\right)}}{\omega+m E +i\Gamma}\right|^2
\label{glocl}
\end{align}
This result, together with \eqref{gloc}, suggests that non-equilibrium Green's functions in the dissipative lattice model can be written as summation over energy states with energies $-\ell E$. This is exactly the potential energy of lattice site $\ell$ in \emph{Coulomb gauge} hamiltonian. This observation inspires us to solve the same problem in Coulomb gauge in order to fully understand the relations among Green's functions. Another advantage of Coulomb gauge is that hamiltonian becomes time-independent for dc-field, and steady-state Green's functions can be diagonalized in frequency domain, which dramatically simplifies Dyson equations. In the following sections, we will discuss scattering-state formalism and how Green's functions can be derived straightforwardly.

\section{Scattering theory formalism}
The following gauge transformation can be applied to hamiltonian \eqref{0inth},
\begin{align}
d_\ell\to\text{e}^{-\ell Et}d_\ell,c_{\ell\alpha}\to\text{e}^{-\ell Et}c_{\ell\alpha}
\label{gauge}
\end{align}
It transforms the hamiltonian to Coulomb gauge where $\phi_\ell=-\ell E,\boldsymbol{A}=0$. The Coulomb-gauge hamiltonian is:
\begin{align}
H_\text{Coul}&=-\gamma\sum_\ell(d^\dag_{\ell+1}d_\ell+H.c.)-\sum_\ell \ell Ed_\ell^\dag d_\ell\nonumber\\
&+\sum_{\ell\alpha}(\epsilon_\alpha-\ell E)c_{\ell\alpha}^\dag c_{\ell\alpha}-g\sum_{\ell\alpha}(c^\dag_{\ell\alpha}d_\ell+H.c.)
\label{0inthc}
\end{align}

\begin{figure}
\centering
\includegraphics[scale=0.9]{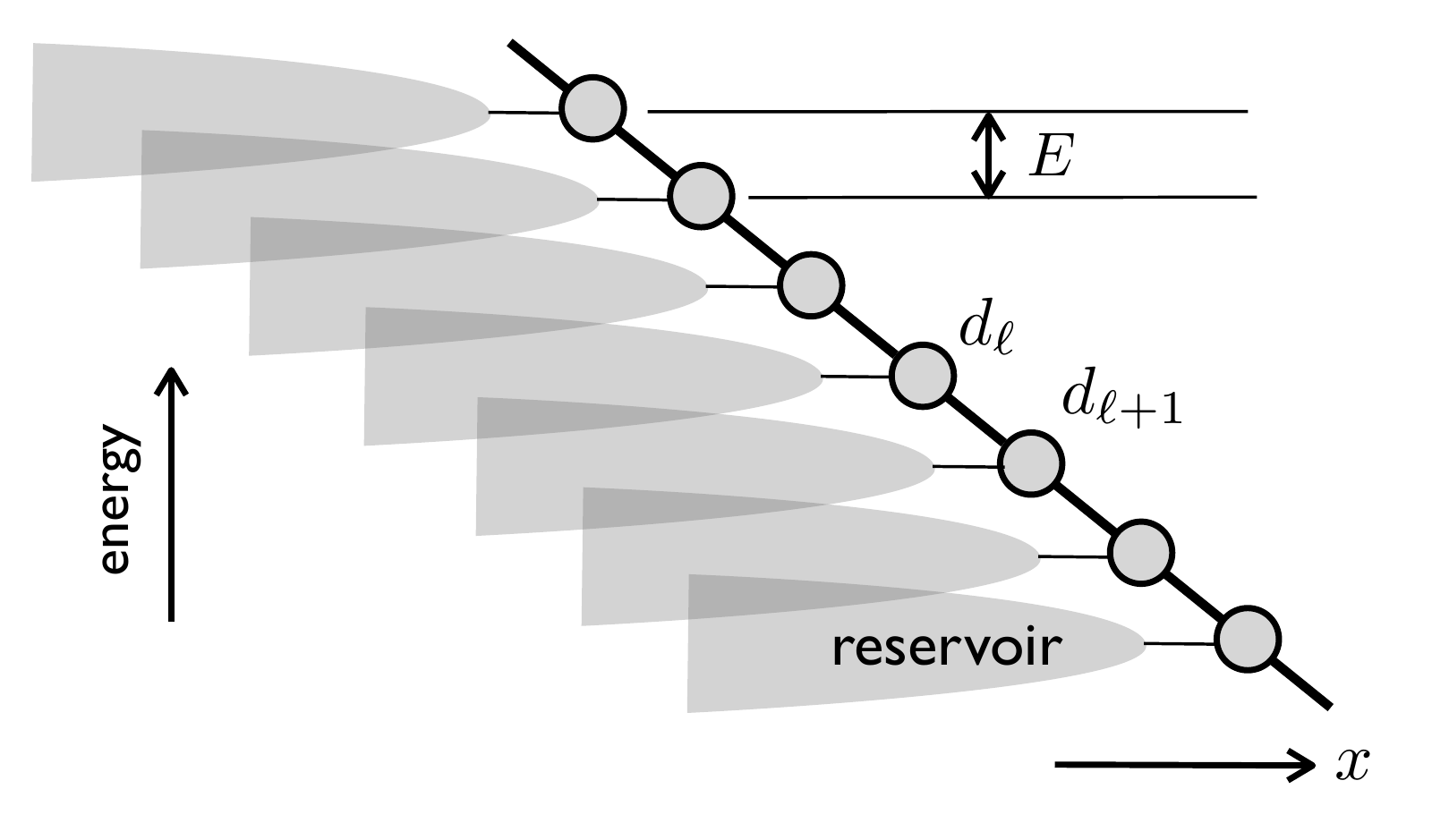}
\caption[Dissipative lattice model under external field]{Dissipative lattice model under electric field in Coulomb gauge. On-site energy is shifted by $-\ell E$ due to electric potential. The Peierls phase $\varphi(t)=0$, therefore the hamiltonian is time-independent.}
\label{coulomb-lattice}
\end{figure}

This hamiltonian is quadratic, and can be analytically diagonalized. Diagonalization can be done by introducing scattering state operators $\psi_{\ell\alpha}$ that satisfy $\{\psi_{\ell\alpha},\psi^\dag_{\ell'\alpha'}\}=\delta_{\ell\ell'}\delta_{\alpha\alpha'}$ and:
\begin{align}
[\psi_{\ell\alpha},H_\text{Coul}]&=(\epsilon_\alpha-\ell E)\psi_{\ell\alpha}\nonumber\\
[\psi_{\ell\alpha}^\dag,H_\text{Coul}]&=-(\epsilon_\alpha-\ell E)\psi^\dag_{\ell\alpha}
\end{align}
To reveal the relation between scattering state operators and original fermion operators, we note
\begin{align}
[H_\text{Coul},\psi_{\ell\alpha}^\dag]-(\epsilon_\alpha-\ell E)\psi^\dag_{\ell\alpha}=0=[H_\text{Coul}-H_g,c^\dag_{\ell\alpha}]-(\epsilon_\alpha-\ell E)c^\dag_{\ell\alpha},
\end{align}
with coupling term $H_g=-g\sum_{\ell\alpha}(c^\dag_{\ell\alpha}d_\ell+H.c.)$. If $g$ vanishes then $\psi_{\ell\alpha}=c_{\ell\alpha}$, and the formalism is invalid as the scattering states become irrelevant. This identity is nothing but Lippmann-Schwinger equation\cite{Gellmann-Goldberger} written with operators:
\begin{align}
\psi_{\ell\alpha}^\dag&=c^\dag_{\ell\alpha}+\frac{1}{\epsilon_\alpha-\ell E-\mathcal{L}+i\eta}[H_g,c^\dag_{\ell\alpha}]\nonumber\\
&=c^\dag_{\ell\alpha}-g\frac{1}{\epsilon_\alpha-\ell E-\mathcal{L}+i\eta}d^\dag_\ell,
\label{LSE}
\end{align}
where $\mathcal{L}A\equiv[H_\text{Coul}, A]$ is the Liouville operator. In terms of scattering state operators, the hamiltonian is written as
\begin{align}
H_\text{Coul}=\sum_{\ell\alpha}(\epsilon_\alpha-\ell E)\psi^\dag_{\ell\alpha}\psi_{\ell\alpha}.
\end{align}
In addition, reservoirs are maintained equilibrium:
\begin{align}
\langle\psi_{\ell\alpha}^\dag\psi_{\ell\alpha}\rangle=f_\text{FD}(\epsilon_\alpha)
\end{align}
With quadratic hamiltonian, which can be diagonalized with unitary transformation, the scattering state operators should be expanded as linear combination of original fermion operators.
\begin{align}
\psi_{\ell\alpha}^\dag=c^\dag_{\ell\alpha}+\sum_{\ell'}d_{\ell'}^\dag C_{\ell\alpha}(d_{\ell'})+\sum_{\ell'\alpha'}c^\dag_{\ell'\alpha'}C_{\ell}\alpha(c_{ell'\alpha'})
\end{align}
Based on Lippmann-Schwinger equation \eqref{LSE}, coefficients are computed using the canonical (anti-)commutation relation:
\begin{align}
C_{\ell\alpha}(A)=\{A,\psi^\dag_{\ell\alpha}-c^\dag_{\ell\alpha}\}=\left\{A,-g\frac{1}{\epsilon_\alpha-\ell E-\mathcal{L}+i\eta}d^\dag_l\right\}
\end{align}
These anti-commutators must be c-numbers for quadratic hamiltonian. In particular, when $A=d_{\ell'}$ or $A=c_{\ell'\alpha'}$, the $C_{\ell\alpha}(A)$'s are correspondingly retarded Green's functions:
\begin{align}
C_{\ell\alpha}(d_{\ell'})&=-g\overline{G}^r_{\ell'\ell}(\epsilon_\alpha-\ell E)\nonumber\\
C_{\ell\alpha}(c_{\ell'\alpha'})&=\frac{g^2}{\epsilon_\alpha-\ell E-\epsilon_{\alpha'}+i\eta}\overline{G}^r_{\ell'\ell}(\epsilon_\alpha-\ell E)=-g\overline{G}^r_{c_{\ell'\alpha'},d_\ell}(\epsilon_\alpha-\ell E),
\end{align}
where $\overline{G}$ is used to denote Coulomb-gauge Green's functions. We then have
\begin{align}
\psi_{\ell\alpha}^\dag=c^\dag_{\ell\alpha}-g\sum_{\ell'}\overline{G}^r_{\ell'\ell}(\epsilon_\alpha-\ell E)d^\dag_{\ell'}+\cdots
\label{scattering-state}
\end{align}
Similar expressions are derived and discussed in quantum dot systems\cite{jong-prb07,jong-prb06}. Up to this point, we have finished diagonalizing hamiltonian in Coulomb gauge by constructing explicitly a complete set of operators which create all energy eigenstates. The next step would be to compute Green's functions and interesting physical quantities with the assistance of our formulation.

\subsection{Green's functions in terms of scattering states}
Diagonalizing the hamiltonian with Scattering state operators assist to compute all Green's functions in the non-equilibrium state. To achieve this goal, we have to express the relevant operators, including all $d_\ell, c_{\ell\alpha}$, in terms of scattering-state operators. Then with the equilibrium-reservoir conditions $\langle\psi^\dag_{\ell\alpha}\psi_{\ell\alpha}\rangle=f(\epsilon_\alpha)$, the lesser/greater Green's functions can be readily related to retarded Greens' functions. To make it concrete, the equation \eqref{scattering-state} can be inverted to obtain
\begin{align}
d_\ell=\sum_{\ell'\alpha'}\tilde{C}_\ell(\ell'\alpha')\psi_{\ell'\alpha'}.
\end{align}
In fact, $\tilde{C}_\ell(\ell'\alpha')=\{\psi_{\ell'\alpha'}^\dag,d_\ell\}$ so that
\begin{align}
d_\ell=-g\sum_{\ell'\alpha'}\overline{G}^r_{\ell\ell'}(\epsilon_\alpha'-\ell'E)\psi_{\ell'\alpha'}.
\label{re-sc}
\end{align}
Although retarded Green's functions appear in this equation so we cannot directly obtain a closed form of them, we do get a self-consistent condition of them by inserting \eqref{re-sc} in the definition of Green's functions:
\begin{align}
\overline{G}^r_{\ell\ell'}(\omega)&=g^2\sum_{m\alpha,m'\alpha'}\frac{\overline{G}_{\ell m}(\epsilon_\alpha-mE)\left[\overline{G}_{\ell' m'}(\epsilon_\alpha'-m'E)\right]^*}{\omega-\epsilon_\alpha+\ell E+i\eta}\nonumber\\
&\times \langle \psi_{m\alpha},\psi^\dag_{m'\alpha'}\rangle\nonumber\\
&=\frac{\Gamma}{\pi}\sum_m\int{d\omega'\frac{\overline{G}_{\ell m}(\omega')\left[\overline{G}_{\ell' m}(\omega')\right]^*}{\omega-\omega'+i\eta}}.
\label{selfc}
\end{align}
When $\ell=\ell'$, this immediately leads to
\begin{align}
\text{Im}\overline{G}^r_{\ell\ell}(\omega)=-\Gamma\sum_m\left|\overline{G}^r_{\ell m}(\omega)\right|^2,
\label{imG}
\end{align}
which is a useful identity. On the other hand, the lesser Green's functions are computed as
\begin{align}
\overline{G}^<_{\ell\ell'}(\omega)&=2\pi ig^2\sum_{m\alpha}\overline{G}^r_{\ell m}(\epsilon_\alpha-mE)\left[\overline{G}^r_{\ell'm}(\epsilon_\alpha-mE)\right]^*\delta(\omega-\epsilon_\alpha+mE)\langle \psi^\dag_{m\alpha}\psi_{m\alpha}\rangle\nonumber\\
&=2i\Gamma\sum_m\overline{G}^r_{\ell m}(\omega)\left[\overline{G}^r_{\ell'm}(\omega)\right]^*f_\text{FD}(\omega+m E).
\label{glss}
\end{align}
This is a transparent relation between retarded and lesser Green's functions. Since infinite flat band is assumed, all reservoirs connected to all lattice sites would contribute to the electron statistics at each site, through quantum correlation given by off-site $G^r_{\ell m}(\omega)$. The lesser Green's function $G^<$'s essentially provide all information about electron distribution in non-equilibrium.

Now to complete our discussion, we need to explicitly compute retarded Green's functions. That amounts to inverting the matrix:
\begin{align}
\left(\boldsymbol{\overline{G}}^r(\omega)\right)^{-1}_{\ell\ell'}=(\omega+\ell E+i\Gamma)\delta_{\ell\ell'}+\gamma\delta_{\langle \ell,\ell'\rangle},
\end{align}
with $-i\Gamma$ being retarded self energy and $-\ell E$ being potential energy of site $\ell$. The solution is found to be 
\begin{align}
\overline{G}^r_{\ell\ell'}(\omega)=\sum_m\frac{J_{\ell-m}\left(\frac{2\gamma}{E}\right)J_{\ell'-m}\left(\frac{2\gamma}{E}\right)}{\omega+mE+i\Gamma},
\label{gret}
\end{align}
which can be verified straightforwardly. When $\ell=\ell'$, these results are identical with \eqref{gloc}, \eqref{glocl}. In addition, we can readily confirm the self-consistency condition \eqref{selfc} by inserting the explicit form of $G^r$ and using residue theorem.

So far we have finished computing all relevant Green's functions in Coulomb gauge. We are to discuss their mathematical properties and then move on to transport theory in the following sections. We will drop the overbar of Coulomb-gauge Green's functions for simplicity, and all Green's functions, unless stated otherwise, should be understood as computed in Coulomb-gauge.

\subsection{Properties of Green's functions}
The first obvious observation, is that Green's functions in Coulomb gauge are time-translational invariant. This justifies the frequency representation we have adopted. It follows from Eq. \eqref{gret} that
\begin{align}
G^r_{\ell+m,\ell'+m}(\omega)=G^r_{\ell,\ell'}(\omega+E),
\end{align}
and therefore,
\begin{align}
G^<_{\ell+m,\ell'+m}(\omega)&=2i\Gamma\sum_n\overline{G}^r_{\ell+m, n}(\omega)\left[\overline{G}^r_{\ell'+m,n}(\omega)\right]^*f_\text{FD}(\omega+n E)\nonumber\\
&=2i\Gamma\sum_n\overline{G}^r_{\ell, n-m}(\omega+mE)\left[\overline{G}^r_{\ell',n-m}(\omega+mE)\right]^*f_\text{FD}(\omega+n E)\nonumber\\
&=2i\Gamma\sum_n\overline{G}^r_{\ell, n}(\omega+mE)\left[\overline{G}^r_{\ell',n}(\omega+mE)\right]^*f_\text{FD}\left(\omega+mE+nE\right)\nonumber\\
&=G^<_{\ell\ell'}(\omega+mE).
\end{align}
These results can also be derived from gauge transformation \eqref{gauge}. When interaction is considered, these identities lead to the same properties for self energies:
\begin{align}
\Sigma^{r,<}_{\ell+m,\ell'+m}(\omega)=\Sigma^{r,<}_{\ell\ell'}(\omega+mE),
\label{transprop}
\end{align}
which are physically expected since the potential slope shifts local spectrum by energy difference $mE$ for two lattice sites separated by $m$ lattice constants. 

\subsection{Electronic transport}
In the regime of weak field, electronic transport is well-documented and explained satisfactorily with semiclassical theories. The simplest among them is Drude theory, which is valid in linear response regime and has usually been the starting point of more sophisticated theoretical studies. In Drude theory, current carriers are accelerated by the external electric field, and is repeatedly scattered and thermalized. Current is given with a linear relation with respect to external field\cite{ashcroft78}:
\begin{align}
\boldsymbol{J}=\left(\frac{nq^2\tau}{m}\right)\boldsymbol{E},
\label{drude}
\end{align} 
with $n,q,m$ the concentration, charge and mass of current carriers. The scattering time $\tau$ is the average time duration that a current carrier is scattered. Despite its oversimplification, Drude theory justifies the Ohm's law with a microscopic model and provides fundamental intuitions for understanding linear transport behavior in solids.

A more quantitative and systematic way to address electronic transport in solids is Boltzmann transport equation(BTE)\cite{ashcroft78}. Boltzmann equation is also based on the semiclassical theory of electrons. Unlike Drude theory, BTE provides detailed information in the distribution $f(\boldsymbol{r},\boldsymbol{p},t)$ of electrons in both real and momentum spaces.
\begin{align}
\frac{\partial f}{\partial t}+\frac{\boldsymbol{p}}{m}\cdot\nabla f+q\boldsymbol{E}\cdot\frac{\partial f}{\partial \boldsymbol{p}}=\left(\frac{\partial f}{\partial t}\right)_\text{col},
\label{bte}
\end{align}
where $\left(\frac{\partial f}{\partial t}\right)_\text{col}$ is the change of distribution due to scatterings. In the relaxation time approximation, this term is approximated to be $(f_0-f)/\tau$, with $f_0$ being equilibrium momentum distribution and $\tau$ being the scattering time. Assuming steady state and homogeneity in space, i.e. $\partial f/\partial \boldsymbol{t}=\partial f/\partial \boldsymbol{p}=0$ , one may establish that $f=f_0(\boldsymbol{p}-q\boldsymbol{E}\tau)$ in linear response regime where $p$ is generally much greater than $qE\tau$. When electrons of quadratic dispersion relation $E=p^2/2m$ is considered, the equilibrium distribution is just a Fermi sphere centered at $\boldsymbol{p}=0$, which is then displaced in non-equilibrium by $q\boldsymbol{E}\tau$. This picture can be easily generalized to arbitrary dimension and shape of Fermi surface. Under this approximation, we still have Eq. \eqref{drude} in one dimension, which usually only differs by a factor from more complicated cases, such as two/three-dimensional systems.

Now we turn to our dissipative lattice model. We firstly consider the momentum distribution $n_k$, which is the electron number at momentum $k$. This can be computed with Fourier transformation with respect to spatial index $n_k=\frac{1}{N}\sum_\ell \text{e}^{ik\ell}\langle d^\dag_\ell d_0\rangle$, and the correlation function $\langle d^\dag_\ell d_0\rangle$ is nothing but $-iG^<_{0\ell}(t,t)=-iG^<_{0\ell}(0,0)$. It can be proven that 
\begin{align}
-i\sum_{\ell}\text{e}^{ik\ell}G^<_{0\ell}(0,0)&=-i\int_0^\infty \frac{d\omega}{2\pi}\sum_{\ell}\text{e}^{ik\ell}G^<_{0\ell}(\omega)\nonumber\\
&=\frac{\Gamma}{\pi}\int d\omega\left|G^r_{k}(\omega)\right|^2f_\text{FD}(\omega),
\end{align}
with $G^r_{k}(\omega)=\sum_{\ell}\text{e}^{ik\ell}G^r_{0\ell}(\omega)$. This can be readily computed with the explicit expression of retarded Green's functions. The analytic formula of $n_k$ is computed as\cite{jong-prb}
\begin{align}
n_k=\frac{\Gamma}{\pi}\sum_{nm}\frac{J_n\left(\frac{2\gamma}{E}\right)J_m\left(\frac{2\gamma}{E}\right)\text{e}^{ik(m-n)}}{-(m-n)E+2i\Gamma}\left[\frac{1}{2}\log\frac{m^2E^2+\Gamma^2}{n^2E^2+\Gamma}+i\chi_{mn}\right],
\end{align}
with
\begin{align}
\chi_{mn}=\pi+\tan^{-1}\frac{mE}{\Gamma}+\tan^{-1}\frac{nE}{\Gamma}.
\end{align}

\begin{figure}
\centering
\includegraphics[scale=1.1]{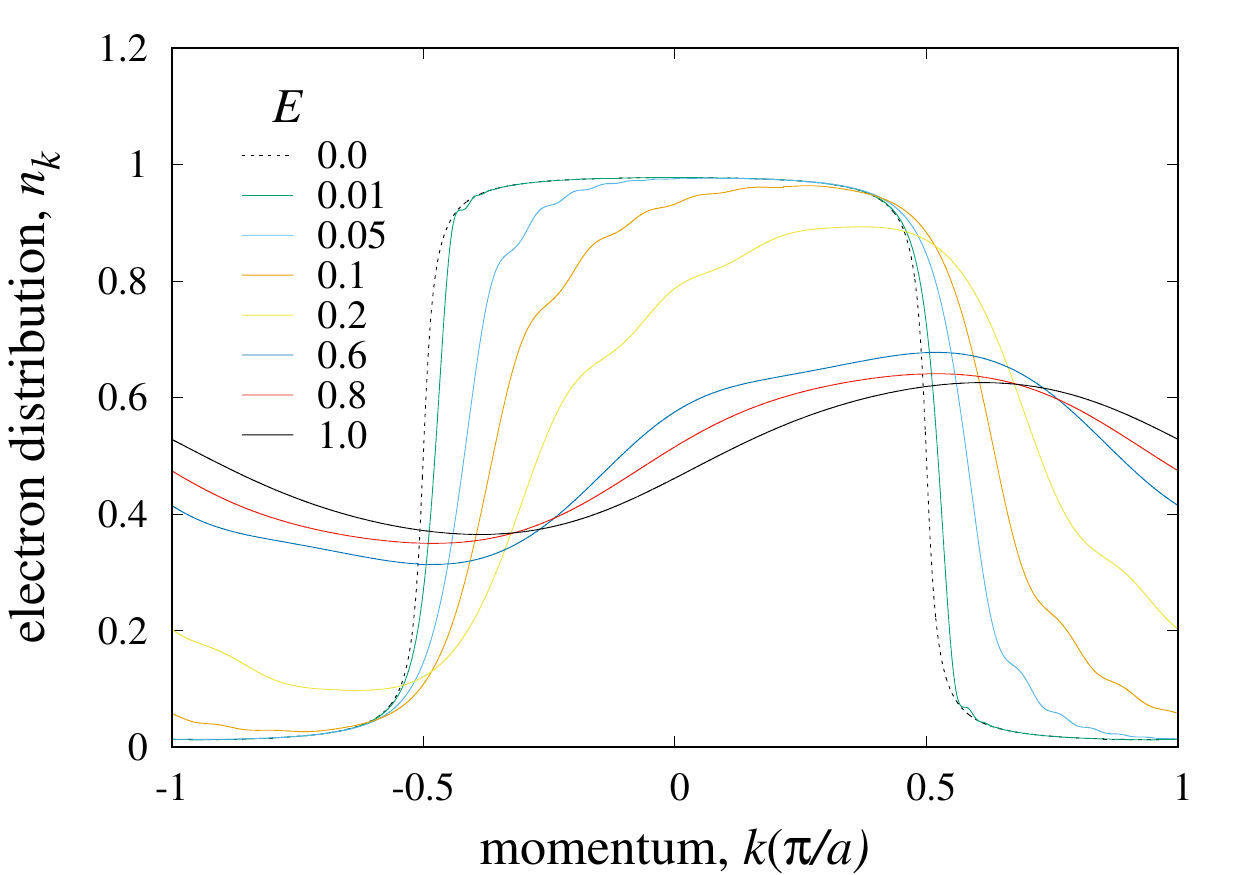}
\caption[Evolution of momentum distribution under increasing electric fields]{Evolution of momentum distribution under increasing electric fields. At zero-field ($E=0$), the electronic distribution in momentum space shows the feature of a Fermi sea with smooth steps due to damping $\Gamma$. As the field increases, the Fermi sea is shifted as expected from Boltzmann transport theory. Finally when $E\gg\Gamma$, the effective temperature becomes high and electrons are spreading all across the First Brillouin Zone.}
\label{nk}
\end{figure}

In the Fig. \ref{nk}, momentum distribution $n_k$ is plotted, showing the picture of displaced Fermi sea for small electric field. Despite the lack of explicit momentum scattering process, the fermion reservoirs do provide the key mechanism by dephasing the electron wavefunction and absorbing excess energy due to constant electric power. It can be shown that the shift of Fermi sea $\delta k\propto E/\Gamma$ as expected in the Boltzmann transport theory. For large electric field $E\gtrsim\Gamma$, the Fermi sea shift deviates from linear relation and the sharp distribution gradually becomes smeared. This suggests a thermal effect due to Joule heating.

\begin{figure}
\centering
\includegraphics[scale=1.1]{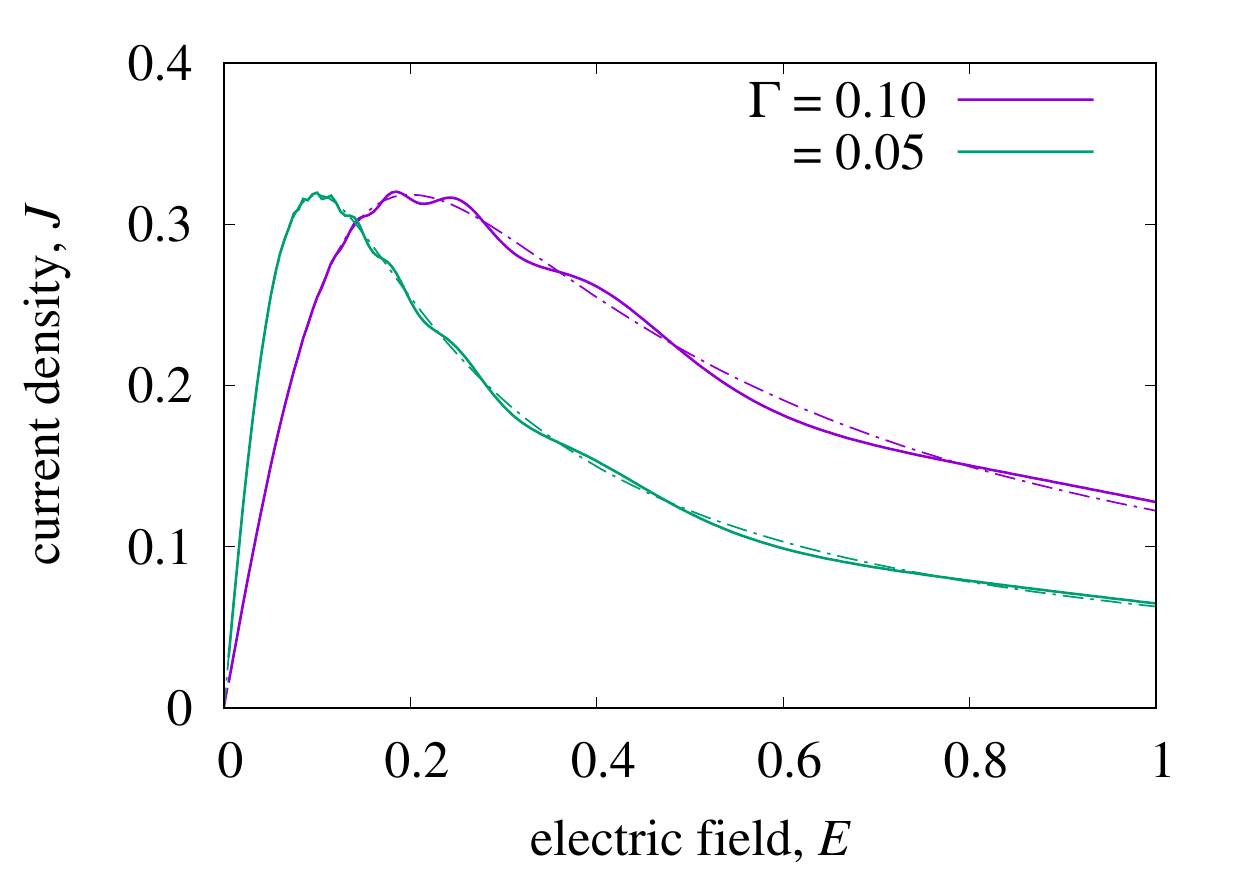}
\caption[DC current versus electric field. ]{DC current versus electric field. For small field, the current has a linear regime with conductivity depending on $\Gamma$. This ohmic behavior is consistent with the physical consequences of bosonic reservoirs. As $E$ increases, Bloch oscillation starts to take effect, and current is reduced. The dashed line represents the approximate formula \eqref{approxcurr}.}
\label{1dcurr}
\end{figure}

To further justify the physical relevance of our model, we calculate the $I-V$ relation. The current can be expressed in terms of lesser Green's function
\begin{align}
J=i\gamma\langle d^\dag_{\ell+1}d_\ell-H.c.\rangle=2\gamma\text{Re}G^<_{\ell,\ell+1}(t,t),
\end{align}
with arbitrary $\ell$. Setting $\ell=0$, the current can be carried out explicitly with the Green's functions:
\begin{align}
J=&\frac{2\gamma\Gamma}{\pi(E^2+4\Gamma^2)}\sum_m J_m\left(\frac{2\gamma}{E}\right)J_{m-1}\left(\frac{2\gamma}{E}\right)\nonumber\\
&\times\left[\Gamma\log\frac{m^2E^2+\Gamma^2}{(m-1)^2E^2+\Gamma^2}+E\chi_{m,m-1}\right].
\label{current}
\end{align}
The current-field relation is plotted in Fig. \ref{1dcurr}. The current follows linear relation in the regime $E\lesssim\Gamma$ and reaches maximum at around $E=2\Gamma$. The current decays slowly for larger electric field. The decaying current is attributed to Bloch oscillation at large electric field. It can also be viewed as a reflection of almost equally occupied distribution at large field in Fig. \ref{nk}.

The expression of current \eqref{current} can be simplified\cite{jong-prb} in the limit of $E,\Gamma\ll\gamma$:
\begin{align}
J\approx\frac{4\gamma\Gamma E}{\pi(E^2+4\Gamma^2)}.
\label{approxcurr}
\end{align}
This expression shows good accuracy for a wide range of parameters, shown as dashed line in the Fig. \ref{1dcurr}. It is worth noting that a similar expression has been found with Boltzmann transport theory\cite{lebwohl70}. And in weak field limit $E\ll\Gamma$, the formula \eqref{approxcurr} reduces to the form of Drude formula,
\begin{align}
J\approx\frac{\gamma E}{\pi\Gamma}\sim\frac{E\tau}{m^*},
\end{align}
where effective mass $m^*\sim 1/\gamma$ and scattering time $\tau\sim1/\Gamma$.

\section{Evolution of wave packet}
\begin{figure}
\centering
\includegraphics[scale=0.4]{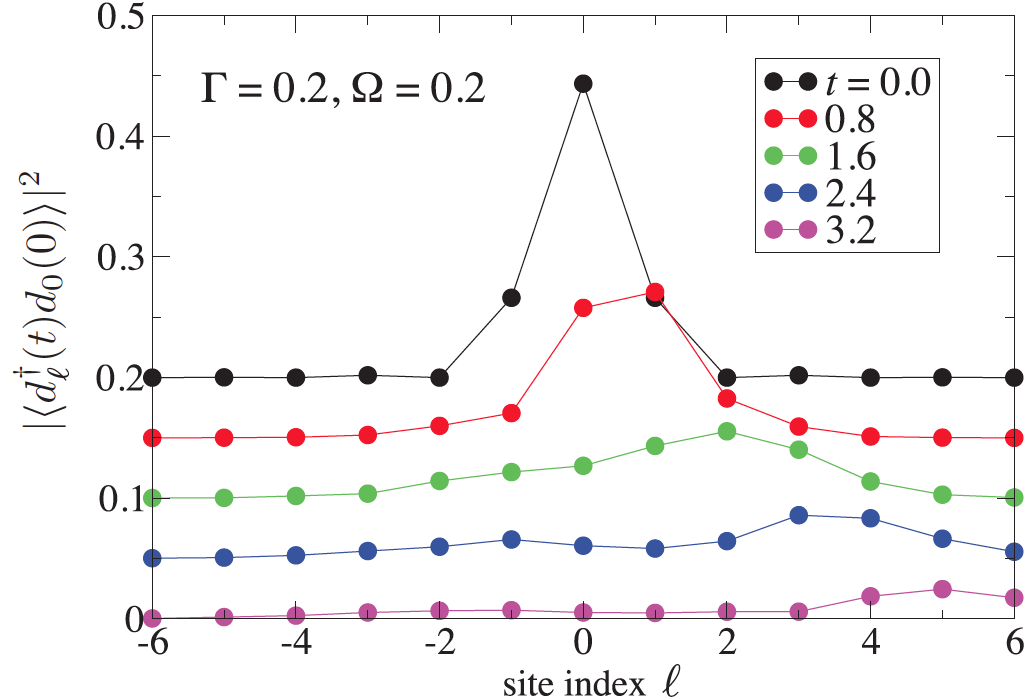}
\caption[Time evolution of a wave packet.]{Time evolution of a wave-packet in the non-equilibrium steady state. The (hole) wave packet is created on $\ell=0$ at time $t=0$. It drifts along the Tight-Binding chain and gradually spreads. The amplitude of wave packet diminishes due to dephasing of reservoir electronic states. To demonstrate the evolution, $\langle d_\ell^\dag(t)d_0(0)$ is plotted for several different $t$'s.}
\label{wavepacket}
\end{figure}

We have verified that the steady-state formalism reproduces the key physics expected to occur in an electronic transport theory. In homogeneous steady state, relevant physical quantities are all stationary and no time evolution of them is expected. However, dynamics actually occurs in steady state and distinguishes a stationary non-equilibrium state from equilibrium. To understand the aspect of time evolution, we now examine how a wave-packet drifts and evolves after it is created. In particular, we create a hole from occupied states at the site $\ell=0$ and measure the probability distribution of positions of the hole after some time $t$. The probability is calculated as follows,
\begin{align}
\mathbb{P}\left[x(t)=\ell|x(0)=0\right]&=|\langle x_h(\ell),t|x_h(0),0\rangle|^2\nonumber\\
&=|\langle d^\dag_\ell(t)d_0(0)\rangle|^2\nonumber\\
&=|G^<_{0\ell}(-t)|^2
\end{align}
The $G^<$ is easily computed in terms of scattering states. Fig. \ref{wavepacket} shows the wave-packet propagates in the direction of the external field, as well as decaying in the time scale of $\Gamma^{-1}$ due to dephasing of fermion reservoirs. When current flows through the tight-binding chain, electrons move down the potential slope, generating particle-hole pairs in the fermion reservoirs that they have passed through. Energy is hence dissipated and transferred to the reservoirs. Since baths are assumed to have infinite bandwidth, the e-h pairs are absorbed deep inside the reservoir and never come back. As a result, the fermion reservoirs play similar roles as bosonic reservoirs, giving rise to inelastic processes to dissipate excess energy.

\section{Effective temperature and energy dissipation}

\subsection{Evaluation of effective temperature}
As we have discussed, the dissipative lattice model can satisfactorily describe non-equilibrium steady state of solids and reproduce the key physics. A central question is to understand how the thermal effect is modifying the physics in the strong-field regime. Therefore, we define the effective local distribution function, 
\begin{align}
f_\text{loc}(\omega)=-\frac{\text{Im}G^r_{00}(\omega)}{2\text{Im}G^r_{00}(\omega)}=\frac{\sum_\ell|G^r_{0\ell}(\omega)|^2f_\text{FD}(\omega+\ell E)}{\sum_{\ell}|G^r_{0\ell}(\omega)|^2},
\end{align}
which is a weighted average of Fermi-Dirac distribution at all lattice sites, with weights being the effective quantum tunneling to the site $\ell=0$. We show the numerically computed distribution function under a variety of electric fields. In the regime of $E, \Gamma\ll \gamma$, $f_\text{loc}(\omega)$ consists of steps coming from the fermion statistics $f_\text{FD}(\omega+\ell E)$ of all sites. And the envelope function will follow a similar shape to equilibrium Fermi Dirac distribution of higher temperature than $T_\text{b}$. 

Although one should not  generally expect the non-equilibrium distribution function mimics Fermi-Dirac function, it is reasonable to expect a similar functional form for $E\ll \Gamma$. Therefore an effective temperature $T_\text{eff}$ can be numerically extracted by curve-fitting. And for more dramatic cases, we will adopt the following definition of effective temperature.
\begin{align}
\frac{\pi^2}{6}T^2_\text{eff}=\int d\omega \omega[f_\text{loc}(\omega)-\theta(-\omega)].
\label{teff}
\end{align}
In this way, $T_\text{eff}$ is defined as square-root of the first moment of $f_\text{loc}(\omega)$. It is consistent with parameter $k_BT=1/\beta$ for Fermi-Dirac distribution $f(\omega)=1/(\text{e}^{\beta\omega}+1)$, and can be in principle carried out for any non-equilibrium distribution function where $0<f_\text{loc}(\omega)<1$. 

\begin{figure}
\centering
\includegraphics[scale=0.25]{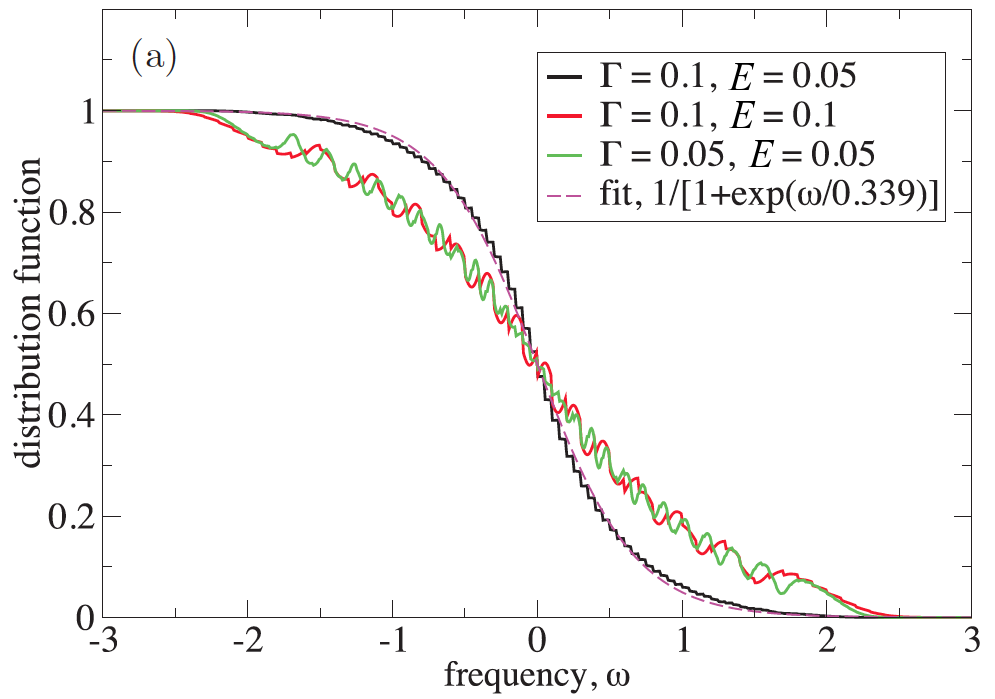}
\includegraphics[scale=0.25]{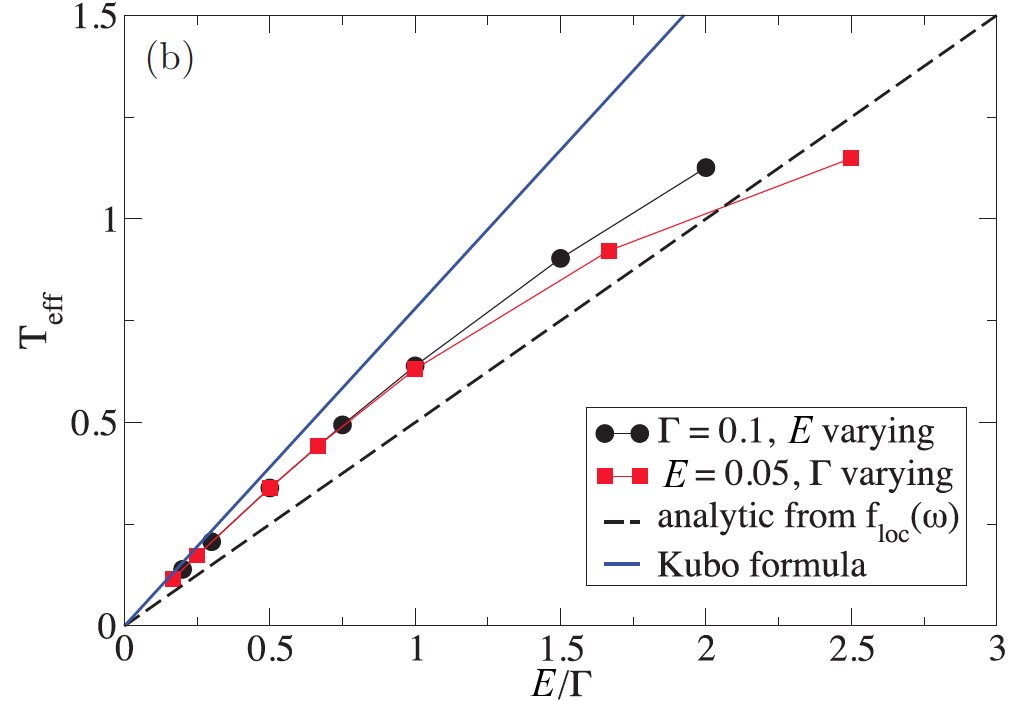}
\includegraphics[scale=0.25]{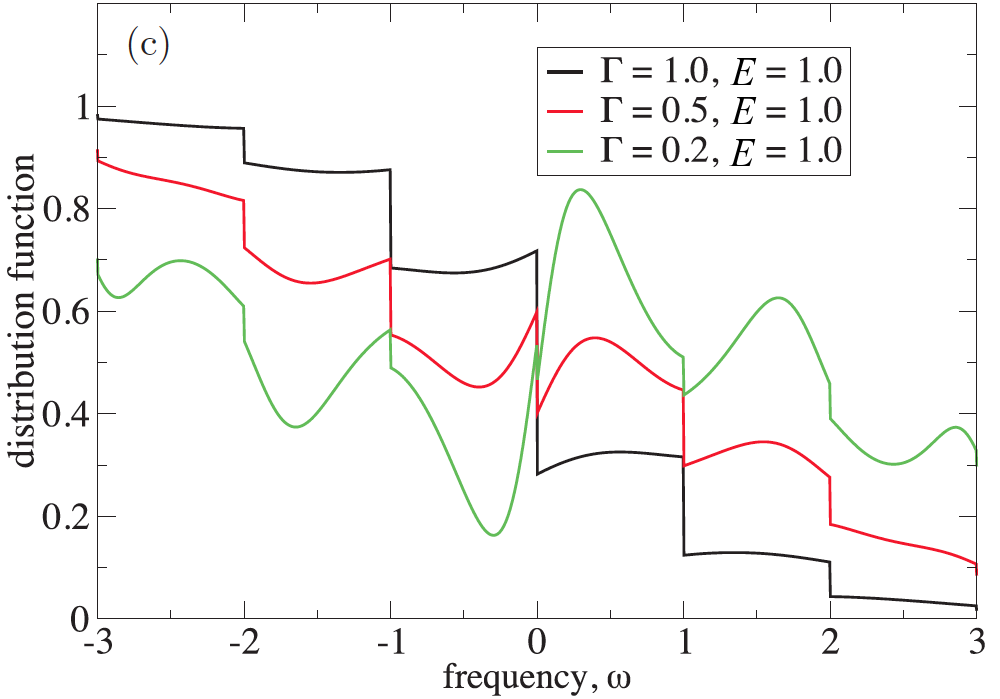}
\caption[Local distribution function and effective temperature]{(a)Local distribution function under a variety of electric fields. Effective temperature is evaluated by fitting the curve with Fermi-Dirac function. (b)Effective temperature scales as $T_\text{eff}\propto E/\Gamma$. The dashed line is obtained by fitting with Fermi-Dirac function, and the blue line is from Kubo formula. (c)For higher electric field $E\gtrsim\Gamma$, distribution function shows prominent steps reflecting the physics of Bloch oscillation. For a large electric field $E=1.0$ with $\Gamma=0.2$,  population inversion happens. }
\label{floc}
\end{figure}

We firstly discuss the $E\ll\Gamma$ regime. We can extract the effective temperature by fitting the slope of $f_\text{loc}(\omega)$ at $\omega=0$. Note the first step of $f_\text{loc}(\omega)$ at $\omega=0$ is
\begin{align}
\Delta=-\frac{\Gamma|G^r_{00}(0)|^2}{\text{Im}G^r_{00}(0)}
\end{align}
In the limit of small $E$, the $G^r_{00}(0)$ is approximated by the equilibrium Green's function
\begin{align}
G^r_{00}(\omega)^{-1}\approx (\omega+i\Gamma)\left[1-\frac{4\gamma^2}{\omega+i\Gamma)^2}\right]^\frac{1}{2}.
\end{align}
Then the zero-frequency slope is approximated as
\begin{align}
-\frac{\Delta}{E}=-\frac{\Gamma}{E\sqrt{4\gamma^2+\Gamma^2}}\approx -\frac{\Gamma}{2\gamma E}.
\end{align}
On the other hand, the slope of a Fermi-Dirac distribution function $[1+\exp(\omega/T_\text{eff})]^{-1}$ is $-4/T_\text{eff}$. Consequently the effective temperature is found to be 
\begin{align}
T_\text{eff}\approx C\gamma\frac{E}{\Gamma},
\end{align}
with a dimensionless constant $C\sim\frac{1}{2}$. This expression is verified both with numerical data in Fig. \ref{floc} and the theoretical result based on Kubo formula in later sections. Although the actual numerical fit overestimates $T_\text{eff}$ due to high-frequency contribution, the functional dependence is quite robust for $\Gamma, E<\gamma$.

It is remarkable to notice that $T_\text{eff}\to\infty$ when damping parameter $\Gamma$ approaches zero. This seemingly counterintuitive conclusion is interpreted as a \emph{short-circuit} effect, when system with negligible resistance becomes extremely hot under finite voltage bias. This is also consistent with previous theoretical studies showing electron temperature reaches infinity in closed driven interacting models.

Caution is necessary to interpret the infinite effective temperature in lattice model. In a lattice model of finite bandwidth, such as single band tight-binding model, the kinetic energy of an electron is bounded and cannot reach infinity like electrons of quadratic dispersion relation. 

As $E$ and $\Gamma$ are comparable to bandwidth, Bloch oscillation begins to dominate the transport physics. As shown in Fig. \ref{floc}, the oscillations in $f_\text{loc}(\omega)$ become more and more dramatic with increasing $E/\Gamma$, even leading to population inversion in small damping limit $\Gamma=0.2$. This makes the evaluation of $T_\text{eff}$ by curve-fitting less robust. In this case, as well as other cases where non-thermal distribution function is observed, definition \eqref{teff} should be used to obtain a well-defined effective temperature.

It is worth noting that in semiconductors, the distribution of electrons is governed by the classical Maxwell distribution and the internal energy $U\sim k_B T$ due to energy equipartition theorem. As we shall see in the next section, this relation results in $T_\text{eff}\propto E^2$, which was speculated in former works\cite{jong-prb}. However, degenerate electron gas in our work has $U\sim (k_BT)^2$, and we obtain the behaviour $T_\text{eff}\propto E$ instead.

\subsection{Dissipation and energy flux}

We now consider the dissipation and energy flux in our model, and look into the definition of effective temperature in general cases. 

First of all, the hamiltonian \eqref{0inth} is divided into three components:
\begin{align}
H_\text{sys}=H_\text{TB}+H_\text{bath}+H_\text{coup},
\end{align}
which correspond to the first three terms in Eq. \eqref{0inth}. When non-equilibrium steady state is considered, the energy stored in lattice and coupling terms $\langle H_\text{TB}\rangle$ and $\langle H_\text{coup}\rangle$ is stationary, i.e.
\begin{align}
\frac{d}{dt}\langle H_\text{TB}\rangle=\frac{d}{dt}\langle H_\text{coup}\rangle=0,
\end{align}
whereas the bath energy $\langle H_\text{bath}\rangle$ can be constantly increasing due to influx of Joule heating. In fact, the reservoirs are assumed to be much larger than the system, so that equilibrium state is maintained even though energy is constantly flowing into them. Specifically, we expect the energy flux into each fermion reservoir is equal to local electric power at the coupled lattice site, or $\frac{d}{dt}\langle H_\text{bath}\rangle/N = JE$, where $N$ is the length of tight-binding chain. To verify it, we firstly note
\begin{align}
\frac{d}{dt}\langle H_\text{TB}\rangle&=i\langle[H_\text{sys},H_\text{TB}]\nonumber\\
&=E\langle \hat{I}\rangle+i\gamma g\sum_\ell\langle (\bar{c}^\dag_{\ell+1}+\bar{c}^\dag_{\ell-1})d_\ell-H.c.\rangle,
\label{flux}
\end{align}
with total current operator $\hat{I}=i\gamma\sum_\ell(d^\dag_{\ell+1}d_\ell-H.c.)$ and $\bar{c}_\ell=\sum_\alpha c_{\ell\alpha}$. The condition of stationarity $\frac{d}{dt}\langle d^\dag_\ell d_\ell\rangle=0$ is used, which can explicitly be derived and generalized to interacting models of steady state. The first term in the RHS of Eq. \eqref{flux} is total Joule heating $E\langle\hat{I}\rangle=NJE$, and the second term is the energy flux from electrons in the lattice to the coupling part $H_\text{coup}$. Defining $\hat{P}=ig\gamma\langle (\bar{c}^\dag_{\ell+1}+\bar{c}^\dag_{\ell-1})d_\ell-H.c.\rangle$  as the energy flux for site $\ell$, one can show that $\langle \hat{P}\rangle=-JE$, hence $\frac{d}{dt}\langle H_\text{TB}\rangle=0$. It can further be shown that $\frac{d}{dt}\langle H_\text{coup}\rangle=0$, as well as $\frac{d}{dt}\langle H_\text{bath}\rangle=JE$. 

We then explicitly prove that no particle flux exists between the lattice and reservoirs and the energy flux actually balances the electric power, hence verify the non-equilibrium steady state is well defined.

We firstly consider the particle number $N_{\text{res},\ell}=\sum_\alpha c^\dag_{\ell\alpha} c_{\ell\alpha}$ in the reservoir. The change rate $d N_{\text{res},\ell}/dt$ reads
\begin{align}
\frac{d}{dt}\sum_\alpha c^\dag_{\ell\alpha} c_{\ell\alpha}&=i[H_\text{sys},\sum_\alpha c^\dag_{\ell\alpha} c_{\ell\alpha}]\nonumber\\
&=g\sum_\alpha(c_{\ell\alpha}^\dag d_\ell-d^\dag_\ell c_{\ell\alpha})\nonumber\\
&=g(\bar{c}^\dag_\ell d_\ell-d_\ell^\dag\bar{c}_\ell).
\end{align}
The $\bar{c}^\dag$ operators can be expressed in terms of scattering-state operators,
\begin{align}
\bar{c}^\dag_\ell=\sum_\alpha\left[ \psi_{\ell\alpha}^\dag+g^2\sum_{\ell'\alpha'} \frac{[G^r_{\ell\ell'}(\epsilon_{\alpha'}-\ell' E)]^*\psi^\dag_{\ell'\alpha'}}{\epsilon_{\alpha'}-\epsilon_\alpha-(\ell'-\ell)E-i\eta}\right],
\label{c_as_psi}
\end{align}
with which one can verify straightforwardly that
\begin{align}
d N_{\text{res},\ell}/dt&=\frac{2\Gamma}{\pi}\int d\epsilon\left[\text{Im}G^r_{\ell\ell}(\epsilon_\ell)f(\epsilon)+\Gamma\sum_m|G^r_{\ell m}(\epsilon_\ell)|^2f(\epsilon_\ell+mE)\right]\nonumber\\
&=\frac{2\Gamma}{\pi}\int d\epsilon\left[\text{Im}G^r_{\ell\ell}(\epsilon)+\Gamma\sum_m|G^r_{\ell -m,0}(\epsilon)|^2\right]f(\epsilon)\nonumber\\
&=0,
\end{align}
where Eq. \eqref{imG} is used in the last step. This shows that particle flux to reservoirs is exactly zero. We conclude that
\begin{align}
\frac{d}{dt}\sum_\alpha\langle c^\dag_{\ell\alpha} c_{\ell\alpha}\rangle=-\frac{d}{dt}\langle d^\dag_\ell d_\ell\rangle=0.
\end{align}
Note that translational invariance is used for deriving the zero-flux conclusion, and when the lattice is finite or disorders are present, non-zero particle flux may flow into the reservoirs. In particular, for a finite TB chain, higher potential sites will have flux into the TB chain whereas the lower potential sites have flux into the reservoirs.

We now turn to the energy flux. We firstly compute the $\langle \hat{P}\rangle$ by inserting Eq. \eqref{c_as_psi}. We let $\ell=0$, and it becomes
\begin{align}
g\langle(\bar{c}^\dag_1+\bar{c}^\dag_{-1})d_0\rangle&=\frac{\Gamma}{\pi}\int d\epsilon[G^r_{01}(\epsilon_1)+G^r_{0,-1}(\epsilon_{-1})]f(\epsilon)+\nonumber\\
&+i\frac{\Gamma}{\pi}\int d\epsilon\sum_m[G^r_{1m}(\epsilon)^*G^r_{0m}(\epsilon)+G^r_{-1m}(\epsilon)^*G^r_{0m}(\epsilon)]f(\epsilon+mE).
\end{align}
The imaginary part of the first term can be evaluated as
\begin{align}
\sum_m\int_{-\infty}^0d\epsilon J_m J_{m-1}\left\{\frac{\Gamma}{[\epsilon+(m-1)E]^2+\Gamma^2}+\frac{\Gamma}{[\epsilon+mE]^2+\Gamma^2}\right\}=\sum_mJ_mJ_{m-1}\chi_{m,m-1},
\end{align}
and the imaginary part of the second term,
\begin{align}
&-2\text{Re}\sum_m\int_{-\infty}^0\frac{J_mJ_{m-1}}{(\epsilon+mE-i\Gamma)[\epsilon+(m-1)E+i\Gamma]}=\nonumber\\
&\frac{1}{E^2+4\Gamma^2}\sum_mJ_mJ_{m-1}\left[\Gamma\ln\frac{m^2E^2+\Gamma^2}{(m-1)^2E^2+\Gamma^2}+E\chi_{m,m-1} \right]
\end{align}
Then the sum of the two terms gives
\begin{align}
g\text{Im}\langle (\bar{c}^\dag_1+\bar{c}^\dag_{-1})d_0\rangle=\frac{\Gamma E}{\pi(E^2+4\Gamma^2)}\sum_m J_m J_{m-1}\left[\Gamma\ln\frac{m^2E^2+\Gamma^2}{(m-1)^2E^2+\Gamma^2}+E\chi_{m,m-1}\right]=-JE.
\end{align}
Finally we have 
\begin{align}
\langle\hat{P}\rangle=-2g\gamma\text{Im}\langle (\bar{c}^\dag_1+\bar{c}^\dag_{-1})d_0\rangle=-JE.
\end{align}

We then consider the total energy per reservoir $\langle h_{\text{bath},\ell}$. Its change rate is the energy flux into the reservoir $\ell$,
\begin{align}
\frac{d}{dt}\langle h_{\text{bath},\ell}\rangle&=i\langle[H_\text{sys},h_{\text{bath},\ell}]\rangle=g\sum_\alpha\int \frac{d\omega}{2\pi}[G^<_{d\alpha}(\omega)-G^<_{\alpha d}(\omega)]
\label{dhdt}
\end{align}
In the last step we have used the known result of $\sum_\alpha\langle c_{\ell\alpha}^\dag d_\ell-d^\dag_\ell c_{\ell\alpha}\rangle=0$ due to zero particle flux.

With Dyson's equation, we can obtain a very useful expression of the energy influx into each fermion reservoir. We start from Eq. \ref{dhdt},
 \begin{align}
 \frac{d}{dt}\langle h_\text{bath}\rangle=g\sum_\alpha\int \frac{d\omega}{2\pi}[G^<_{d\alpha}(\omega)-G^<_{\alpha d}(\omega)],
 \end{align}
 with $G^<_{d\alpha}(t)=i\langle c^\dag_{0\alpha}(0)d_0(t)\rangle$ and $G^<_{\alpha d}(t)=i\langle d^\dag_0(-t)c_{0\alpha}(0)\rangle$. From Dyson equation, one can prove that
 \begin{align}
 G^<_{d\alpha}(\omega)-G^<_{\alpha d}(\omega)&=-2\pi ig\delta(\omega-\epsilon_\alpha)\left\{ G^<_{00}(\omega)-f(\omega)[G^a_{00}(\omega)-G^r_{00}(\omega)]\right\}\nonumber\\
 &=4\pi^2g\delta(\omega-\epsilon_\alpha)A_\text{loc}(\omega)[f_\text{loc}(\omega)-f_\text{FD}(\omega)].
 \end{align}
 where the local spectrum function 
\begin{align}
A_\text{loc}(\omega)=-\frac{1}{\pi}\text{Im}G^r_\text{loc}(\omega).
\end{align}
Then we have 
 \begin{align}
 \frac{d}{dt}\langle h_\text{bath}\rangle =2\Gamma\int{d\omega\omega A_\text{loc}(\omega)[f_\text{loc}(\omega)-f_\text{FD}(\omega)]}
 \end{align}
Due to energy conservation we should have $d\langle h_\text{bath}\rangle/dt=JE$, therefore we obtain the equation,
\begin{align}
JE=2\Gamma\int d\omega \omega A_\text{loc}(\omega)[f_\text{loc}(\omega)-f_\text{FD}(\omega)].
\label{eflux}
\end{align}
This equation relates Joule heating with the local distribution function and equilibrium Fermi-Dirac distribution. The energy flux has a simple interpretation: at each energy level, the flux is proportional to internal energy of electron gas measured from equilibrium value, where the integrand is energy $\omega$ multiplied by particle number $ A_\text{loc}(\omega)[f_\text{loc}(\omega)-f_\text{FD}(\omega)]$. On the other hand, this formula provides a direct way to compute current out of local quantities. The argument of energy conservation generalizes to interacting model, so the formula can be conveniently utilized within the DMFT formulation with a conserving approximation. 

It seems paradoxical at the first glance that the total energy $\langle H_\text{sys}\rangle$ is non-stationary. However, when we include another part of the hamiltonian of closed system, the problem is satisfactorily resolved: that is the battery generating electric field as well as providing electric power. Considering a tight-binding chain with length $N$, then the voltage bias is $V_\text{battery}=NE$. On the other hand, the battery loses electric charge $Q$ with the rate $\dot{Q}=-J$, therefore $\frac{d}{dt}\langle H_\text{battery}\rangle=\dot{Q}V_\text{battery}=-JNE$ and the total energy $H_\text{tot}=H_\text{sys}+H_\text{battery}$ is a constant in steady state. 

So far we have proven the Coulomb-gauge formalism is self-consistent and produce identical results as temporal-gauge calculations. The Coulomb-gauge dissipative lattice model reproduces all crucial physical consequences of Boltzmann transport theory and introduces no unphysical effects. The discussion above confirms the fermion baths act as energy reservoirs and no net electron flux is flowing into the reservoirs. When current flows through the main lattice, particle-hole excitations are created and play the role of bosonic baths, despite the possible difference due to a different dispersion relation and nonlinear effects of the bosonic statistics.

Finally, we are going to discuss effective temperature using Eq. \eqref{eflux} in different cases.

\subsection{Effective temperature and Kubo formula}
In linear response regime $E\ll\Gamma$, effective temperature can be computed by fitting local distribution function. On the other hand, we can evaluate effective temperature with the explicit expression of energy flux \eqref{eflux}. 
\begin{align}
JE=2\Gamma\int d\omega \omega A_\text{loc}(\omega)[f_\text{loc}(\omega)-f_\text{FD}(\omega)]\nonumber
\end{align}
In the case $A_\text{loc}(\omega)$ is smooth around $\omega=0$, we approximate the RHS of \eqref{flux} with Sommerfeld expansion, assuming $f_\text{loc}(\omega)$ has the form of Fermi-Dirac distribution and bath $T_\text{b}=0$:
\begin{align}
JE=\frac{2\Gamma}{E}\frac{\pi^2}{6}T_\text{eff}^2 A_\text{loc}(0),
\end{align}
In linear response regime, the current $J=\sigma_0 E$ is evaluated with Kubo formula(see appendix \ref{apkubo}), giving
\begin{align}
T_\text{eff}=\sqrt{\frac{3\sigma_0}{\pi^2\Gamma A_\text{loc}(0)}}E.
\label{linteff}
\end{align}
In the case of one-dimensional tight-binding lattice, the conductivity and local spectrum function can be computed:
\begin{align}
\sigma_0&=\frac{2\gamma^2}{\pi\Gamma\sqrt{\Gamma^2+4\gamma^2}}\\
A_\text{loc}(0)&=\frac{1}{N}\sum_k \frac{\Gamma/\pi}{4\gamma^2\cos^2k+\Gamma^2}=(\pi\sqrt{\Gamma^2+4\gamma^2})^{-1},
\end{align}
The effective temperature is then calculated as
\begin{align}
T_\text{eff}=\sqrt{\frac{6}{\pi^2}}\gamma\frac{E}{\Gamma}\approx 0.7796\gamma\frac{E}{\Gamma},
\end{align}
which justifies the result obtained from curve-fitting.

As we shall see in the following chapters, effective temperature is one of the central quantities for interpreting non-equilibrium physics, so it is worthwhile to discuss it in some more cases. The first one is two-dimensional tight-binding lattice, where $A_\text{loc}(\omega)$ has a van Hove singularity in $\omega\sim0$. There are no analytic expressions of conductivity $\sigma_0$ and $A_\text{loc}(\omega)$ for two-dimensional tight-binding model. Using Kubo formula, the conductivity is approximately $\sigma_0^\text{2D}\sim 1/\Gamma$ for $\Gamma\ll t$. An approximate formula is obtained for local spectrum function in Appendix \ref{apkubo},
\begin{align}
A_\text{loc}^{2D}(\omega)&\approx \frac{1}{t}\log\left(4\sqrt{\frac{t}{|\omega|}}\right),\quad \omega\ll t\nonumber\\
&\sim-\frac{1}{2t}\log|\omega|+\mathcal{O}(\omega)
\end{align}
The integration of internal energy is then evaluated. Here we assume $T_\text{eff}$ is small so that $f_\text{loc}(\omega)\approx 1/[\exp(\omega/T_\text{eff})+1]$,
\begin{align}
\sigma_0^\text{2D}E^2&=4\Gamma\int d\omega \omega\log\omega[f_\text{loc}(\omega)-f_\text{FD}(\omega)].
\end{align}
Using approximated formula $\sigma_0^\text{2D}\propto \Gamma^{-1}$, we obtain an equation of $T_\text{eff}$:
\begin{align}
\left(\frac{E}{\Gamma}\right)^2=-T_\text{eff}^2(a+b\log T_\text{eff}),
\end{align}
where $a,b$ are positive constants. Based the equation, we conclude that $T_\text{eff}$ approaches zero when $E/\Gamma\to0$. Even though $E/\Gamma\ll1$ is assumed, it is still difficult to get a closed formula of effective temperature and the function $T_\text{eff}\left(\frac{E}{\Gamma}\right)$ is expectedly complicated.  This calculation demonstrates how singularity of spectral function $A_\text{loc}(\omega)$ can affect the functional form of $T_\text{eff}$. It also suggests the rich behaviors of $T_\text{eff}$ for different energy structures.

The second example is linear dispersion relation $E=c|\boldsymbol{p}|$. In particular, for two-dimensional system this is the case of Dirac electron in graphene, with states below zero energy (the lower half of Dirac cone) ignored. When $\mu$ is large enough, the lower half of Dirac cone should be inactive and the following calculation is expected to describe faithfully the effective temperature.

Using Kubo formula, we have 
\begin{align}
\sigma_0=\frac{1}{4\pi^2}+\frac{\mu}{8\pi\Gamma}.
\label{condgrph}
\end{align}
The first term is the minimum conductivity and the second is the regular term proportional to chemical potential $\mu$. Note the linear conductivity of \emph{intrinsic} graphene is found to be ambiguous in literature\cite{dassarma11}. When valley/spin degeneracy and electron-hole symmetry are considered, the first term in \eqref{condgrph} is nothing but the universal quantum-limited conductivity $\sigma_0=2\pi^2=4e^2/\pi h$, whose relevance is already justified in experiments\cite{Miao1530}.

In the weak damping limit, the density of states is calculated as
\begin{align}
A(\mu)=\frac{\mu}{2\pi c^2}.
\end{align} 
Therefore, using Eq. \eqref{linteff} we can calculate the effective temperature
\begin{align}
T_\text{eff}=c\sqrt{\frac{3\Gamma}{2\pi^3\mu}+\frac{3}{4\pi^2}}\left(\frac{E}{\Gamma}\right).
\end{align}
When $\mu$ is large, we have 
\begin{align}
T_\text{eff}\approx c\frac{\sqrt{3}}{2\pi}\frac{E}{\Gamma},
\end{align}
which is similar to the one-dimensional case. However, a superficial singularity arises when $\mu\to0$, giving infinite effective temperature. This is when the system is at Dirac point, and DoS is zero. In this limit, we should compute the RHS of Eq. \eqref{eflux} explicitly with the $A(\omega)\propto\omega$,
\begin{align}
\frac{1}{4\pi^2}E^2&=\int d\omega\omega A(\omega)[f_\text{loc}(\omega)-f_\text{FD}(\omega)]d\omega\nonumber\\
&=\frac{2\Gamma}{\pi c^2}T_\text{eff}^3\int{dx\frac{x^2}{\text{e}^x+1}},
\end{align}
which gives $T_\text{eff}\propto E^{\frac{3}{2}}$. Fully addressing this problem requires the introduction of the full ``Dirac-cone hamiltonian" for Dirac electrons. As we shall see in the last chapter of the thesis, an external field drives electrons in the lower cone to tunnel to the ``upper cone", creating current-carrying electrons and holes. The actual charge-carrier density is thus always non-zero and effectively controlled by the external field. The non-equilibrium dynamics of Dirac electrons is discussed in chapter \ref{graphene}, where the physical picture is dramatically changed. 

\section{Conclusion}
We have discussed the electronic transport in a tight-binding model connected to fermion reservoirs in both temporal and Coulomb gauges. The time-dependent hamiltonian in Coulomb gauge can be exactly solved with scattering-state formalism, which provides an intuitive interpretation as well as an instructive computational framework. Moreover, Hershfield has suggested that non-equilibrium statistics can be naturally expressed with scattering-state operators, which allows exploration towards interacting theories.

In this work, we have shown that the fermion bath model provides the necessary dissipative mechanism to establish non-equilibrium steady state and reproduce the key physics of Boltzmann transport theory. The external electric field drives electrons to drift and form finite electric current in steady state. The linear response regime is confirmed in the model. Beyond the linear response, the electric power is balanced by the energy flux into reservoirs, and an effective non-equilibrium temperature is maintained higher than the bath(or ambient) temperature. As a result, the effective temperature depends strongly on electric field and damping parameter $\Gamma$, in the form of $T_\text{eff}\propto E/\Gamma$, and approaches infinity for $\Gamma=0$ as the short-circuit effect. This result verifies a variety of numerical calculations in previous theoretical works on isolated lattice models. Our finding demonstrates the importance of calculating effective temperature \emph{as a result of} non-equilibrium steady state, instead of inserting it as an external parameter in Gibbsian distribution. In addition, a general relation between energy flux and local quantities \eqref{flux} is derived, which can be viewed as a generalization of Meir-Wingreen formula\cite{meir-wingreen}. 

The simple fermion bath model can be used as an ideal building block for constructing an interacting model. Based on this time-independent formalism, it would be convenient to examine strong-correlation physics in field-driven dissipative lattice. In particular, DMFT calculation can be readily implemented within Coulomb-gauge hamiltonian. This would be the topic of the following chapters. We will see that dissipation strongly interplays with interaction effect in non-equilibrium steady state, and the thermal effect would be a key to understanding non-equilibrium phase transitions.

\chapter{Field-driven phase transition in strongly correlated materials}
\label{prl}
As discussed in the introduction, the resistivity of some correlated materials change sharply under strong electric field, which is termed as resistive switching (RS). The change of resistivity can be up to 5 orders of magnitude and its threshold electric field $E_\text{th}\sim10^{4--6}$ V/m is within the experimentally accessible regime. The time scale of the RS can be as short as 10$\mu$s. In addition, hysteresis and spatial inhomogeneities are ubiquitously found in $I-V$ characteristics during resistive switching. One of our main goals in the thesis is to establish a microscopic theory of the RS phenomenon.

\begin{figure}
\centering
\includegraphics[scale=0.6]{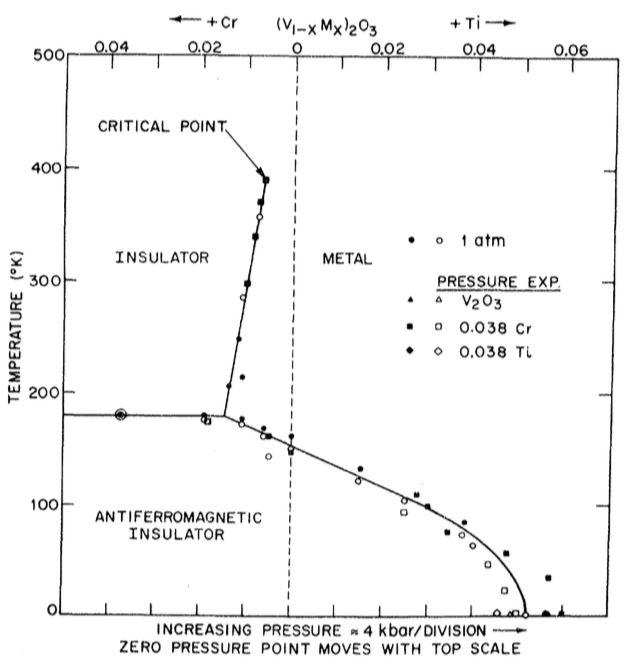}
\repeatcaption{RSmcwhan}{Equilibrium phase diagram of V$2$O$3$. Left to the critical point, a regime exists where system undergoes metal-to-insulator transition driven by the electric field.}
\end{figure}

In this chapter, we will construct an interacting theory based on the driven-dissipative lattice model to describe the correlated metal in non-equilibrium and is driven to a metal-insulator transition by electric field. In particular, we will verify that the thermal scenario of the resistive switching effect in the model. Recalling the equilibrium phase diagram \ref{RSmcwhan}, we will concentrate on the metal-to-insulator transition from a metallic state in low temperature to an insulating state in high temperature.

\section{Dynamical Mean Field Theory}
Dynamical mean field theory (DMFT) is one of the most powerful tools dealing with strongly correlated lattice systems. It approximately maps the interacting lattice model to an Anderson impurity model which is self-consistently determined in the numerical procedure. We will review the procedure in the real-time Green's function formalism, and refer the reader to the literature for more details\cite{kotliar-rmp, RMP-NEQDMFT}.

\subsection{Equilibirum DMFT}
In Dynamical Mean Field Theory, we make the local approximation that self energy $\Sigma_{ij}\propto\delta_{ij}$ where $i,j$ are site indices. The lattice model is then mapped to an Anderson impurity model and is solved self-consistently. To be concrete, let us consider a $d$-dimensional square Hubbard lattice
\begin{align}
H=\sum_{i,j}t_{ij}d^\dag_{i\sigma} d_{j\sigma} + U\sum_i n_{i\uparrow}n_{i\downarrow},
\end{align}
with $n_{i\sigma}=d^\dag_{i\sigma}d_{i\sigma}$. Defining matrix $\hat{h}_{ij}=t_{ij}$, the retarded Green's functions can be computed as
\begin{align}
G^r_{ij}(\omega)=\left(\omega-\hat{h}-\Sigma_U^r(\omega)\mathbb{I}+i0^+\right)^{-1}_{ij},
\end{align}
where $\Sigma^r_U(\omega)$ is the self energy which is uniform at all lattice sites. The matrix-inversion is the easiest in the momentum space where $\hat{h}$ is diagonalized,
\begin{align}
G^r_{\bm{k}}(\omega)=1/\left(\omega-\epsilon_{\bm{k}}-\Sigma^r_U(\omega)+i0^+\right),
\end{align} 
with $\epsilon_{\bm{k}}$ being the dispersion relation, or the eigenvalues of $\hat{h}$ indexed by Bloch momentum $\bm{k}\in \text{F.B.Z.}$ Note the momentum space is defined through Fourier Transform $d_{\bm{k}}=\sum_{\bm{r}}\exp(i\bm{k}\cdot\bm{r})d_{\bm{r}}/\sqrt{N}$. Due to fluctuation-dissipation theorem, the lesser/greater Green's functions are computed as follows,
\begin{align}
&G^<_{\bm{k}}(\omega)=-2i\text{Im}G^r_{\bm{k}}(\omega)f_\text{FD}(\omega),\nonumber\\
&G^>_{\bm{k}}(\omega)=-2i\text{Im}G^r_{\bm{k}}(\omega)[1-f_\text{FD}(\omega)].
\label{fdt}
\end{align}
Then the Green's functions with spatial indices can be obtained with inverse Fourier Transforms. We do not know $\Sigma^r_U(\omega)$ before solving this model, so we have to implement the above procedure self-consistently. To complete the self-consistent procedure, we consider an Anderson impurity model associated with the lattice model, where the local site $i=\bm{0}$ is the impurity and other parts are regarded as the environment. Then the non-interacting Green's functions of electrons at the impurity, or the \emph{Weiss-field} Green's functions, are defined by switching off interaction only at the local site,
\begin{align}
\mathcal{G}^r(\omega)^{-1}=G^r_{\bm{0}\bm{0}}(\omega)^{-1}+\Sigma^r_U(\omega).
\end{align}
We start the iterations with $\Sigma^r_U(\omega)=0$ and compute Weiss-field Green's functions. Then the new self energies are updated using the Weiss-field Green's functions combined with interaction term $H_{U,\text{loc}}=Un_{\bm{0}\uparrow}n_{\bm{0}\downarrow}$\cite{georges96}. The iterations are repeated until convergence. Note that the equilibrium DMFT is usually done with Matsubara Green's functions, but we use real-time Green's functions here to show its relation with the non-equilibrium DMFT.

\subsection{Non-equilibrium Green's functions}
The DMFT method is generalized to non-equilibrium systems by directly considering the real-time dynamics\cite{RMP-NEQDMFT}. In general cases, the Green's functions are $G^{r,\lessgtr}(t,t')$ without time-translational invariance. Since fluctuation-dissipation theorem does no hold in non-equilibrium, we need to write down Dyson's equations separately for the Green's functions,
\begin{align}
\mathbf{G}^{r}&=\left(\mathbf{G}^{r,0}-\mathbf{\Sigma}^r\right)^{-1},\nonumber\\
\mathbf{G}^{\lessgtr}&=(\mathbb{I}+\mathbf{G}^{r}\mathbf{\Sigma}^r)\mathbf{G}^{\lessgtr,0}(\mathbb{I}+\mathbf{\Sigma}^a\mathbf{G}^{a})+\mathbf{G}^{r}\mathbf{\Sigma}^\lessgtr\mathbf{G}^{a},
\end{align}
where the matrix indices include both spatial and time indices. In the case of a general time-dependent hamiltonian, it is necessary to solve for $G^{r,\lessgtr}(t,t')$ for all $t,t'$ self-consistently. This formulation has been established and applied to different physical systems\cite{RMP-NEQDMFT}. In this thesis, we will concentrate on steady-state physics. As will be shown below, in a convenient gauge (Coulomb gauge), all Green's functions are time-translationally invariant so that $G^{r,\lessgtr}(t,t')=G^{r,\lessgtr}(t-t')$. Hence the Dyson equations can be Fourier-transformed to frequency domain. 

\subsection{Time-independent hamiltonian in Coulomb gauge}
We study a dissipative Hubbard model, which is the dissipative lattice mode with Hubbard interaction term added for each site. The lattice is driven by a homogeneous dc-external electric field and Coulomb gauge is chosen. In one dimension, the non-interacting hamiltonian is a direct generalization of Eq. \eqref{0inthc} with spin indices inserted, i.e.
\begin{align}
H_0&=-\gamma\sum_\ell(d^\dag_{\ell+1,\sigma}d_{\ell\sigma}+H.c.)\nonumber\\
&+\sum_{\ell\alpha\sigma}\epsilon_\alpha c_{\ell\alpha\sigma}^\dag c_{\ell\alpha\sigma}-g\sum_{\ell\alpha\sigma}(c^\dag_{\ell\alpha\sigma}d_{\ell\sigma}+H.c.)\nonumber\\
&-\sum_{\ell\sigma}\ell E\left(d_{\ell\sigma}^\dag d_{\ell\sigma}+c_{\ell\alpha}^\dag c_{\ell\alpha}\right),
\label{uinth}
\end{align}
with $d^\dag_{\ell\sigma}$ creating electrons in the tight-binding chain and $c^\dag_{\ell\alpha\sigma}$ creating those in the fermion reservoirs. The difference between the above hamiltonian and Eq. \eqref{0inthc} is that spin $\sigma=\uparrow,\downarrow$ is considered here. $g$ is again the coupling of TB chain and the fermion reservoirs, and a flat density of states (bandwidth is infinite) for reservoirs is assumed. The solution of hamiltonian \eqref{uinth} is essentially identical to that of \eqref{0inthc}, thus is consistent with Boltzmann transport theory. We define the damping parameter $\Gamma=\pi g^2 N(0)$ where $N(0)$ is the constant DoS of the fermion reservoir. In the following discussions, we will scale energies in units of TB bandwidth, which is $W=4\gamma=1$ for 1D and $W=12\gamma=1$ for 3D. After the Hubbard term $H_U$ is added in the model, the full hamiltonian reads $H=H_0+H_U$, where
\begin{align}
H_U=U\sum_\ell\left(d^\dag_\uparrow d_\uparrow-\frac{1}{2}\right)\left(d^\dag_\downarrow d_\downarrow-\frac{1}{2}\right).
\label{uterm}
\end{align}
We always assume particle-hole symmetry in this chapter. 

\subsection{Formulating the dynamical mean field theory}
We will solve the interacting model with dynamical mean-field theory (DMFT). The self energies contributed from many-body interaction are self-consistently computed with a local approximation. Note that the total self energy is a sum of many-body term and reservoir term, e.g.
\begin{align}
\Sigma^r(\omega)&=-i\Gamma+\Sigma^r_U(\omega),\nonumber\\
\Sigma^<(\omega)&=2i\Gamma f_\text{FD}(\omega)+\Sigma^<_U(\omega),
\end{align}
where Fermi-Dirac function $f_\text{FD}(\omega)=1/[1+\exp(\omega/T_\text{b})]$. Under the approximation of DMFT, all self energies are local and identical besides energy shift due to the potential slope. In other words, we have
\begin{align}
\Sigma^{r,\lessgtr}_{\ell\ell'}(\omega)=\Sigma^{r,\lessgtr}(\omega+\ell E)\delta_{\ell\ell'} 
\label{localself}
\end{align}
This is a direct consequence of Eq. \eqref{transprop}. We then describe the self-consistent loop of DMFT.

\begin{figure}
\centering
\includegraphics{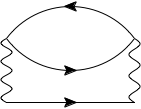}
\caption[Feynman diagram for self energy]{The second order 1-PI Feynman diagram in the particle-hole symmetric Anderson impurity model. Solid lines are electrons and wiggled lines are photons. The Hubbard interaction involves four electron-lines and have coupling constant $U$.}
\label{feyn-diag}
\end{figure}

First of all, we suppose that local self energies are already computed, then the full Green's functions can in principle be constructed for the whole lattice via Dyson's equation. In Coulomb gauge, the Green's functions are all time-translationally invariant and can be Fourier-transformed to frequency domain, hence the convolutions in time domain converts to direct multiplications. The interacting Green's functions are
\begin{align}
\mathbf{G}^r(\omega)^{-1}_{\ell\ell'}&=\left(\omega+\ell E+i\Gamma+\Sigma^r_U(\omega)\right)\delta_{\ell\ell'}+\gamma\delta_{\langle\ell\ell'\rangle}\nonumber\\
\mathbf{G}^<(\omega)&=\mathbf{G}^r(\omega)\mathbf{\Sigma}^<(\omega)\mathbf{G}^a(\omega),
\label{dyson}
\end{align}
with matrix indices being only lattice site indices. The local Green's functions are $G^{r,\lessgtr}_\text{loc}(\omega)=\mathbf{G}^{r,\lessgtr}(\omega)_\text{0,0}$. Now we divide the lattice into two parts: the local site ($\ell = 0$) and the ``environment" consists of all sites with $\ell\ne0$. To implement the DMFT formulation, we get to the Anderson impurity model by integrating out the environmental part of lattice and interpreting the local site as the impurity. Then the effective action of local lattice site becomes:
\begin{align}
S=\int dt'dt d^\dag_0(t') \mathcal{G}_\text{loc}(t-t') d_0(t)-U\Delta n_{0\uparrow}\Delta n_{0\downarrow},
\end{align}
where $\Delta n_{0\sigma}=n_{0\sigma}-1/2$, and $\mathcal{G}$ is the Weiss-field Kadanoff-Baym Green's function:
\begin{align}
\mathcal{G}=\begin{pmatrix}\mathcal{G}^t&\mathcal{G}^<\\ \mathcal{G}^>&\mathcal{G}^{\tilde{t}}
\end{pmatrix},
\label{imp-model}
\end{align}
which are obtained by switching off the interaction \emph{only} on the local site $\ell=0$. This can be implemented by setting on-site self energies and applying Dyson's equation \eqref{dyson}, 
\begin{align}
\mathcal{G}^r(\omega)^{-1}_{\ell\ell'}&=\mathbf{G}^r(\omega)^{-1}_{\ell\ell'}+\Sigma^r_U(\omega)\delta_{\ell0}\delta_{\ell'0},\nonumber\\
\mathcal{G}_\text{loc}^{\lessgtr}(\omega)&=|\mathcal{G}_\text{loc}^r(\omega)|^2\left(\frac{G^\lessgtr_\text{loc}(\omega)}{|G^r_\text{loc}(\omega)|^2}-\Sigma^{\lessgtr}_{U,\text{loc}}(\omega)\right)
\end{align}
To deal with the Anderson impurity model, we will iteratively use second order perturbation theory in $U$,
\begin{align}
\Sigma^{\lessgtr}_U(t)=U^2[\mathcal{G}^\lessgtr(t)]^2\mathcal{G}^\gtrless(t),
\end{align}
which is shown diagrammatically in Fig. \ref{feyn-diag}.

After local self energies are computed, we can subsequently find the complete self energy matrix by using the translation property \eqref{localself},
\begin{align}
\boldsymbol{\Sigma}^{r,\lessgtr}_U(\omega)_{\ell\ell'}=\Sigma^{r,\lessgtr}_U(\omega+\ell E)\delta_{\ell\ell'}.
\end{align}
Note that off-site self energies are zero due to the assumption of dynamical mean-field theory. The self energies are then used to compute new Green's functions with Dyson's equations \eqref{dyson}. The procedure is repeated self-consistently until convergence.

After convergence is achieved, interesting physical quantities, such as local distribution function and electric current are computed. In particular, the electric current per spin can be measured as follows,
\begin{align}
J&=\frac{i}{2}\langle d^\dag_{1\sigma}d_{0\sigma}-d^\dag_{0\sigma}d_{1\sigma}+d^\dag_{0\sigma}d_{-1\sigma}-d^\dag_{-1\sigma}d_{0\sigma}\rangle\nonumber\\
&=\gamma\text{Re}\int\frac{d\omega}{2\pi}[G^<_{01}(\omega)-G^<_{0-1}(\omega)]\nonumber\\
&=-\gamma^2\text{Re}\int\frac{d\omega}{2\pi}\{G^<_\text{loc}(\omega)[F^a_+(\omega+E)-F^a_-(\omega-E)]-\nonumber\\
&\hspace{22mm}-G^r_\text{loc}(\omega)[F^<_+(\omega+E)-F^<_-(\omega-E)]\}.
\end{align}

\subsection{Recursion relations}
One of the key steps in DMFT calculation is to find retarded Green's functions by inverting a large matrix in Eq. \eqref{dyson}. One option to implement this step is to truncate the infinite lattice to finite chain $\ell=-N,-N+1,\ldots,N$, and actually invert the truncated matrix. Here we introduce an efficient method via a couple of recursion relations. We firstly divide the chain to three parts: the central local site $\ell=0$, the left semi-infinite chain with $\ell<0$ and the right semi-infinite chain $\ell>0$. Suppose we isolate the right semi-infinite chain, it has a property of self-similarity, that the chain is almost the same besides all on-site energies shifted by $E$ if the \emph{left}-end site is deleted. Consequently the local retarded Green's function at its left-end site $F^r_+(\omega)$ should satisfy
\begin{align} 
F^r_+(\omega)=\omega -\Sigma^r(\omega)-\gamma^2F^r_+(\omega+E).
\end{align}
Note that integrating out the rest of the semi-infinite chain (without the left-end site) results in the hybridization function $\gamma^2 F^r_+(\omega+E)$ which is essentially the retarded Green's function itself shifted by $E$ due to potential slope. The same argument can be applied to lesser/greater Green's functions, and we have
\begin{align}
F^{\lessgtr}_+(\omega)=|F^{r}_+(\omega)|^2[\Sigma^{\lessgtr}(\omega)+\gamma^2F^{\lessgtr}_+(\omega+E)].
\end{align}
Similar results are obtained for left semi-infinite chain
\begin{align}
F^r_-(\omega)=\omega -\Sigma^r(\omega)-\gamma^2F^r_-(\omega-E)\\
F^{\lessgtr}_-(\omega)=|F^{r}_-(\omega)|^2[\Sigma^{\lessgtr}(\omega)+\gamma^2F^{\lessgtr}_-(\omega-E)].
\label{recursion}
\end{align}
The local site $\ell=0$ connects to both right/left semi-infinite chains. An electron at the local site couples to the right/left semi-infinite chains through hopping $\gamma$. In terms of Feynman diagrams, the electron may hop to each of the semi-infinite chains and then hop back to the local site, leading to a self energy term proportional to $\gamma^2F_{\pm}$. The on-site Green's functions for $\ell=0$ then obey the following Dyson's equations:
\begin{align}
G^r_\text{loc}(\omega)^{-1}&=\omega-\Sigma^r(\omega)-\gamma^2 F^r_
\text{tot}(\omega)\\
G^<_\text{loc}(\omega)&=|G^r_\text{loc}(\omega)|^2[\Sigma^<(\omega)+\gamma^2 F^<_\text{tot}(\omega)],
\end{align}
where $\gamma^2F^{r,\lessgtr}_\text{tot}$ are ``self energies" due to hybridizing with both left/right semi-infinite chains, $F^{r,\lessgtr}_\text{tot}(\omega)=F^{r,\lessgtr}_+(\omega+E)+F^{r,\lessgtr}_-(\omega-E)$. Then the Weiss-field Green's functions are obtained straightforwardly.

The advantage of recursion relations is clear. The $F_{\pm}^{r,\lessgtr}$'s are very efficient to evaluate, and the computed Green's functions are intrinsically of an infinite lattice and free of truncation errors due to finite length.

\subsection{Higher dimensions}
So far we have been discussing one-dimensional case. Now we generalize the method to higher dimensions. Consider a lattice model of 2 or more dimensions. Suppose the electric field is applied in one of the principal axes, say $\boldsymbol{E}=E\boldsymbol{\hat{x}}$. Then in the directions perpendicular to $\boldsymbol{\hat{x}}$, the hamiltonian is translationally invariant, thus can be diagonalized in momentum space. Then the hamiltonian becomes independent one-dimensional pieces of different transverse momenta $\boldsymbol{k}_\perp$, and each of which can be solved separately via the method used in the 1-$d$ case. The range of transverse momenta is just the $d-1$ dimensional Brillouin zone. 

In particular, the hypercubic TB lattice results in dispersion relation $\epsilon_{\boldsymbol{k}_\perp}=-2\gamma\sum_i\cos(k_{\perp,i})$, where $k_{\perp,i}$'s are components in the perpendicular directions. Each transverse mode of $\boldsymbol{k}_\perp$ is nothing but a one-dimensional tight-binding model with $\epsilon_{\boldsymbol{k}_\perp}$ added to the on-site energy. After solving the 1-$d$ model, we obtain the Green's functions $G^{r,\lessgtr}_{\boldsymbol{k}_\perp}(\omega)$, and the full local Green's functions can be computed by summing over transverse momenta,
\begin{align}
G^{r,\lessgtr}_\text{loc}(\omega)=\int_\text{BZ}\frac{d^{d-1}\boldsymbol{k}_\perp}{(2\pi)^{d-1}}G^{r,\lessgtr}_{\boldsymbol{k}_\perp}(\omega)=\int d\epsilon_\perp D_{d-1}(\epsilon_\perp)G^{r,\lessgtr}(\epsilon_\perp,\omega),
\end{align}
where $D_{d-1}(\epsilon_\perp)$ is the $d-1$ dimensional density of states. The Weiss-field Green's functions can then be calculated and the DMFT self-consistent procedure is continued until convergence. 

\section{Linear response regime}
First of all, we discuss the linear response regime of the interacting model. Kubo formula can be used to compute the dc-conductivity under zero electric-field. In zero-field limit, the one-electron hamiltonian is translationally invariant and diagonalized in momentum space. Then Kubo formula reads,
\begin{align}
\sigma_\text{dc}&=\lim_{\omega\to0}\sum_{\boldsymbol{k}}\int d\nu \rho_{\boldsymbol{k}}(\nu)\rho_{\boldsymbol{k}}(\nu+\omega)\frac{f_\text{FD}(\nu)-f_\text{FD}(\nu+\omega)}{\omega}\nonumber\\
&=\sum_{\boldsymbol{k}}\int d \nu[\rho_{\boldsymbol{k}}(\nu)]^2\delta(\nu),
\label{intkubo}
\end{align}
where $\rho_{\boldsymbol{k}}(\nu)$ is the spectral function of momentum $\boldsymbol{k}$. Within the approximation of DMFT, the self energy $\Sigma^r_U(\omega)$ is local and spatially uniform, thus it has no dependence of $\boldsymbol{k}$ in momentum space. As a result, the spectral function can be written as
\begin{align}
\rho_{\boldsymbol{k}}(\nu)=-\frac{1}{\pi}\text{Im}\left(\frac{1}{\nu-\epsilon_{\boldsymbol{k}}+i\Gamma-\Sigma_U^r(\nu)}\right)
\end{align}

\begin{figure}
\centering
\includegraphics[scale=0.4]{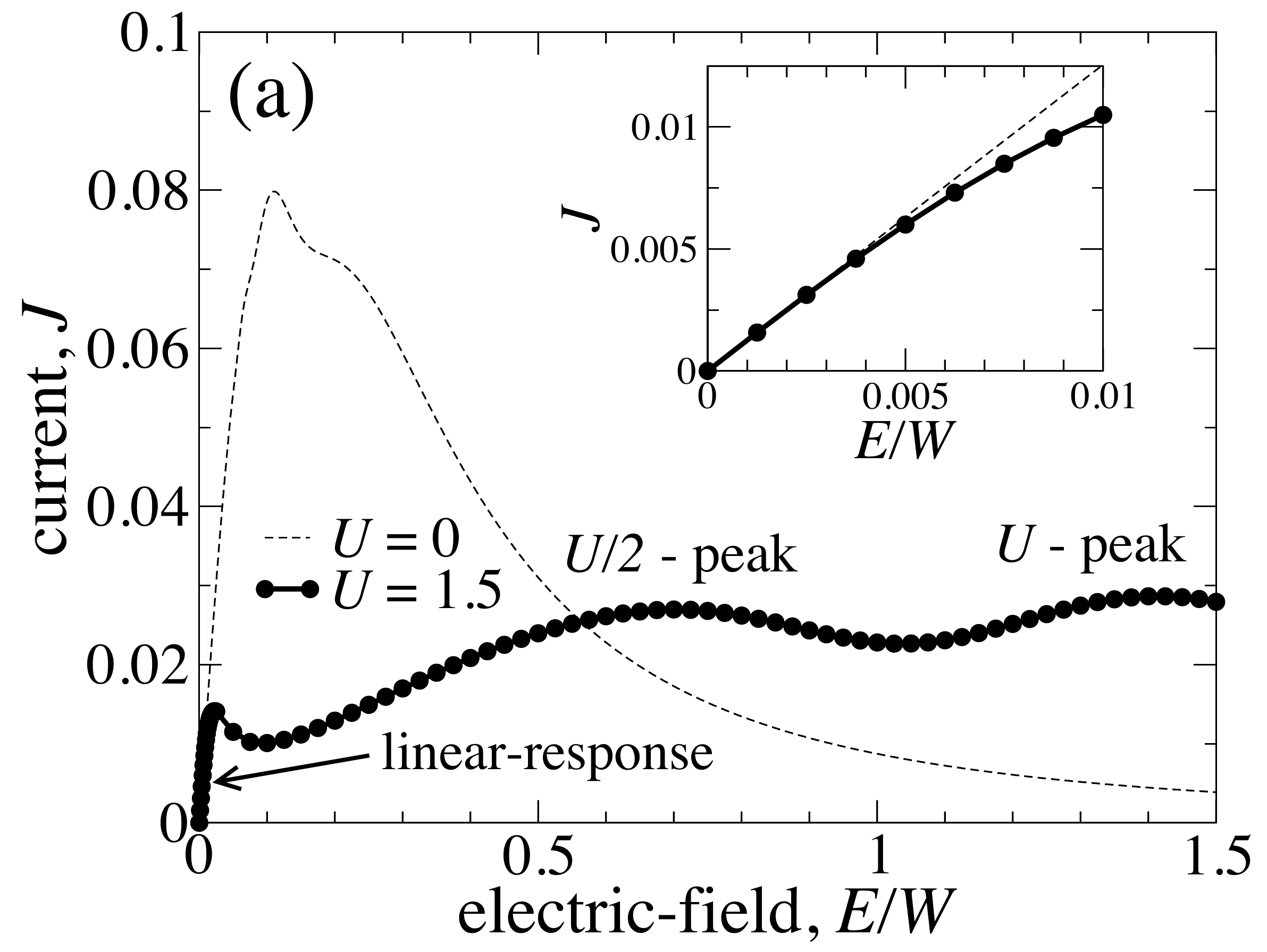}
\includegraphics[scale=0.4]{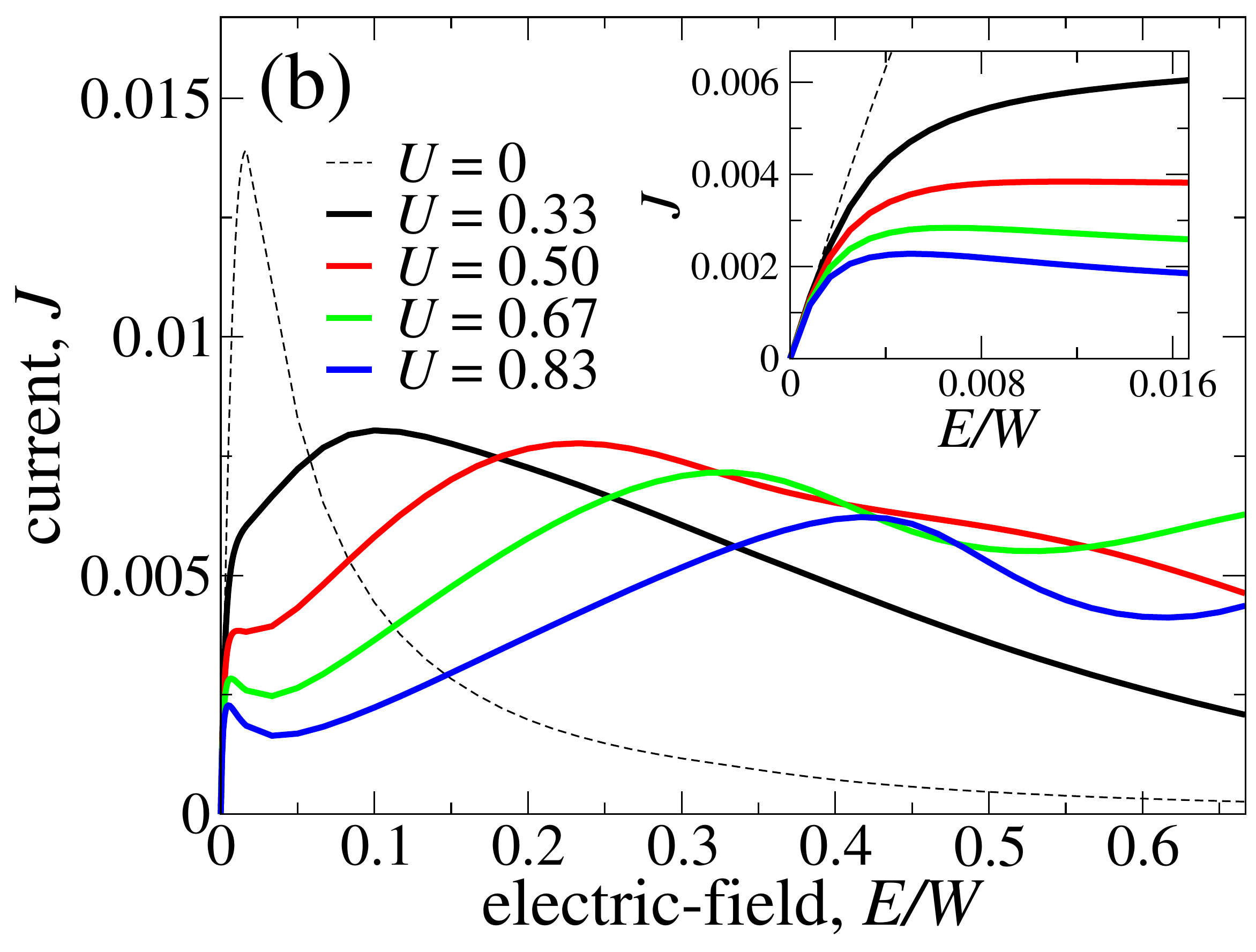}
\caption[Electric current $J$ versus electric field $E$]{Electric current $J$ versus electric field $E$. (a) one-dimensional chain with damping $\Gamma=0.0625W$ and fermion bath temperature $T_\text{b}=0.00125W$. The 1D TB bandwidth is $W=4\gamma$. The linear conductivity in the zero-field limit is the same for non-interacting ($U=0$) and interacting $U=1.5W$ models. Current deviates from linear behavior under higher electric fields, showing peaks at $E=U/2$ and $E=U$. (b) three-dimensional TB lattice with $\Gamma=0.0083W$ and $T_\text{b}=0.00042W$. The 3D TB bandwidth is $W=12\gamma$. The main features are the same for 1D and 3D results. }
\label{prl-fig1}
\end{figure}

According to Eq. \eqref{intkubo}, the dc-conductivity only depends on spectral function at zero energy $\rho(0)$. And the interaction self energy $\Sigma^r_U(\nu)\to0$ when $\nu\to0,T\to0$. Consequently, the dc-conductivity has no dependence on electronic interaction\cite{prange-kadanoff}. In recent theoretical calculations, the linear response regime independent of interaction is not addressed\cite{Tsuji08, Aron-prl12, Amaricci12}. Fig. \ref{prl-fig1} confirms the linear response theory. As the figure shows, the slope of $J-E$ curve at $E=0$ is independent of interaction parameter $U$, in both (a) one and (b) three dimensions. Interestingly, the linear behaviour deviates at the field $E_\text{lin}\approx0.003$ in Fig. \ref{prl-fig1}(a). This field is orders of magnitude smaller than the renormalized quasi-particle (QP) bandwidth $W^*=zW\approx0.5$, where the equilibrium renormalization factor $z=1/[1-\text{Re}\partial\Sigma^r_U(\omega)/\partial\omega]^{-1}_{\omega=E=T_b=0}$. 

As the electric field increases, the current-field curve shows features reflecting the physics of tunneling between neighboring sites. At $E=U/2$, a peak in current appears due to overlap between in-gap QP states (Abrikosov-Suhl resonance) and upper/lower Hubbard bands at neighboring sites. And when $E=U$, current reaches a second peak since Hubbards at neighboring sites overlap\cite{Aron-prb12, joura08}.

\begin{figure}
\centering
\includegraphics[scale=0.4]{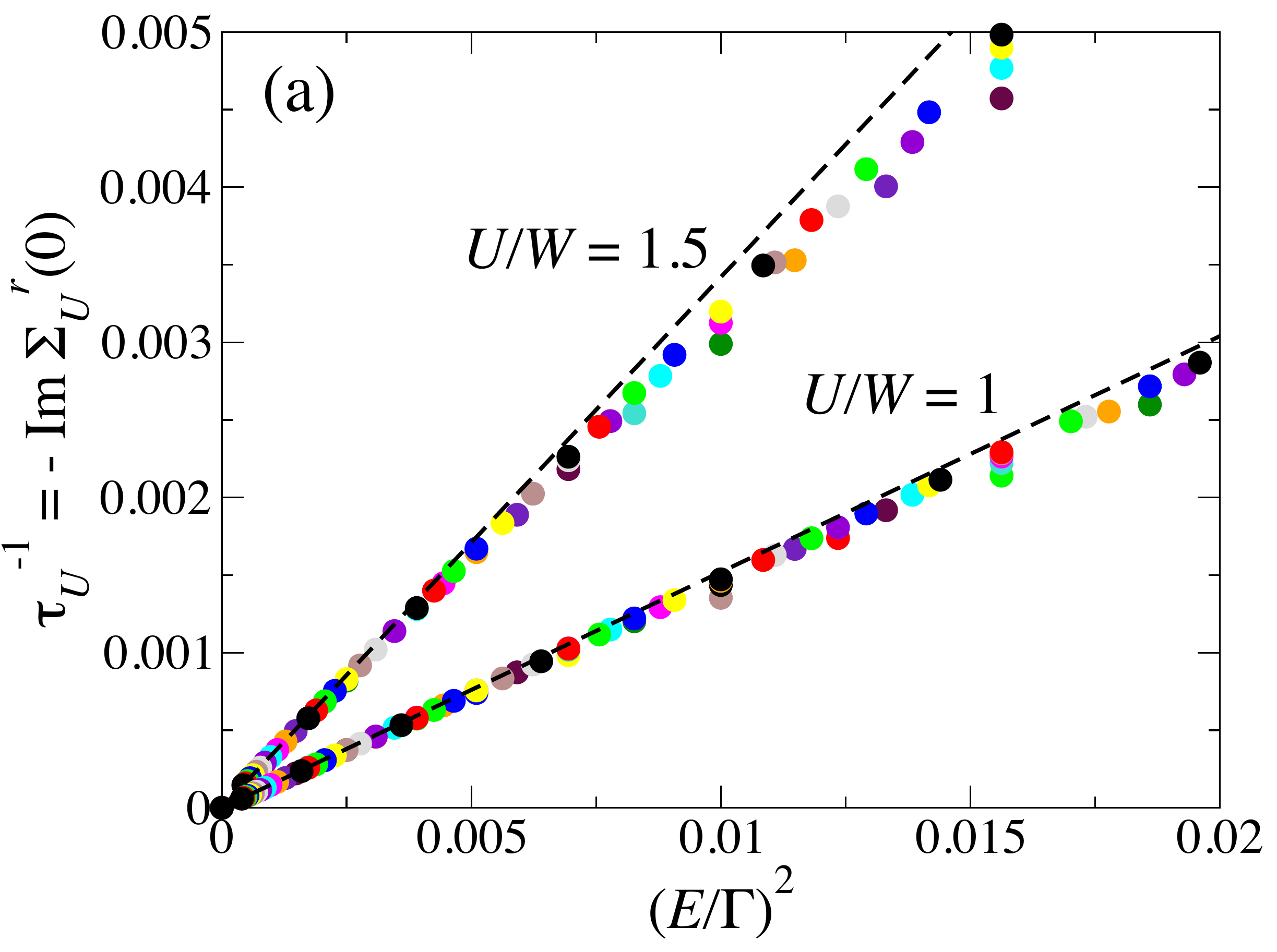}
\includegraphics[scale=0.4]{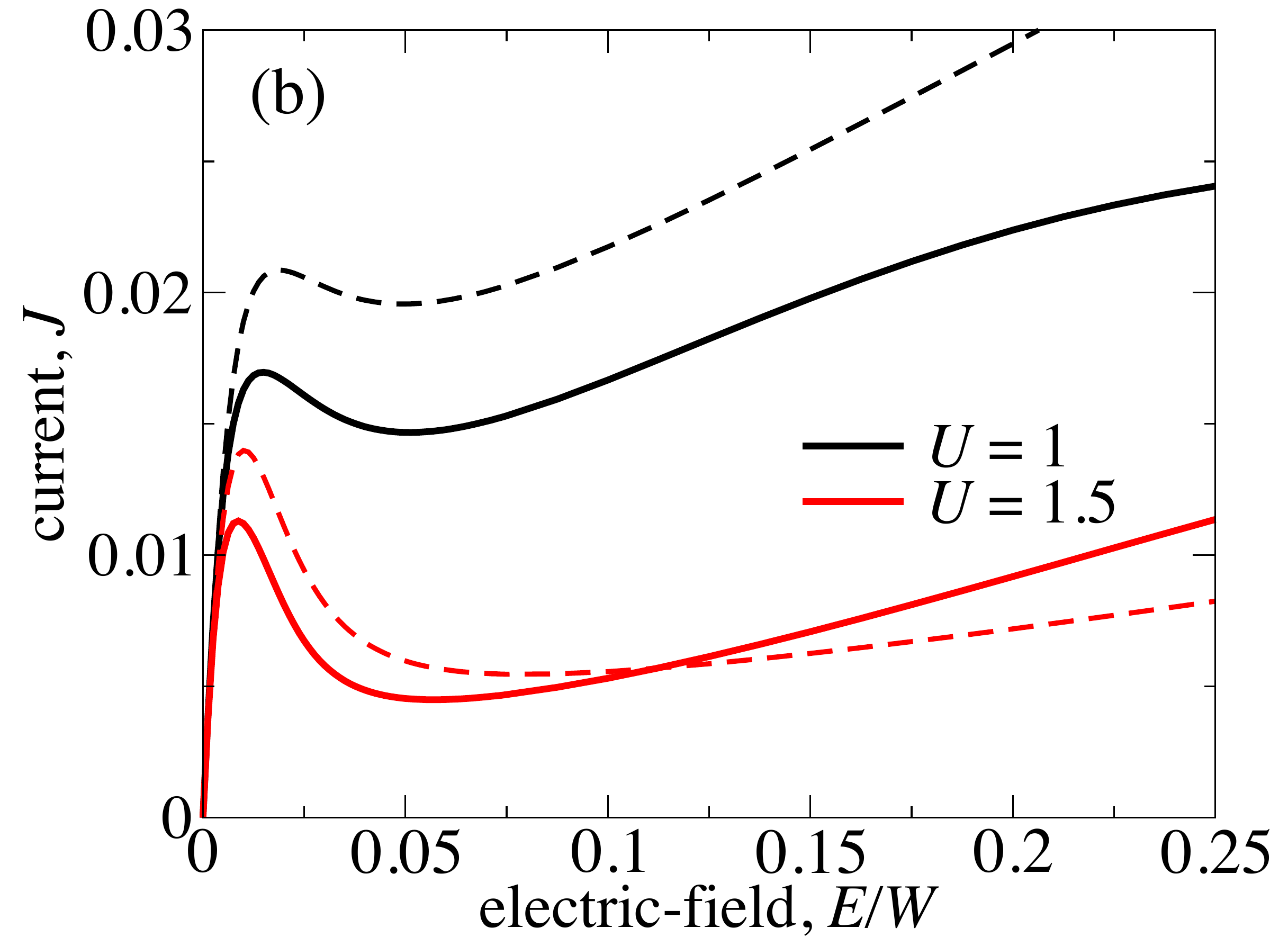}
\caption[Scattering rate and Drude formula]{(a) Scattering rate due to electronic interaction versus $(E/\Gamma)^2$. Different colors correspond to different damping $\Gamma=0.0125,...,0.06$ with the interval of $0.0025$. Data points are calculated in 1D chain, collapsing to the same straight lines for $U=1$ and $1.5$ for small fields. The dashed lines are predicted with Eq. \eqref{sret}. (b) Comparison of the current with the approximated results from Drude formula. The total scattering rate $\tau_\text{tot}^{-1}=\Gamma+\tau^{-1}_U$ is used.}
\label{prl-fig2}
\end{figure}

To theoretically understand why electric current deviates from linear behavior at very small fields, we need to go beyond the limit of $\Sigma^r_U(0)=0$, and at least include the next-order contributions in $E$ to the self energy. To obtain an approximated expression, we note that Joule heating raises the effective temperature of the system very quickly, according to the formula \eqref{teff}
\begin{align}
T_\text{eff}=\frac{\sqrt{6}}{\pi}\gamma\frac{E}{\Gamma}.\nonumber
\end{align}
With this thermal effect considered, the non-equilibrium self energy $\Sigma^r_U(0)$ should be approximately expressed as the equilibrium expression with a raised effective temperature \eqref{teff}\cite{yamada75}. Hence in the weak field limit, we obtain the formula of scattering time $\tau_U$ for electronic interaction, which is nothing but the imaginary part of interaction self energy,
\begin{align}
\tau^{-1}_U=-\text{Im}\Sigma^r_U(\omega)\approx\frac{\pi^2}{2}A_0(0)^3U^2T^2_\text{eff},
\label{sret}
\end{align}
where $A_0(0)=\left(\pi\sqrt{\Gamma^2+4\gamma^2}\right)^{-1}$ is the density of states at $\omega=0$ in non-interacting model. This theory is tested against numerical data in Fig. \ref{prl-fig2}. The dashed line is predicted by Eq. \eqref{sret} and fits numerical data quantitatively well. 

The discussion above shows the linear response regime is dominated by the Joule heating. In addition, the effective temperature $T_\text{eff}$ of the interacting system is given by the non-interacting formula \eqref{teff} in linear response regime. However, as we shall see below, the $T_\text{eff}$ will strongly deviate from the simple $E/\Gamma$ behavior beyond the linear response limit, especially in the case $\tau^{-1}_U$ dominates $\Gamma$. And its actual functional form has profound effect on the properties of the system.

We now relate the scattering rate $\tau^{-1}_U$ to the $J-E$ curve by using the Drude formula,
\begin{align}
\sigma_\text{dc}(E)=\frac{\tau_\text{tot}}{\tau_\Gamma}\sigma_\text{0,dc}=\frac{\Gamma}{\Gamma+\tau^{-1}_U}\sigma_\text{0,dc},
\label{intdrude}
\end{align}
where $\sigma_\text{0,dc}=2\gamma^2/(\pi\Gamma\sqrt{\Gamma^2+4\gamma^2})$ is the zero-field conductivity. As shown in Fig. \ref{prl-fig2}, the prediction of the Drude formula qualitatively agrees with the numerical data over a wide range. Furthermore, we can use the formula of self energy Eq. \eqref{sret} and obtain an approximate expression for the current,
\begin{align}
J=\frac{\sigma_\text{0,dc}E}{1+E^2/E^2_\text{lin}}.
\end{align}
The departure from linear behavior happens when $E\sim E_\text{lin}=(8\pi^2/3)^{1/2}\gamma^{1/2}\Gamma^{3/2}/U$. This condition is satisfied when $\tau^{-1}_U\sim \Gamma$. This formula is valid over a wide range of $U$, and only fails at $U\sim0$ where Bloch oscillation is responsible for the deviation as well as the metal-insulator-transition regime where $U/\gamma$ is extremely large. Note that the exact form of $E_\text{lin}$ depends on the type of dissipative mechanism. For instance, impurity scattering happens in a more realistic model and becomes dominant in weak-field limit. We will in that case have $E_\text{lin}\sim\tau^{-1}_\text{imp}\Gamma/U$. 

Although studies typically show negative-differential-resistance (NDR) in a lattice model, they are usually due to Bloch oscillation\cite{lebwohl70, jong-prb}, as shown in the dashed line of Fig. \ref{prl-fig1}. However, the NDR occurs in our dissipative Hubbard model is due to strong electronic scattering enhanced by effective temperature.

\section{Metal-insulator transition and thermal scenario}
Now we examine the parameter regime where $\Gamma$ is small and $U$ is large. Due to strong scattering and weak dissipation, the non-equilibrium effects become more dramatic. In this situation, the effective temperature rises sharply due to a small $\Gamma$. And with a narrow renormalized bandwidth, the system deviates immediately from linear behavior to avoid overheating. This results in a very narrow linear response regime or very small $E_\text{lin}$.

Moreover, if the system is close to a quantum phase transition in equilibrium, this dramatic behavior may result in a non-equilibrium phase transition. In this section, we will discuss the electric-field-driven metal-insulator transition. We will see that in a region of $U$ and $E$, non-equilibrium DMFT calculation finds both metallic and insulating solution, revealing the existence of a first-order transition. 

\begin{figure}
\centering
\includegraphics[scale=0.9]{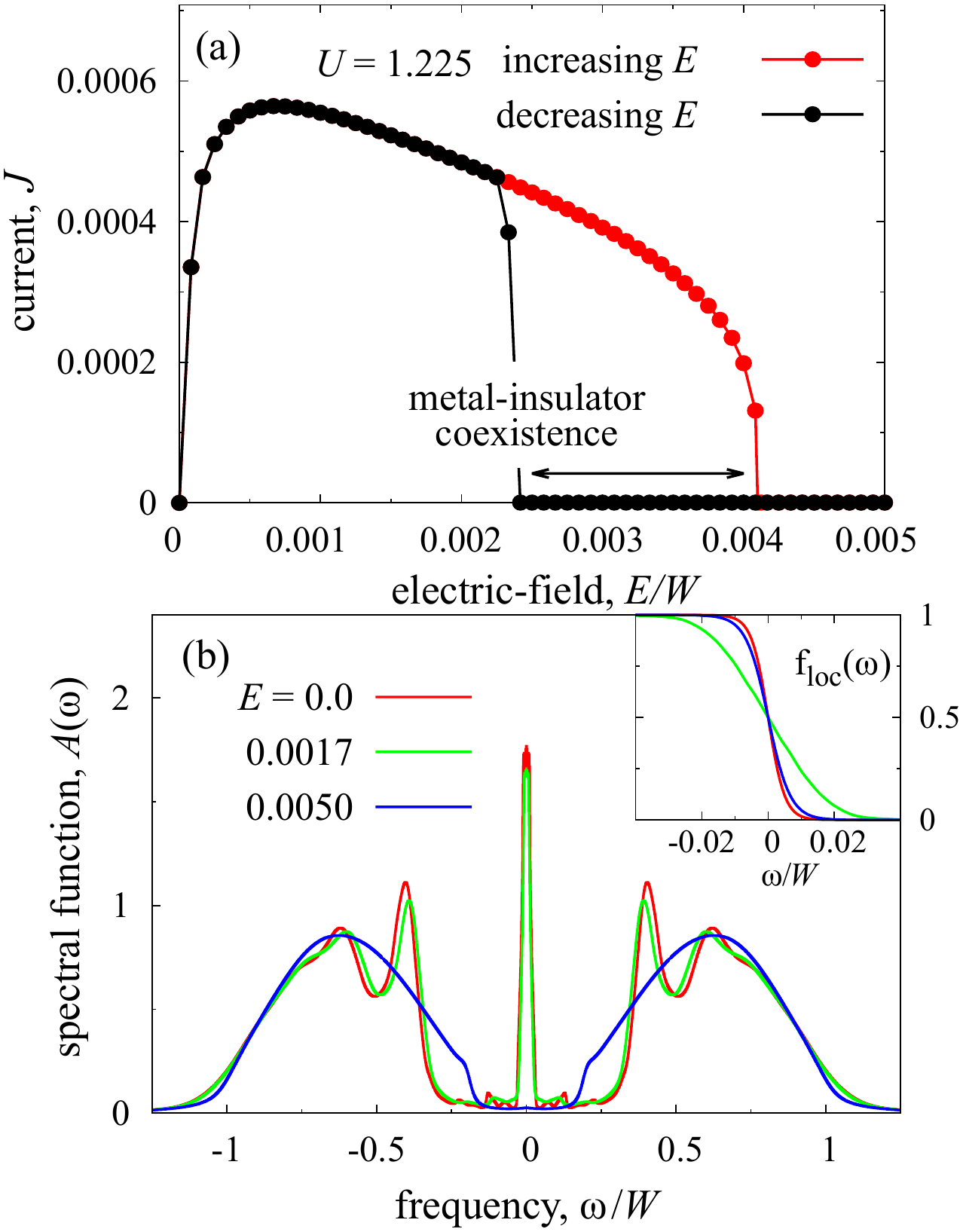}
\caption[Resistive Switching driven by electric field]{(a) Electric-field-driven metal-to-insulator transition (MIT). The equilibrium system is in the vicinity of a Mott insulator at $U=1.225,\Gamma=0.00167$ and $T_\text{b}=0.0025$ in a 3D cubic lattice with the electric field applied in $x$-direction. The system undergoes electric-field-driven MIT at an electric field that is orders of magnitude smaller than bare energy scales. Depending on whether the electric field is increased or decreased, different critical fields are obtained for MIT and IMT, showing phase coexistence inside the hysteresis loop. (b) Evolution of spectral function and distribution function under electric fields, with an increasing electric field. The quasiparticle (QP) spectral weight sharply disappears at the MIT, forming an insulating gap. The non-equilibrium distribution function indicates that the system becomes hot due to Joule heating before MIT and drops back to a cold state after the transition.}
\label{prl-fig3}
\end{figure}

In Fig. \ref{prl-fig3}(a), the system is a correlated metal in the vicinity of equilibrium Mott insulator transition with $U=1.225$. We then increase electric field and compute the self-consistent solution with DMFT. The solution at a certain electric field would be used as the ``initial guess" of the next $E$-field calculations. As demonstrated above, the NDR behavior of $J-E$ relation follows the very narrow linear response regime, with $E_\text{lin}\sim10^{-4}$. A metal-to-insulator transition, or resistive switching, occurs at $E_\text{MIT}\sim0.004$, where the current suddenly drops to nearly zero. From the opposite direction, if we start from the strong-field insulating phase and gradually decrease $E$, the system abruptly transits to metallic state at a different critical field $E_\text{IMT}$. In general the insulator-to-metal transition (IMT) happens at different electric fields from the metal-to-insulator transition (MIT), i.e. $E_\text{MIT}\ne E_\text{IMT}$. The bistability of metallic/insulating solutions suggests a first-order non-equilibrium phase transition. Experiments have observed such strong non-linear $J-E$ behaviors in transition metal oxides, particularly in V$_2$O$_3$\cite{chudnovskii98} and NiO\cite{sblee}. 

In Fig. \ref{prl-fig3}(b) we plotted the spectral functions in different electric fields. As shown in the plot, the spectral function gradually changes from $E=0.0$ to $E=0.0017$. The in-gap quasiparticle is nearly unaffected by non-equilibrium effect within this range. With increasing electric field, the renormalized bandwidth $W^*$ is unchanged. However, an insulating gap suddenly opens as the metal-to-insulator transition occurs. The abrupt disappearance of QP peak verifies that a non-equilibrium Mott transition has happened due to external electric field, as suggested by the sudden drop of current in Fig. \ref{prl-fig3}. On the other hand, the local distribution function $f_\text{loc}(\omega)=-\frac{1}{2}\text{Im}G^<(\omega)/\text{Im}G^r(\omega)$ evolves from low-temperature F-D distribution to a distribution with higher $T_\text{eff}$ before resistive switching happens. After the transition, the system becomes insulating and the current is reduced by orders of magnitude. The termination of Joule heating causes the distribution function to come back to low-temperature shape. Note that although the system is cold after RS, a small residual current is flowing through it and self-consistently generating Joule heating to support the coexistence of metallic/insulating solutions. This can be seen in Fig. \ref{prl-fig4}(b), as the $T_\text{eff}$ in non-equilibrium coexistence regime should be mapped to the lower temperature boundary of the mixed phase in equilibrium phase diagram.

Note the hierarchy of energy scale, 
\begin{align}
E_\text{lin}\ll E_\text{MIT}\ll W^*,
\end{align}
which is remarkably different from quantum dot transport. We emphasize that dissipation happens at every lattice site in our model, resulting in a balance between electric power and dissipation into reservoirs. This differs from the case of quantum dot where dissipation only happens inside the electrodes and the threshold field is about the order of magnitude of QP energy scale\cite{goldhaber-gordon98, cronenwett98}.

The critical field $E_\text{MIT}\approx 0.004$ at $U=1.225$ is converted to $E_\text{MIT}=10^7-10^8$ V/m with $U=1-10$ eV. In the next section, we derive a scaling law that $E_\text{MIT}\propto\sqrt{\Gamma}$. Therefore to reach the experimental critical fields, a $\Gamma\sim 10^{-3}$ meV is required. So far the driven-dissipative Hubbard model satisfactorily captures the qualitative features of resistive switching phenomenon. A better modeling of the dissipative mechanism would be required for more quantitative calculations.

\section{Non-equilibrium phase diagram}

\begin{figure}
\centering
\includegraphics[scale=0.7]{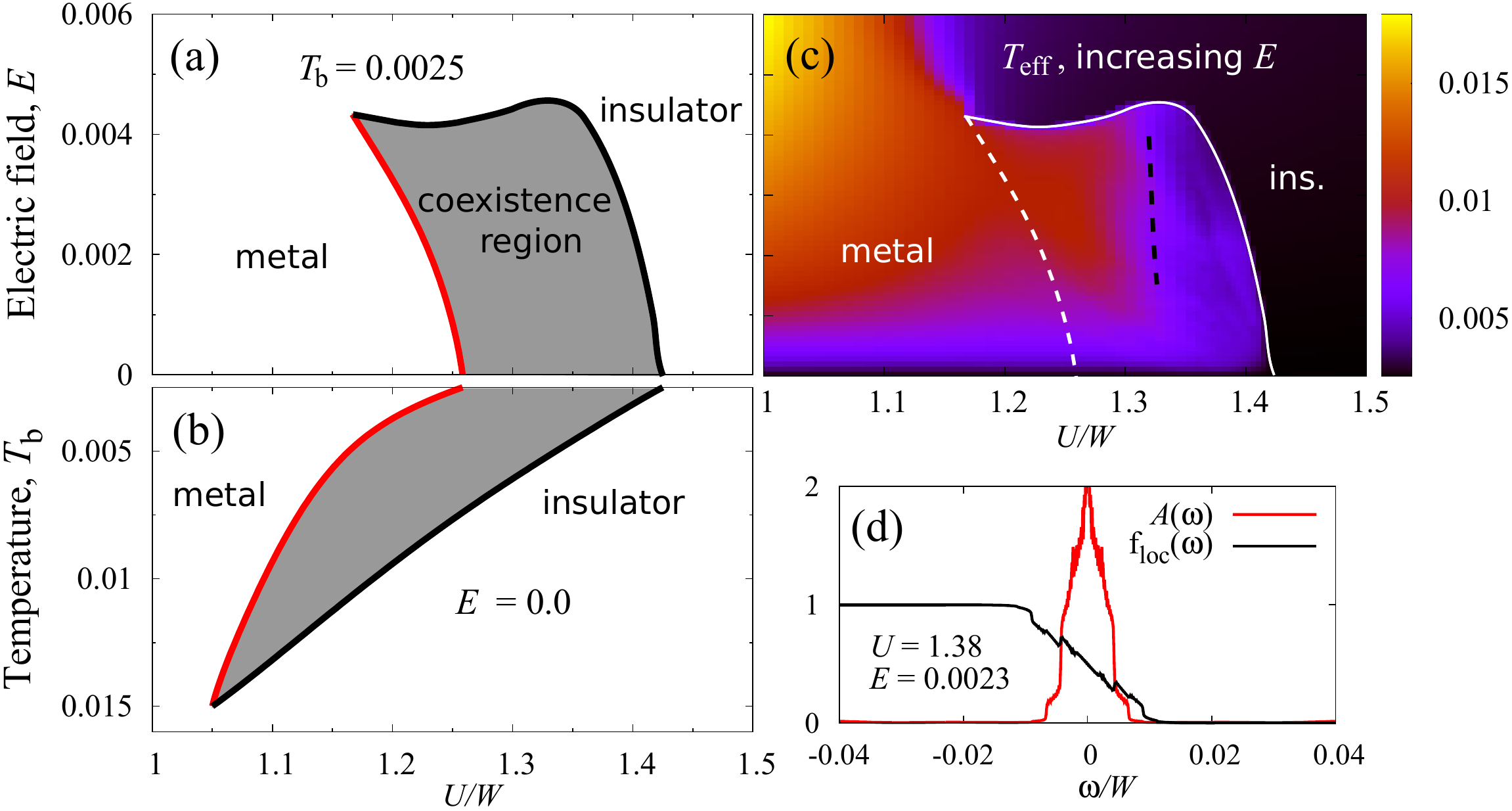}
\caption[Phase diagram of the metal-to-insulator transitions]{Phase diagram of the metal-to-insulator transitions in (a) non-equilibrium driven by electric field and (b) equilibrium driven by the bath temperature $T_\text{b}$. Calculations are done on a 3D cubic lattice, and $\Gamma=0.00167$. (c) Effective temperature map with increasing $E$ (MIT). The white dashed line is the phase boundary of IMT with a decreasing field. The black dashed line is the crossover line of different behaviors of $T_\text{eff}$. (d) Spectral and distribution functions for strong U beyond the crossover line. Quasiparticle states are disconnected from incoherent spectra and the bandwidth becomes extremely narrow. The distribution function shows strongly non-thermal properties.}
\label{prl-fig4}
\end{figure}

Fig. \ref{prl-fig4} shows the non-equilibrium phase diagram against the equilibrium case. Note that effective temperature is measured by fitting the distribution function with Fermi-Dirac distribution for data satisfying $f_\text{loc}(\omega)-0.5|<0.25$. It is seen that the non-equilibrium phase diagram looks like a reflection of that of equilibrium. Fig. \ref{prl-fig4}(c) shows effective temperature increases with increasing electric field, which is consistent with our observation in distribution function $f_\text{loc}(\omega)$. This clearly shows that Joule heating and the resulted $T_\text{eff}$ is the key concepts to understand electric-field-driven resistive switching. In addition, the seemingly counterintuitive upturn of the upper critical $E$-field ($E_\text{MIT}$ as black curve in Fig. \ref{prl-fig4}(a)) with increasing $U$ can be explained with different behaviors of effective temperature. 

\subsection{Effective temperature in interacting model}
We now discuss the effective temperature in interacting system, and use it to explain the non-equilibrium phase diagram \eqref{prl-fig4}.

First of all, we recite the equation \eqref{eflux}
\begin{align}
JE=2\Gamma\int d\omega \omega A_\text{loc}(\omega)[f_\text{loc}(\omega)-f_\text{FD}(\omega)],\nonumber
\end{align}
which also holds in interacting case. Suppose $W^*>T_\text{eff}\gg T_\text{b}$ when $U<U_\text{cross}$, the Sommerfeld expansion gives
\begin{align}
JE=\frac{\pi^2}{3}\Gamma A(0)T_\text{eff}^2,
\end{align}
which agrees with the phenomenological equation suggested by other groups\cite{altshuler09}. Physically, it tells that electric power in LHS of the equation, is balanced by energy flux in the RHS. And the energy flux is proportional to $T_\text{eff}^2$ as the internal energy of degenerate electron gas is generally proportional to $T^2$. 

\begin{figure}
\centering
\includegraphics[scale=1.1]{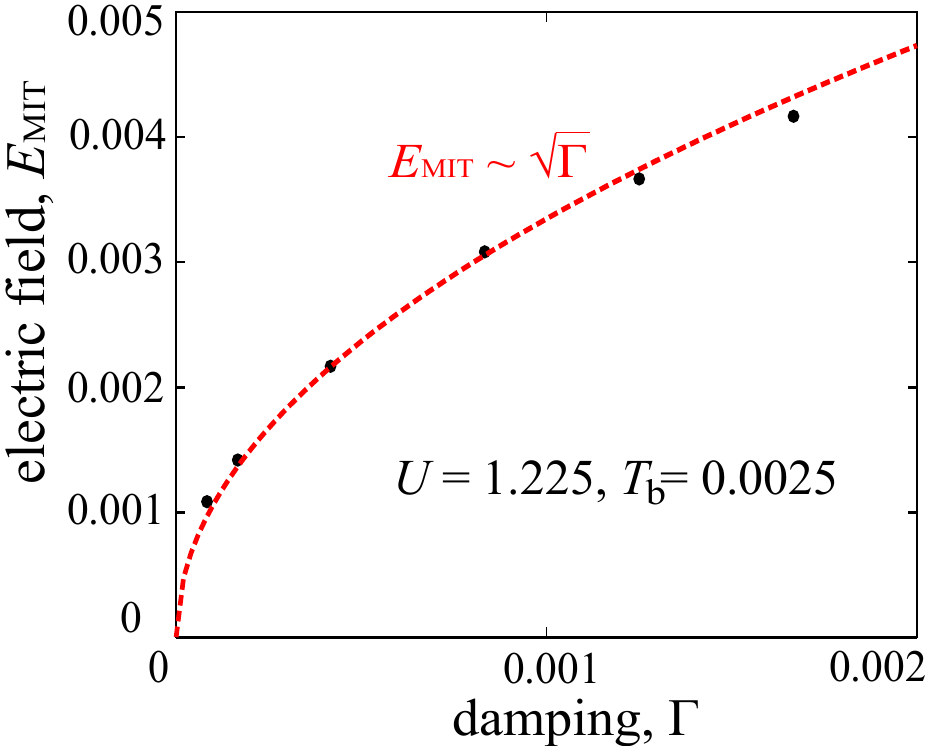}
\caption[Scaling relation of $E_\text{MIT}$ versus $\Gamma$]{Scaling of $E_\text{MIT}$ versus $\Gamma$. The $E\propto\sqrt{\Gamma}$ relation is predicted by the argument in the main text. }
\label{prl-figS2}
\end{figure}

Away from linear response regime, the scattering rate $\tau^{-1}_U$ due to electronic interaction dominates that of fermion dissipation $\Gamma$, or $\tau^{-1}_U\gg\Gamma$, then from Drude formula,
\begin{align}
J\propto\gamma\tau_UE,
\end{align}
hence we obtain
\begin{align}
\frac{E^2}{\Gamma}=C\frac{T^2_\text{eff}}{\tau_U W^2}.
\label{smflux}
\end{align}
Note that this equation only holds when $T_\text{eff}\lesssim W^*$ so that Sommerfeld expansion is valid. As Eq. \eqref{sret} suggests, the scattering time should approximately be a function of $T_\text{eff}$, i.e. $\tau_U=\tau_U(T_\text{eff})$. This would be an intuitive consequence if the thermal mechanism is actually responsible for the MIT. To make the approximation more convincing, we note the external field $E$ has been quite small compared with bandwidth $W$ in our all discussions. Moreover, the spectral weight at $\omega=0$ really does not change much before MIT as in Fig. \ref{prl-fig3}, hence the major non-equilibrium effect should be attributed to $f_\text{loc}(\omega)$. Then we conclude that $T_\text{eff}$ should only depend on $W$ and $E^2/\Gamma$,
\begin{align}
T_\text{eff}=\Theta\left(\frac{E^2}{\Gamma}\right),
\label{intteff}
\end{align}
where $\Theta$ is an unknown function depending on the functional form of $\tau_U(T_\text{eff})$. As MIT happens, the $T_\text{eff}$ should match the equilibrium transition temperature $T_\text{MIT}$, therefore $E_\text{MIT}/\Gamma^2=\Theta^{-1}(T_\text{MIT})$ leads to $E_\text{MIT}\propto\sqrt{\Gamma}$. We will discuss later the possible specific forms of $\Theta$, but Eq. \eqref{intteff} suffices to reach the conclusion that $E_\text{MIT}\propto \sqrt{\Gamma}$, as verified in Fig. \ref{prl-figS2}. This provides further supports for our conclusion of Joule heating inducing the non-equilibrium transition.

We now derive the explicit form of $T_\text{eff}$ in two cases: $T_\text{eff}\lesssim W^*$ and $T_\text{eff}\gg W^*$. When $T_\text{eff}$ is less than $W^*$, the Eq. \ref{smflux} holds, and we can insert the approximate $\tau^{-1}_U=\pi^3 A_0(0)^3U^2T^2_\text{eff}/2$ as in Eq. \ref{sret}. We arrive at the scaling relation,
\begin{align}
E^2\propto\Gamma U^2T_\text{eff}^4/W^5,\text{ or }T_\text{eff}\propto \left(\frac{E^2}{U^2\Gamma}\right)^\frac{1}{4}.
\end{align}
In this limit, effective temperature scales as $\sqrt{E}$ while electric field is increased. System will become hot and then undergo the metal-to-insulator transition upon $T_\text{eff}\sim T_\text{MIT}$. But in a different limit where $T_\text{eff}\gg W^*$, the narrow QP peak in spectral function does not allow Sommerfeld expansion, therefore the integrand in the RHS of Eq. \eqref{eflux} is effectively non-zero only within half-QP-bandwidth $\pm W^*/2$. Then we obtain that,
\begin{align}
JE\propto \frac{\Gamma W^{*2}}{W}, \text{ and } \tau^{-1}_U\propto \frac{U^2 W^{*2}}{W^3}.
\end{align}
Now $T_\text{eff}$ drops out in these equations, so that the effective temperature becomes insensitive to $E$-field.

With these cases in mind, we then examine the features of the non-equilibrium phase diagram. From Fig. \ref{prl-fig4}(c) we see that the upturn of $E_\text{MIT}(U)$ curve occurs around the crossover line of different $T_\text{eff}$ behaviors at $U_\text{cross}/W\approx1.32$. For $U<U_\text{cross}$, the QP bandwidth $W^*$ is greater than $T_\text{eff}$ and we have the scaling relation $T_\text{eff}\propto\sqrt{E/U}$. In this regime effective temperature is quickly raised by electric field until MIT occurs. But for $U>U_\text{cross}$, the QP bandwidth becomes very narrow and we have $W^*\lesssim T_\text{eff}$. The effective temperature is controlled by the bandwidth $W^*$ and depends very weakly on $E$-field. As shown in Fig. \ref{prl-fig4}(d), the distribution function in this regime shows strong non-thermal behavior controlled by the narrow QP bandwidth. The weak dependence of $T_\text{eff}$ on $E$ leads to higher critical electric field and results in the maximum around $U\sim U_\text{cross}$ in the $E_\text{MIT}(U)$ curve.

\section{Conclusion}
We have conducted calculations on dc-electric-field-driven dissipative Hubbard model to study the resistive switching phenomenon in non-equilibrium state. Our theoretical calculations successfully access both linear response regime and the strong-field limit. In particular, we find the effective temperature being the critical quantity for understanding the deviation from linear behavior as well as the field-driven metal-to-insulator transition. The $E_\text{lin}$ and $E_\text{MIT}$ are controlled by the damping $\Gamma$ rather than renormalized bandwidth. The result indicates the RS is triggered by thermal effect due to Joule heating. Coexistence of non-equilibrium metallic/insulating solutions is revealed by the DMFT calculations. Our simple model is applicable in NiO\cite{sblee} and V$_2$O$_3$\cite{mcwhan73, hansmann13}, where the material undergoes MIT as temperature increases. Phases with long range order, such as antiferromagnetism, are not considered by far, and will be considered in the next chapter. In addition, generalizations to cluster DMFT and multi-band models could address the resistive switching in more complex materials such as VO$_2$.

Although the calculations are done on uniform lattice and only homogeneous solutions are computed, the coexistence of metallic/insulating solutions implies possible segregation of different phases in the lattice. The thermodynamic state would be complex and permits inhomogeneous temperature distribution. For electric field in the coexistence regime, or $E_\text{IMT}<E<E_\text{MIT}$, we can imagine that filamentary metallic phase forms out of an insulator and orients in the direction of field, which is widely observed in experiments. 

\chapter{Microscopic Theory of Resistive Switching: Filament Formation}
\label{nano}

\section{Filament formation in Resistive Switching}
In last chapter, resistive switching is examined in a uniform lattice model. The model reproduces many realistic features of the RS and convincingly justifies the thermal scenario, but fails to capture a key observation: filament formation in real systems. In fact, it is widely observed in many systems that a conductive filament forms when a strong voltage bias provokes the RS. This phenomenon is found in ordered insulators such as VO$_2$ (with dimerized vanadium pairs) and V$_2$O$_3$ (with antiferromagnetism)\cite{zimmers13, duchene71, berglund69}. In addition, the dynamics of the conductive filament has been interpreted as electrical instabilities related to the peculiar S-shaped $I-V$ relation. 

\begin{figure}
\centering
\includegraphics[scale=0.5]{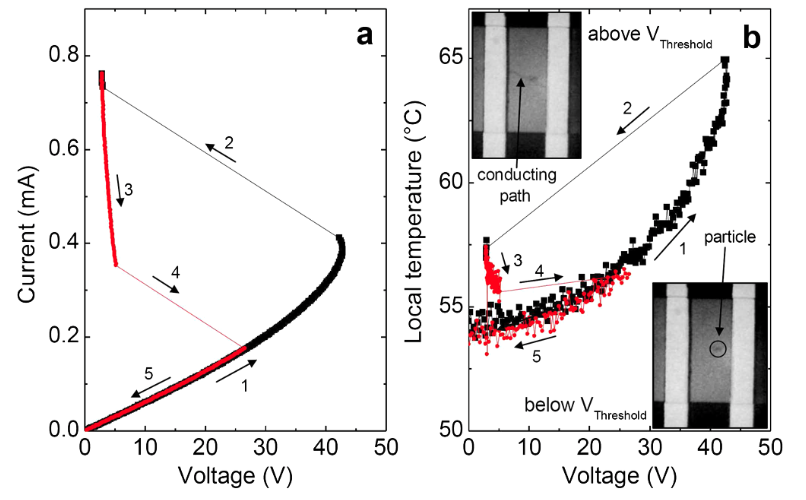}
\repeatcaption{zimmers}{Current and temperature of the VO$_2$ sample under external voltage bias. Negative differential resistance is shown in panel (a), while the formation of a conductive filament is observed in the mean time, as shown in the insets of (b). Non-equilibrium temperature is plotted against sample voltage, showing strong evidence of thermal scenario.}
\end{figure}

In this chapter, we will explain and reproduce the main features of the RS as discussed above, with a generic microscopic model. This quantum mechanical modeling is based on the dissipative lattice model we discussed in previous chapters, and internally includes broken symmetry, energy dissipation, strong correlation physics and spatial inhomogeneities. Non-equilibrium phase transition and phase segregation naturally emerges in our calculations. We will systematically explain the phenomenology observed in Fig. \ref{zimmers} and relate it to the microscopic calculations. Our results will provide crucial information about the underlying mechanism of the RS and show how thermal and electronic scenarios are connected.

\subsection{Microscopic model of a finite sample}
To consider spatial inhomogeneities in the RS, we need to do calculations on a finite-size lattice model. We consider a two-dimensional dissipative Hubbard lattice of length $L$, which is placed between two metallic electrodes. The sample is connected to an external resistor $R$ and a dc-voltage generator $V_t$. The voltage across the sample is then $V_s=V_t-IR$, where $I$ is total current through the sample. A homogeneous dc-electric field $E=V_s/L$ is established across the Hubbard lattice, pointing from one lead to the other. We introduce the fermion reservoirs connected to each lattice site, providing dissipative mechanism in the bulk. As we shall see soon, the bulk dissipation and external resistor are key ingredients to model RS, but are usually ignored in previous theoretical studies. The external resistor is critical to reveal the negative differential resistance (NDR) regime, and the dissipation is necessary to maintain a finite effective temperature and prevent the sample from overheating.

As we discussed in previous chapters, the hamiltonian is now divided into three parts, 
\begin{align}
H=H_\text{lat}+H_\text{bath+leads}+H_E,
\label{finiteinth}
\end{align}
where the three terms on the RHS are correspondingly for the Hubbard lattice, the fermion reservoirs plus electrodes and the potential slope due to dc-electric field. Note that Coulomb gauge is again chosen for expressing electric field. Specifically, the lattice part reads,
\begin{align}
H_\text{lat}=-\gamma\sum_{\langle\boldsymbol{r}\boldsymbol{r}'\rangle\sigma}(d^\dag_{\boldsymbol{r}\sigma}d_{\boldsymbol{r'}\sigma}+H.c.)+\sum_{\boldsymbol{r}\sigma}\Delta\epsilon_{\boldsymbol{r}}d^\dag_{\boldsymbol{r}\sigma}d_{\boldsymbol{r}\sigma}+U\sum_{\boldsymbol{r}}\Delta n_{\boldsymbol{r}\uparrow}\Delta n_{\boldsymbol{r}\downarrow},
\end{align}
where $d^\dag_{\boldsymbol{r}\sigma}$ creates a fermion with spin $\sigma=\uparrow,\downarrow$ in the orbital at site $\boldsymbol{r}$, and $\Delta n_{\boldsymbol{r}\sigma}=d^\dag_{\boldsymbol{r}\sigma}d_{\boldsymbol{r}\sigma}-1/2$. The overlap between electron orbitals at neighboring sites is $\gamma$, and only nearest-neighbor hopping is considered. $U$ is the strength of onsite Coulomb interaction. In realistic samples exists impurities, defects and grain boundaries, and they are modeled as site-dependent energy levels $\Delta \epsilon_{\boldsymbol{r}}$ in the hamiltonian.

As our usual strategy, the dissipative mechanism is provided by fermion reservoirs coupled to every lattice site. They are used to model the environment of the sample, including acoustic phonons. In addition, two non-interacting leads are connected at the boundaries of the sample in addition to above-mentioned reservoirs to provide voltage bias. All of the fermion reservoirs, including the two leads, are maintained in equilibrium with temperature $T_\text{bath}$. In this chapter, $T_\text{bath}$ is set to zero unless otherwise stated. This part of hamiltonian is written as follows:
\begin{align}
H_\text{bath+leads}=\sum_{\boldsymbol{r}\alpha\sigma}\epsilon_\alpha c^\dag_{\boldsymbol{r}\alpha\sigma}c_{\boldsymbol{r}\alpha\sigma}-\sum_{\boldsymbol{r}\alpha\sigma}g_{\boldsymbol{r}}(d^\dag_{\boldsymbol{r}\sigma}c_{\boldsymbol{r}\alpha\sigma}+H.c),
\end{align} 
where $c^\dag_{\boldsymbol{r}\alpha\sigma}$ creates a fermion in the orbital of the reservoir at $\boldsymbol{r}$, and $\alpha$ is the continuum index of orbitals in each reservoir. $g_{\boldsymbol{r}}$ is the local coupling to the reservoir. The reservoirs include the two leads at boundaries. We will again consider infinite flat band for reservoirs, so that damping parameter $\Gamma_{\boldsymbol{r}}=\pi|g_{\boldsymbol{r}}|^2\sum_\alpha\delta(\omega-\epsilon_\alpha)$ is defined to describe the strength of local electron relaxation. We have set $\Gamma_\text{leads}=1.0$ for electrodes and $\Gamma=0.01$ for bulk dissipation. We emphasize that the RS is a bulk non-equilibrium effect. Without bulk dissipation, the local effective temperature would be unrealistically high except for lattice sites very close to the leads.

At last is the hamiltonian for dc-field. we align the leads as well as the electric field along the $y$-direction, and define electrostatic scalar potential $\phi(\boldsymbol{r})=-yE$,
\begin{align}
H_E=\sum_{\boldsymbol{r}\sigma}\phi(\boldsymbol{r})\left(d^\dag_{\boldsymbol{r}\sigma}d_{\boldsymbol{r}\sigma}+\sum_\alpha c^\dag_{\boldsymbol{r}\alpha\sigma}c_{\boldsymbol{r}\alpha\sigma}\right)
\end{align}

To study the strong correlation physics in inhomogeneous non-equilibrium steady state, we employ the Hartree-Fock (HF) approximation, that is to introduce mean field $\langle\Delta n_{\boldsymbol{r}\sigma}\rangle$ in the Hubbard term and solve for the solution self-consistently. It is well-known that HF approximation predicts a continuous phase transition from a low-temperature/large-U anriferromagnetic insulator (AFI) to a high-temperature/small-U paramagnetic metal (PM). The order parameter is a staggered local field $\Delta_{\boldsymbol{r}}$ defined by $(-1)^{n_x+n_y}\Delta_{\boldsymbol{r}}=U\langle n_{\boldsymbol{r}\uparrow}-n_{\boldsymbol{r}\downarrow}\rangle/2$. It is generally inhomogeneous in finite sample calculations.

To solve the model self-consistently, we firstly assume the mean-field $\langle\Delta n_{\boldsymbol{r}\sigma}\rangle$ is already calculated and the associated electric current is $I$. We then conclude that the electric field $E=V_s/L=(V_t-IR)/L$ and write down the Dyson equations for non-equilibrium Green's functions in steady state,
\begin{align}
\mathbf{G}^r_\sigma(\omega)^{-1}_{\boldsymbol{r}\boldsymbol{r}'}&=[\omega-\Delta\epsilon_{\boldsymbol{r}}-\phi(\boldsymbol{r})-U\langle\Delta n_{\boldsymbol{r},-\sigma\rangle}+i\Gamma_{\boldsymbol{r}}]\delta_{\boldsymbol{r}\boldsymbol{r}'}+\gamma\delta_{\langle\boldsymbol{r}\boldsymbol{r}'\rangle},\nonumber\\
G^<_{\boldsymbol{r}\boldsymbol{r},\sigma}(\omega)&=\sum_{\boldsymbol{s}}|G^r_{\boldsymbol{r}\boldsymbol{s},\sigma}(\omega)|^2\Sigma^<_{\boldsymbol{r}}(\omega),
\label{latsum}
\end{align}
where the summation is over all lattice sites $\boldsymbol{s}$. The lesser self energy $\Sigma^<_{\boldsymbol{r}}(\omega)=2i\Gamma_{\boldsymbol{r}} f_\text{FD}(\omega-\mu_{\boldsymbol{r}})$, where $\mu_{\boldsymbol{r}}=\phi(\boldsymbol{r})$ is the local chemical potential. Using the lesser Green's functions, the local mean field can be evaluated as
\begin{align}
\langle\Delta n_{\boldsymbol{r}\sigma}\rangle=\langle\Delta d^\dag_{\boldsymbol{r}\sigma} d_{\boldsymbol{r}\sigma}\rangle-\frac{1}{2}=-iG^<_{\boldsymbol{rr},\sigma}(t,t)-\frac{1}{2},
\end{align}
and the electric current distribution per spin 
\begin{align}
I_{\boldsymbol{r},\boldsymbol{\hat{e}}}=i\gamma\sum_\sigma\langle d^\dag_{\boldsymbol{r}+\boldsymbol{\hat{e}}\sigma}d_{\boldsymbol{r}\sigma}-H.c.\rangle=2\text{Re}G^<_{\boldsymbol{r}+\boldsymbol{\hat{e}},\boldsymbol{r}\sigma}(t,t),
\end{align}
for two neighboring sites $\boldsymbol{r}$ and $\boldsymbol{r+e}$. The total current $I$ is then obtained by summing over $I_{\boldsymbol{rr}'\sigma}$ on any cross section of the sample and both spin orientations. With the newly calculated mean-field and current $I$, we can then update the sample voltage $V_S$ and repeat the procedure above until convergence is reached.

\subsection{Current leak in finite sample calculation}

In Chap. \ref{prb}, we proved that current leak is exactly zero in a homogeneous infinite lattice. However in finite sample calculation, especially with disorder, there is no guarantee that current leak would be locally zero everywhere. Instead, one can derive the current leak,
\begin{equation}
I_{{\rm leak},\sigma \boldsymbol{ r}} = 
-\sum_{\alpha\sigma}g_{\boldsymbol{ r}}(\langle d^\dagger_{\boldsymbol{r}\sigma}
c_{\boldsymbol{ r}\alpha\sigma}\rangle-{\rm H.c.})
 =  2\Gamma_{\boldsymbol{r}}\int \!
\omega \,
A_{\sigma \boldsymbol{ r}}(\omega)[f_{\sigma \boldsymbol{
r}}(\omega)-f_0(\omega-\mu_{\boldsymbol{ r}})],
\label{leak}
\end{equation}
with the local density of states $A_{\sigma\boldsymbol{r}}(\omega)=-\pi^{-1}\mbox{Im } G^{\rm r}_{\boldsymbol{r}\boldsymbol{ r}\sigma}(\omega)$ and the local
nonequilibrium distribution function $f_{\sigma \boldsymbol{
r}}(\omega)=G^<_{\boldsymbol{
r}\boldsymbol{ r}\sigma}(\omega)/(2\pi
 A_{\sigma \boldsymbol{ r}}(\omega))$.
 
To prevent current leak in the finite sample case, we adjust the local chemical potential In each HF iteration to satisfy the zero-leak condition, $I_{{\rm leak},\sigma \boldsymbol{ r}} =0$. Practically we find that current leak is always very small (up to 2\% to 3\% of local current density).

\section{Landau-Zener tunneling versus thermal effect}
\label{LZthermal}

\subsection{Recursion relation in the presence of long range order}
We firstly discuss the underlying mechanism that brings the system to an RS with external field and bulk dissipation. In a purely thermal scenario, it may be conjectured that non-equilibrium effect will only enter the picture as raised $T_\text{eff}(E)$. However, it turns out to be more subtle than a direct modification of temperature. As observed in experiments, the RS is strongly discontinuous and has a clear hysteretic $I-V$ curve, whereas mean-field theory predicts a continuous temperature-controlled transition. To resolve this puzzle, we start with examining resistive switching in infinite uniform lattice, where $L\to\infty$ and $\Delta \epsilon_{\boldsymbol{r}}=0$. In this case, a set of recursion relations similar to Eq. \eqref{recursion} can be derived. These formulations are crucial to efficiently compute Green's functions. 

Firstly, we consider the case that the sample is cut along the (10) lattice orientation, and the electric field as well as $y$-axis is along (10) direction. With staggered field $\Delta_{\boldsymbol{r}}=(-1)^{n_x+n_y}\Delta$, the TB lattice decomposes into sublattices A and B, according to whether $n_x+n_y$ is even or odd. In the case of $E=0$, the one-electron hamiltonian can be transformed to momentum space separately for A-sites and B-sites,
\begin{align}
h_{\text{TB}.\sigma}=\begin{pmatrix}\Delta_\sigma&\epsilon_{\boldsymbol{k}}\\\epsilon_{\boldsymbol{k}}&-\Delta_\sigma\end{pmatrix},
\end{align}
where spin-dependent 
\begin{align}
\Delta_\sigma=\begin{cases}\Delta, &\sigma=\uparrow\\
-\Delta, & \sigma=\downarrow\end{cases},
\end{align}
and the range of $\boldsymbol{k}$ is the reduced First Brillouin zone $[-\pi/2a,\pi/2a]\times[-\pi/2a,\pi/2a]$ because translational invariance holds only for period $2a$ now. The single-band energy $\epsilon_{\boldsymbol{k}}=-2\gamma\sum_{k_i=x,y}\cos(2k_ia)$. Then the energy spectrum is solved to be $\epsilon_{\pm,\boldsymbol{k}}=\pm\sqrt{\epsilon_{\boldsymbol{k}}^2+\Delta^2}$.

Now with finite $E>0$, the hamiltonian has similar translational invariance with period $2a$ in the direction perpendicular to $\boldsymbol{E}=E\boldsymbol{\hat{e}}_y$, i.e. $x$-direction. Therefore the hamiltonian is divided into rows of different $n_y$, each of whom is a one-dimensional lattice consists of alternating A/B sites. We can Fourier-transform the hamiltonian in $x$-direction for all $n_y$,
\begin{align}
H_{TB}(k_x,\sigma)&=\sum_{n_y}\begin{pmatrix}d_{An_yk_x\sigma}\\d_{Bn_yk_x\sigma}\end{pmatrix}^\dag \begin{pmatrix} (-1)^{n_y}\Delta_\sigma & \epsilon_{k_x}\\\epsilon_{k_x} & (-1)^{n_y+1}\Delta_\sigma\end{pmatrix}\begin{pmatrix}d_{An_yk_x\sigma}\\d_{Bn_yk_x\sigma}\end{pmatrix}+\nonumber\\
&+\gamma\sum_{n_y}\left(d^{\dag}_{An_yk_x\sigma}d_{Bn_y-1k_x\sigma}+d^{\dag}_{Bn_yk_x\sigma}d_{An_y-1k_x\sigma}+H.c.\right),
\end{align}
with $d_{s n_yk_x\sigma}$ annihilating the fermion of spin $\sigma$ in $s$-lattice with $x$-momentum $k_x$ and row-index $n_y$. $s$ can be either A or B. Note that inter-row coupling is only between sites of different sublattices. 

Other parts of the total hamiltonian, including those of coupling with fermion reservoirs, are all local products of operators, and in momentum representation they simply become products of the same $k_x$ and $\sigma$. In summary, we now have independent pieces of one-dimensional hamiltonian with fixed $k_x$. They are separately connected to fermion reservoirs, 
\begin{align}
H_\text{bath}(k_x,\sigma)=-g\sum_{sn_y\alpha}(c^\dag_{sn_yk_x\alpha}d_{sn_yk_x}+H.c.)+\sum_{sn_y\alpha}\epsilon_\alpha c^\dag_{sn_yk_x\alpha}c_{sn_yk_x\alpha}
\end{align}
and have potential slope terms,
\begin{align}
H_E(k_x,\sigma)=-\sum_{n_y}yE\left(d^\dag_{sn_yk_x}d_{sn_yk_x}+\sum_\alpha c^\dag_{sn_yk_x\alpha}d_{sn_yk_x\alpha}\right).
\end{align}
We can then generalize the recursion relations defined in Eq. \eqref{recursion} to the case with antiferromagnetic order. A key difference in this case is that all of the Green's functions and self energies should have four components, and should be written as matrix
\begin{align}
\mathbf{G}_{n_yk_x\sigma}=\begin{pmatrix}G_{n_yk_x\sigma,AA}&G_{n_yk_x\sigma,AB}\\G_{n_yk_x\sigma,BA}&G_{n_yk_x\sigma,BB}\end{pmatrix}\nonumber\\
\mathbf{\Sigma}_{n_yk_x\sigma}=\begin{pmatrix}\Sigma_{n_yk_x\sigma,AA}&\Sigma_{n_yk_x\sigma,AB}\\\Sigma_{n_yk_x\sigma,BA}&\Sigma_{n_yk_x\sigma,BB}\end{pmatrix}.
\end{align}
And to account for the fact that electrons only hop between sites in opposite sublattice, we define
\begin{align}
\hat{T}&=\begin{pmatrix}0&&1\\1&&0\end{pmatrix},\text{ and}\\
\tilde{A}&=\hat{T}A\hat{T}.
\end{align}
We also define the single-particle hamiltonian,
\begin{align}
h(k_x,\sigma)=\begin{pmatrix}\Delta_\sigma&\epsilon_{k_x}\\\epsilon_{k_x}&-\Delta_\sigma\end{pmatrix}.
\end{align}
Now we can write down the recursion relations,
\begin{align}
\mathbf{F}^r_{\pm,k_x\sigma}(\omega)^{-1}=\omega-h(k_x,\sigma) -\mathbf{\Sigma}^r(\omega)-\gamma^2\mathbf{\tilde{F}}^r_{\pm,k_x\sigma}(\omega\pm E)\\
\mathbf{F}^{\lessgtr}_{\pm,k_x\sigma}(\omega)=\mathbf{F}^{r}_-(\omega)[\mathbf{\Sigma}^{\lessgtr}(\omega)+\gamma^2\mathbf{\tilde{F}}^{\lessgtr}_-(\omega-E)]\mathbf{F}^{r,\dag}_-(\omega),
\label{afrecursion}
\end{align}
where $\mathbf{\Sigma}^r(\omega)=-i\Gamma \mathbb{I}$ and $\mathbf{\Sigma}^<(\omega)=2i\Gamma f_\text{FD}(\omega)\mathbb{I}$. The local Green's functions are then computed,
\begin{align}
\mathbf{G}^r_{\text{loc},\sigma}(\omega)^{-1}&=\omega-h(k_x,\sigma)-\Sigma^r(\omega)-\gamma^2 \mathbf{\tilde{F}}^r_
{\text{tot},\sigma}(\omega)\\
\mathbf{G}^<_{\text{loc},\sigma}(\omega)&=\mathbf{G}^r_{\text{loc},\sigma}(\omega)[\Sigma^<(\omega)+\gamma^2 \mathbf{\tilde{F}}^<_{\text{tot},\sigma}(\omega)]\mathbf{G}^{r,\dag}_{\text{loc},\sigma}(\omega),
\label{gloc22}
\end{align}
With total hybridization functions $\mathbf{F}_{\text{tot},\sigma}$ similarly defined.

Now we have derived the recursion relations for the electric field aligned in (10)-direction. When we apply the field in (11)-diagonal, recursion relations is derived in a similar manner. The set of relations is actually simpler in that case, where in the same row (of index $n_y$) all atoms are in the same sublattice. So the Green's function matrices again reduce to scalar functions $G^{r,\lessgtr}_{sk_x\sigma}(\omega)$ with $s=A,B$.

\begin{figure}
\centering
\includegraphics[scale=0.7]{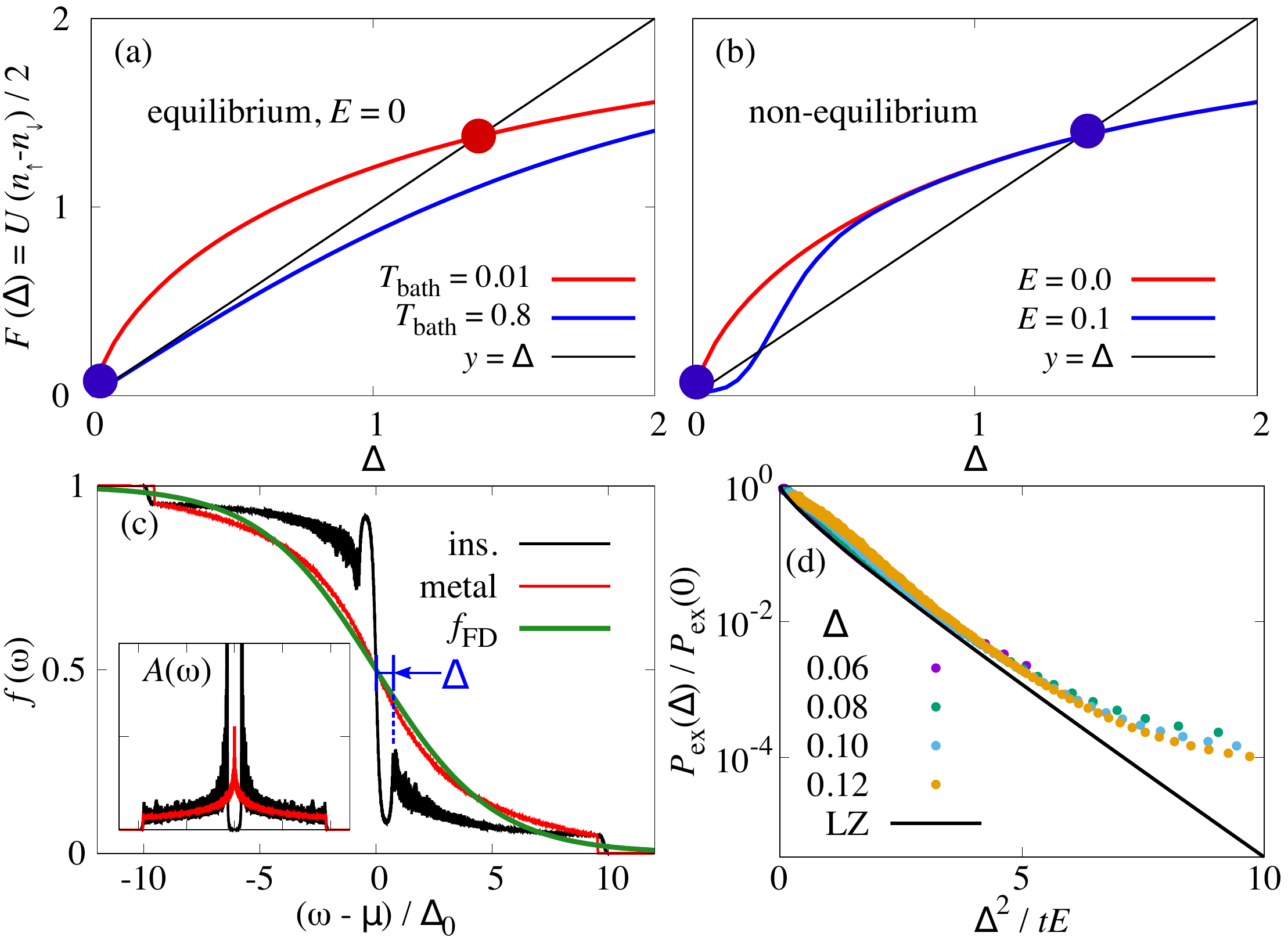}
\caption[Mechanism of the RS in an infinite system]{(a-b)self-consistent condition in HF approximation in (a) equilibrium and (b) non-equilibrium. The mean-field condition is $\Delta=F(\Delta;E,T_\text{bath})=\frac{1}{2}U\langle n_\uparrow-n_\downarrow\rangle$ for an antiferromagnetic order parameter. The order parameter continuously evolves to zero when $T_\text{bath}$ increases in equilibrium. In non-equilibrium, Joule heating in the metallic side increases $T_\text{eff}>T_\text{bath}$, resulting in stable PM solution. Bistable regime of both metallic/insulating solutions thus emerges due to the non-equilibrium physics. (c) Local distribution function in metallic and insulating phases under electric field. The Fermi-Dirac function with $T_\text{eff}=1.05$ is shown for comparison. (d) Total number of non-equilibrium excitations above the chemical potential, $P_\text{ex}(\Delta)$, compared with Landau-Zener tunneling rate. Numerical data of $P_\text{ex}(\Delta)$ match well the Fermi-surface averaged Landau-Zener tunneling rate, with a damping $\Gamma=0.001$.}
\label{nano-fig1}
\end{figure}

Using the recursion relations, we implement HF approximation and solve for the model self-consistently with different parameters. The solution is determined with a self-consistent condition on the order parameter,
\begin{align}
\Delta=F(\Delta;E,T_\text{bath},\Gamma)=\frac{U}{2}\langle n_\uparrow-n_\downarrow\rangle.
\end{align}
In equilibrium, the Slater HF theory predicts a second-order phase transition between AFI and PM as temperature approaches the N\'{e}el temperature $T_\text{bath}=T_\text{N}$. As shown in Fig. \ref{nano-fig1}(a), there is only one AFI solution with finite gap $\Delta=\Delta_0$ when $T_\text{bath}<T_\text{N}$. The gap continuously evolves to zero as $T_\text{bath}$ goes beyond $T_\text{N}$. 

In non-equilibrium, however, the situation is dramatically different. Two stable solutions now exist at finite electric field: an AFI solution with $\Delta=\Delta_0$ and a PM solution with $\Delta=0$ which was unstable in equilibrium. In non-equilibrium state, strong Joule heating occurs in metallic state and results in high effective temperature, so that the metallic state is stabilized when $T_\text{eff}\propto E/\Gamma\sim T_\text{N}$. But in the AFI solution large resistance due to the insulating gap strongly reduces Joule heating and the effective temperature, so that the ordered phase is still stable. The intermediate solution is unstable. In short, the same electric field induces dramatically different Joule heating in AFI and PM solutions, stabilizing both solutions in the appropriate parameter regime. Therefore the resistive switching is discontinuous due to bistable high/low temperature states, with different critical fields for insulator-to-metal transition (IMT) upon increasing electric field and metal-to-insulator transition (MIT) upon decreasing electric field. Just as we discussed in Chapter \ref{prl}, it opens the possibility of heterogeneous phases in RS. 

To reveal the underlying mechanism of the RS, we plot the local distribution function 
\begin{align}
f_{\boldsymbol{r}}(\omega)=-\frac{\text{Im}G^<_{\boldsymbol{rr}\sigma}(\omega)}{2\text{Im}G^r_{\boldsymbol{rr}\sigma}(\omega)},
\end{align}
in Fig. \ref{nano-fig1}(c). The non-equilibrium distribution functions of both metallic/insulating solutions deviate  from the Fermi-Dirac form at finite electric field. In particular for the insulating solution, significant non-equilibrium excitations are created beyond the insulating gap, forming a peak at the bottom of upper AF band as well as a symmetric valley in the lower band. In metallic solution, the distribution function has a similar overall form as F-D distribution (green curve), but actually has a different functional form. Fig. \ref{nano-fig1}(d) shows total number of non-equilibrium excitations in insulating phase, with a variety of values of the gap. It matches the prediction of Landau-Zener tunneling rate very well, indicating the electronic mechanism is responsible for the RS. The deviation in the small $E$ (large $\Delta/tE^2$) regime is due to damping from fermion baths. This finding shows that quasi-particles are accelerated by the external electric field, and are tunneling across the insulating gap to the upper band, rendering the system metallic. Note that the insulating gap $\Delta$ is self-consistently determined in HF iterations, resulted from the balance between the electronic interactions, the external driving field, and the bulk dissipation. As we shall see below, the electronic mechanism is still compatible with the thermal description featuring the effective temperature. And the non-equilibrium excitations can be interpreted satisfactorily as (non-equilibrium) thermal excitations. 

\subsection{Distribution function and LZ mechanism}
We have seen in Chapter \ref{prb} and \ref{prl} that the local distribution function provides rich information on the non-equilibrium dynamics of a lattice system. In the inhomogeneous case, we can expand the definition of the local distribution with Eq. \eqref{latsum}. Then $f_\text{loc}(\omega)$ can be written in terms of retarded Green's functions and equilibrium Fermi-Dirac function of the fermion reservoirs coupled to all lattice sites,
\begin{align}
f_{\boldsymbol{r}}(\omega)&=-\frac{1}{2}\frac{\text{Im}G^<_{\boldsymbol{rr}\sigma}(\omega)}{\text{Im}G^r_{\boldsymbol{rr}\sigma}(\omega)}\nonumber\\
&=\frac{\sum_{\boldsymbol{r'}}|G^r_{\boldsymbol{rr'}}(\omega)|^2f_\text{FD}(\omega+\boldsymbol{r'}\cdot \boldsymbol{E})}{\sum_{\boldsymbol{r'}}|G^r_{\boldsymbol{rr'}}(\omega)|^2}.
\label{flocsum}
\end{align}
In homogeneous infinite lattice, we concentrate on the local distribution function $f_{\boldsymbol{00}}(\omega)$. As we see, the final expression of (non-equilibrium) local distribution function is nothing but a weighted average over all bath F-D distributions maintained at temperature $T_\text{bath}$. The weights are just retarded Green's functions from site $\boldsymbol{r}$ to all lattice sites $\boldsymbol{r'}$. Intuitively, the quantities $|G^r_{\boldsymbol{rr'}}(\omega)|^2$ express the quantum correlation between two distant lattice sites $\boldsymbol{r},\boldsymbol{r'}$. In our model, they are responsible for conveying statistical information across the whole lattice in the non-equilibrium.

In non-interacting model with $E,\Gamma\ll \gamma$, the correlation $|G^r_{\boldsymbol{rr'}}(\omega)|^2$ is quite smooth and symmetric around sites $\boldsymbol{r'}\sim\boldsymbol{r}$. And we have known that the non-equilibrium distribution function is a superposition of small steps coming from Fermi-Dirac distributions of all lattice sites with chemical potential $-\phi(\boldsymbol{r})=-\boldsymbol{r}\cdot\boldsymbol{E}$. Further in the limit $E/\Gamma\ll1$, the non-equilibrium distribution function is smooth enough to have a well-defined effective temperature. In more extreme cases, the distribution function becomes dramatic, but its overall profile and first moment can still be used to define $T_\text{eff}$. In the AFI state, on the contrary, the non-equilibrium distribution function becomes essentially different from equilibrium Fermi-Dirac function. As shown in Fig. \ref{nano-fig1}(c), quasi-particles are excited from the lower AF band to the upper AF band. Based on the discussion above, we stress that this excited distribution is also a result of the superposition in Eq. \eqref{flocsum}, thus is totally due to quantum-mechanically electronic mechanism. In particular, the time-scale to establish the non-equilibrium distribution is much shorter than that of any thermal diffusion dynamics.

\begin{figure}
\centering
\includegraphics[scale=0.4]{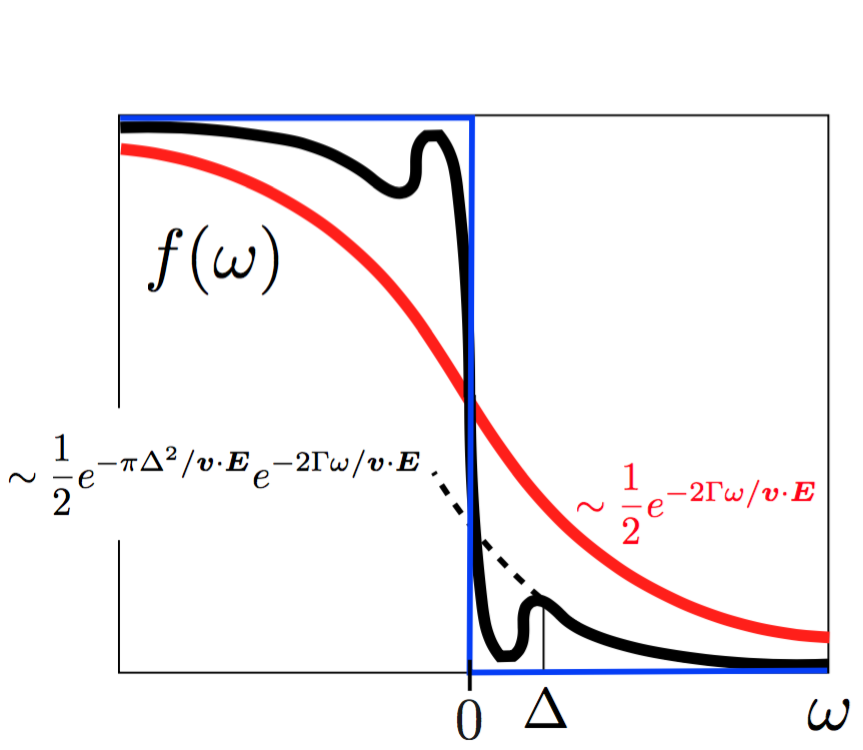}
\caption[Demonstration of non-equilibrium distribution function in AFI state]{Schematic plot of the electric-field-driven distribution function in AFI state. The exponential decay in the tail of distribution function in metallic state is due to dephasing of fermion baths. On top of this tail, the non-equilibrium distribution function in AFI state is further reduced by Landau-Zener tunneling probability and is depleted inside the AF gap.}
\label{nano-figS3a}
\end{figure}

To further elaborate on the non-equilibrium distribution function in the presence of AF gap and electric field, we show a schematic decomposition of function in Fig. \ref{nano-figS3a}. For metallic state, electrons freely tunnel through the lattice and are dephased by fermion reservoirs with damping parameter $\Gamma$. This dissipation leads to an exponential decay in tunneling probability between distant sites, or $\ln|G^r_{\boldsymbol{0r}}|^2\sim -\Gamma|\boldsymbol{r}|$. And for $\omega>0$, assuming $T_\text{bath}=0$ and $\boldsymbol{E}\parallel \boldsymbol{\hat{e}}_y$, the Fermi-Dirac functions $f_\text{FD}(\omega+\boldsymbol{r}'\cdot\boldsymbol{E})$ are non-zero only when $y<-\omega/E$. This results in the $\exp(-2\Gamma/\boldsymbol{v}\cdot\boldsymbol{E})$ in local distribution function. The scaling relation $T_\text{eff}\sim E/\Gamma$ can also be viewed as a conclusion of this tunneling behavior. In AFI state, this argument is still valid, and something else happens on top of it: Landau-Zener tunneling. Due to the AFI gap, the exponential-decay factor is further reduced by the LZ-tunneling probability $\exp(-\pi\Delta^2/\boldsymbol{v}\cdot\boldsymbol{E})$. And the electrons are depleted in the range within the range of $\Delta$ because propagation inside the gap is forbidden. 
 
The panel (d) of Fig. \ref{nano-fig1} supports the above discussions. It compares the numerical data on the total number of non-equilibrium excitations above the chemical potential ($P_\text{ex}(\Delta)$) with the Landau-Zener tunneling rate,
\begin{align}
P_\text{ex}(\Delta)=\int_0^\infty A(\omega;\Delta)f_\text{loc}(\omega;\Delta)d\omega,
\end{align}
with local spectral function $A(\omega;\Delta)$ of gap $\Delta$. This quantity is computed numerically. And the Landau-Zener tunneling is the result of two competing processes: tunneling with a rate $\gamma_\text{LZ}(\boldsymbol{k})\sim E\exp(-\pi\Delta^2/|\boldsymbol{v}_{\boldsymbol{k}}\cdot\boldsymbol{E}|)$ and the electronic relaxation with a rate $\Gamma$. Therefore the stationary condition gives $P_\text{ex}(\boldsymbol{k},\Delta)\gamma_\text{LZ}(\boldsymbol{k})=[1-P_\text{ex}(\boldsymbol{k},\Delta)]\Gamma$ for net tunneling rate $P_\text{ex}(\boldsymbol{k},\Delta)$ at momentum $\boldsymbol{k}$. We then obtain the total rate by summing over the Fermi surface, 
\begin{align}
P_\text{ex}(\Delta)=\frac{1}{\mathcal{S}_\text{FS}}\int_{\boldsymbol{k}\in FS}d\boldsymbol{k}\frac{\gamma_\text{LZ}(\boldsymbol{k})}{\gamma_\text{LZ}(\boldsymbol{k})+\Gamma}.
\end{align}
In the two-dimensional case where $\boldsymbol{E}$ is along the (11) diagonal, the Fermi surface integral is reduced to 
\begin{align}
P_\text{ex}(\Delta)=\frac{1}{\mathcal{S}_\text{FS}}\int_{\boldsymbol{k}\in FS}d\boldsymbol{k}\frac{\gamma_\text{LZ}(\boldsymbol{k})}{\gamma_\text{LZ}(\boldsymbol{k})+\Gamma}.
\end{align}
This is directly compared with numerical data in Fig. \ref{nano-fig1}(d), which demonstrates excellent agreement for a variety of different $\Delta$'s. This provides robust numerical evidence that the RS is triggered by electronic mechanism, i.e. Landau-Zener tunneling.

\section{Filament formation and negative differential resistance}
\begin{figure}
\centering
\includegraphics[scale=0.7]{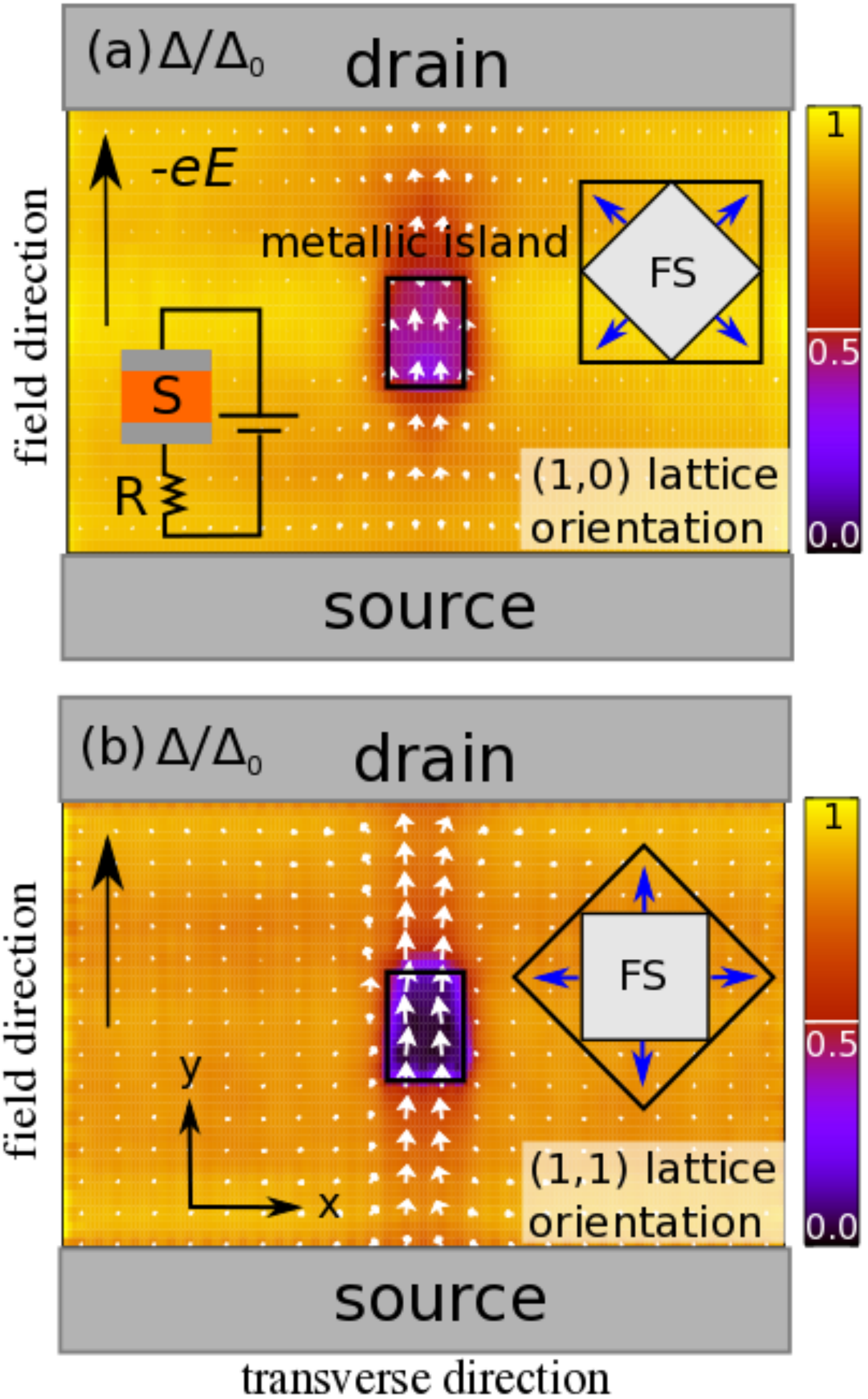}
\caption[Conductive pattern formation and anisotropy]{Formation of conductive pattern in an insulating sample under dc-electric field. A $5\times5$ metallic island is created in the center of the sample. The color map shows the magnitude of AF order parameter, and the white arrows indicate the direction and magnitude of the local current. (a) The sample is cut along (10)-direction. No filament is observed, and the current barely flows outside the impurity. (b) The sample is cut along (11)-direction. A robust conductive filament forms through the impurity along the field-direction. Current flows between the leads through the filament. This is attributed to the anisotropy of Fermi surface in half-filling lattice. The filament forms easily when $E$-field is aligned with Fermi velocity $\boldsymbol{v}_F$. $E=0.252\Delta_0$, with the equilibrium gap $\Delta_0=1.35, U=4.0$ and $T_\text{bath}=0.3$.}
\label{nano-fig2}
\end{figure}

After discussing the mechanism of the RS in infinite system, let us consider now a more realistic device sample, which is modeled by a finite Hubbard lattice connected to source/drain leads. The finite lattice is of 1200 lattice sites with size $(80a/\sqrt{2})\times(30a/\sqrt{2})$. To make the model realistic and to investigate the roles of spatial inhomogeneity, we create a $5\times5$ metallic island with $\Delta \epsilon_{\boldsymbol{r}}=1.5\gamma$ at the center of the insulating sample. In Fig. \ref{nano-fig2}, local order parameter $\Delta_{\boldsymbol{r}}$ and local current are plotted. Two lattice orientations are considered; (10)-direction as $y$-axis and (11)-diagonal as $y$-axis. It turns out that cutting the sample in different lattice orientations really change its behavior during the RS. In the former case, the RS occurs almost uniformly without noticeable pattern formation at electric fields close to the switching field obtained in infinite lattice; but in the latter case, the electric field is along (11)-direction and a strong and collimated conducting filaments form at much weaker electric fields. 

The anisotropy is attributed to the different orientations of Fermi surface. In a half filled square lattice, the Fermi velocity at Fermi surface $\boldsymbol{v}_\text{F}$ is actually along the (11)-direction in real space. Therefore, the (11)-direction is the ``easy direction" for electrons to move and for the filament to form. In weak-field and non-interacting limit, we can actually derive an expression of field-direction-dependent $T_\text{eff}(\boldsymbol{E})$,
\begin{align}
T_\text{eff}\sim \frac{|\boldsymbol{v}_\text{F}\cdot\boldsymbol{E}|}{\Gamma},
\end{align}
which supports the anisotropy in the aspect of Joule heating. We leave the detailed derivation in the appendix. In real samples, we usually expect a polycrystalline structure, hence the filaments should be globally collimated with the external field, but with microscopic domain walls aligned with $\boldsymbol{v}_\text{F}$.

\begin{figure}
\centering
\includegraphics[scale=0.9]{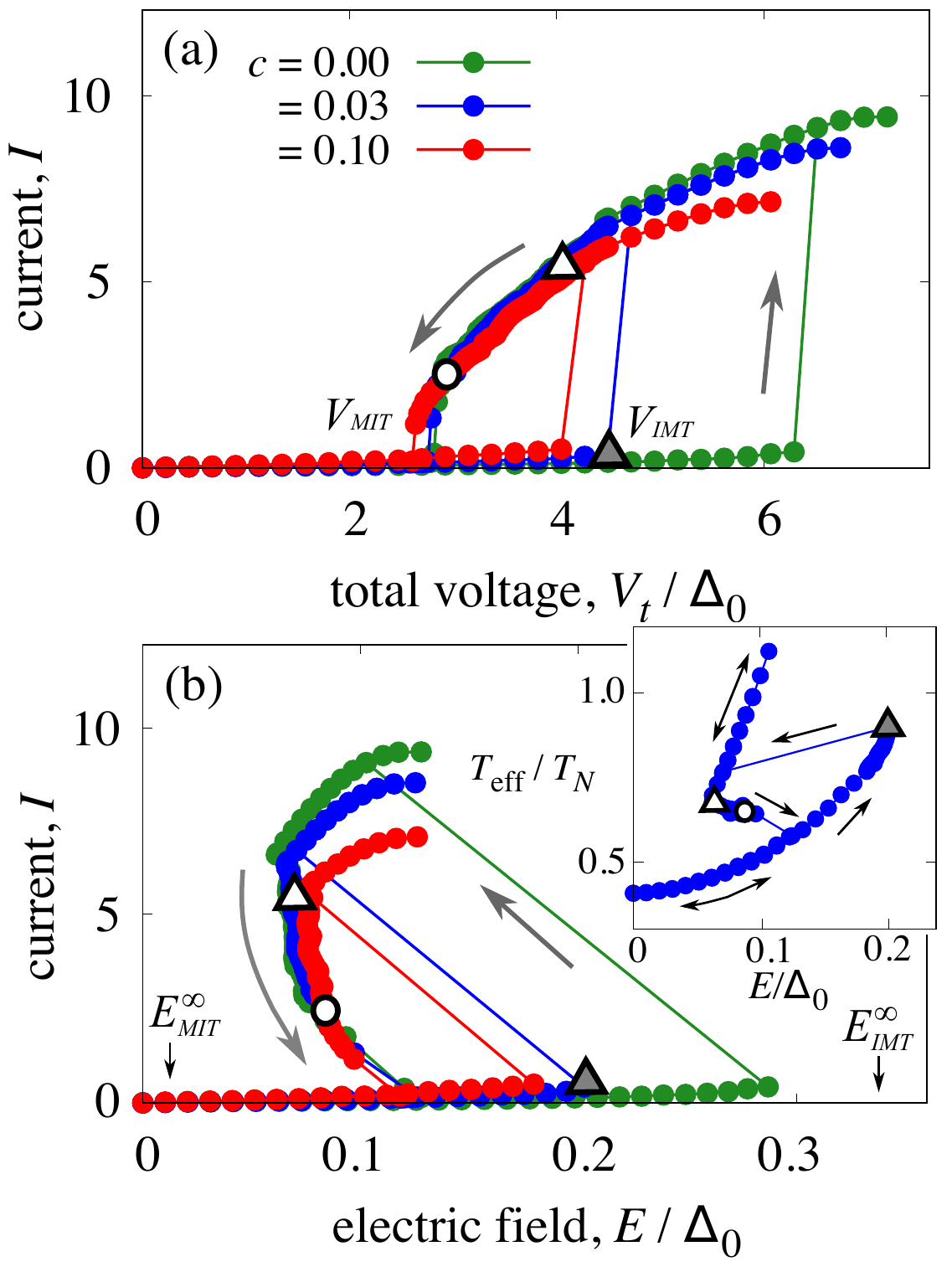}
\caption[$I-V$ relation in finite samples]{$I-V$ relations of a rectangular sample cut along the (11)-lattice orientation. Increasing $V_t$ in insulating phase leads to IMT when $E= E_\text{IMT}$, which is much reduced by bulk disorders. Decreasing $V_t$ from metallic phase causes the sample to enter an NDR branch ($dI/dV_s<0$) and undergoes IMT, resulted from narrowing of a conductive filament. $E^\infty_\text{IMT}$ and $E^\infty_\text{MIT}$ are the threshold fields in infinite uniform lattice. The inset of (b) plots the average effective temperature versus electric field for $c=0.03$. It shows hysteretic behavior and average $T_\text{eff}$ approaches $T_\text{N}$ around the IMT. The upward kink near the white triangle is a finite-size effect and will disappear in the limit of a large system.}
\label{nano-fig3}
\end{figure}

With the understanding of the prototypical model above, we now discuss a model with randomly distributed metallic impurities. For different concentration of impurities $c$, the hysteretic $I-V$ curves are plotted in Fig. \ref{nano-fig3} as a function of (a) total voltage $V_t$ and (b) the electric field $E=V_s/L$. We find a sharp MIT upon increasing electric field and the IMT upon decreasing bias. During the MIT, the corresponding electric field is increasing as current decreases in (b). This negative-differential-resistance (NDR, $dI/dV_s<0$) behavior will be discussed in more details later.

The critical electric field $E_\text{IMT}$ is found to be fractions of the equilibrium order parameter $E_\text{IMT}\sim0.2\Delta_0$, and is strongly reduced when impurities are present in the sample. On the other hand, the threshold field of MIT is generally insensitive to impurities. As we shall see below, this difference is explained with the distinct nature of the two transitions. 

\begin{figure}
\centering
\includegraphics[scale=0.7]{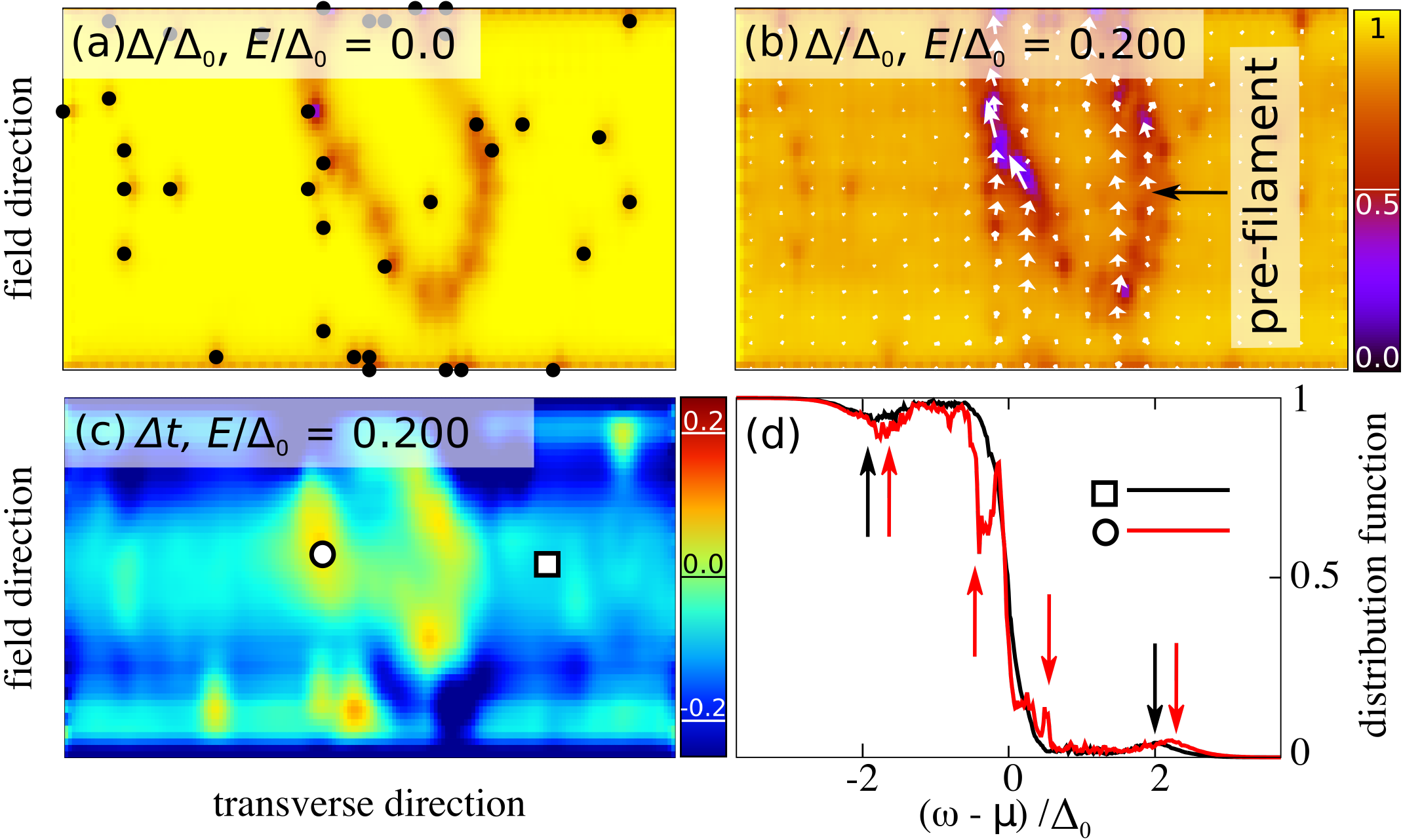}
\caption[Pattern formation in IMT]{Pattern formation right before the IMT in a sample with impurities randomly distributed and $c=0.03$. The sample is cut along (11)-diagonal. (a-b) AF order parameter (color map) and current (white arrows) for increasing electric fields. (a) At $E=0$, the impurities (black dots) create patterns with low-$\Delta$, which will become precursors of filament formation. (b) At $E=0.200\Delta_0$ (grey triangle in Fig. \ref{nano-fig3}), a conductive path (pre-filament) forms around the low-$\Delta$ region, triggering the IMT. (c) Distribution of effective temperature $\delta t \equiv (T_\text{eff}-T_\text{N})/T_\text{N}$. Note its similarity with current pattern in (b). (d) Non-equilibrium distribution function at sites marked with white circle and square in (c). The arrows point to strong non-equilibrium excitations.}
\label{nano-fig4}
\end{figure}

We firstly look at the IMT. Fig. \ref{nano-fig4}(a-b) demonstrates how order parameter and current evolves under electric fields. At $E=0$, extended metallic inhomogeneities (darker regions) exist in the sample. It connects the two leads even when the concentration $c=0.03$ is far below the classical 2$d$ percolation threshold. The lower-$\Delta$ pattern is formed due to distributed impurities, but does not exactly follow the positions of impurities. This is because the coherence length is larger than the impurity spacing in our model. Under weak electric field, the sample is not metallic enough to have any non-zero conductivity as well as linear response regime. But when electric field is high, these low-$\Delta$ paths become precursors for conductive filaments.

At electric field close to IMT point (grey triangle in Fig. \ref{nano-fig3}). The insulating gap around the low-$\Delta$ paths is strongly suppressed, and a sizeable current is now flowing through the paths. This pattern of current paths acts as catalyst of resistive switching, and reduces the critical electric field $E_\text{IMT}$ quite dramatically.

The local effective temperature is evaluated sitewisely with Eq. \eqref{teff}, and its relative difference from N\'{e}el temperature $\delta t(\boldsymbol{r})=[T_\text{N}-T_\text{eff}(\boldsymbol{r})]/T_\text{N}$ is plotted in FIg. \ref{nano-fig4}(c). In the pre-filament regions, the effective temperature is slightly hotter than $T_\text{N}$ whereas in other regions it is slightly cooler. The overall temperature is about $T_\text{N}$ as shown in the inset of Fig. \ref{nano-fig3}(b). In (d), a non-Fermi-Dirac shaped distribution function is observed, showing non-equilibrium excitations in the upper band. And the hotter region has stronger excitations as expected.

\begin{figure}
\centering
\includegraphics[scale=0.7]{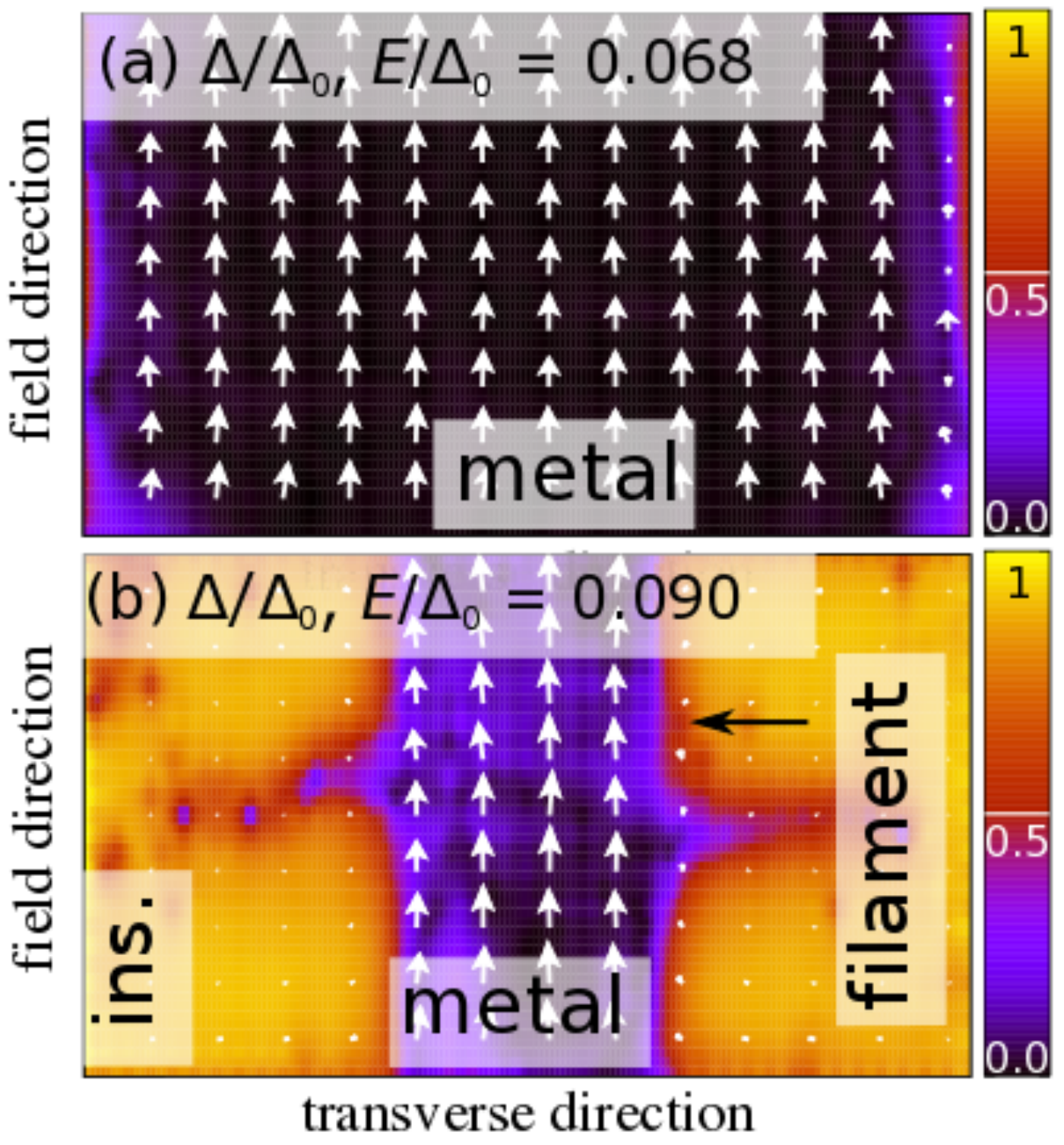}
\caption[Narrowing of metallic filament in MIT]{Narrowing of conductive filament during MIT. (a) At $E=0.068\Delta_0$, the insulating phase starts to nucleate at the the edges and is not much affected by bulk disorders. (b) On the NDR branch, the metallic filament narrows progressively until the IMT at $E=0.090\Delta_0$, shown as white circle in FIg. \ref{nano-fig3}. }
\label{nano-fig5}
\end{figure}

Now we look at the MIT under decreasing external bias $V_t$. As shown in Fig. \ref{nano-fig5}, it is triggered by the shrinking of conductive filaments upon decreasing bias. In Fig. \ref{nano-fig5}(a), the insulating phase just starts to nucleate from the edges of the sample at $E=0.068\Delta_0$. And in (b), the insulating phase keeps accumulating and the conductive filament in the center shrinks as total current decreases. As a result, the MIT depends on the sample boundary geometry, and is rather insensitive to the disorders in the bulk. 

\subsection{Filament dynamics and NDR behavior}
We have seen in Fig. \ref{nano-fig3} the NDR behavior during the MIT. It intrinsically originates from the non-equilibrium evolution of the conductive filament in the ordered solids. And it is crucial to have external resistor $R$ to reveal this regime. It is worth noting that although the NDR branch is revealed only when $R>0$, it is an intrinsic property of the sample. This can be proven numerically. In fact, one can stop simulation as soon as one solution on the NDR branch is reached. And once the solution is found, its convergence can be verified with the external resistor disconnected. We generally find the filamentary solutions are still convergent without external resistor. Moreover, it is possible to reproduce the same NDR $I-V$ curve by changing voltage $V_s$, starting from any convergent filamentary solution on the NDR branch. This procedure is confirmed on the reported $I-V$ curves.

\begin{figure}
\centering
\includegraphics[scale=1.0]{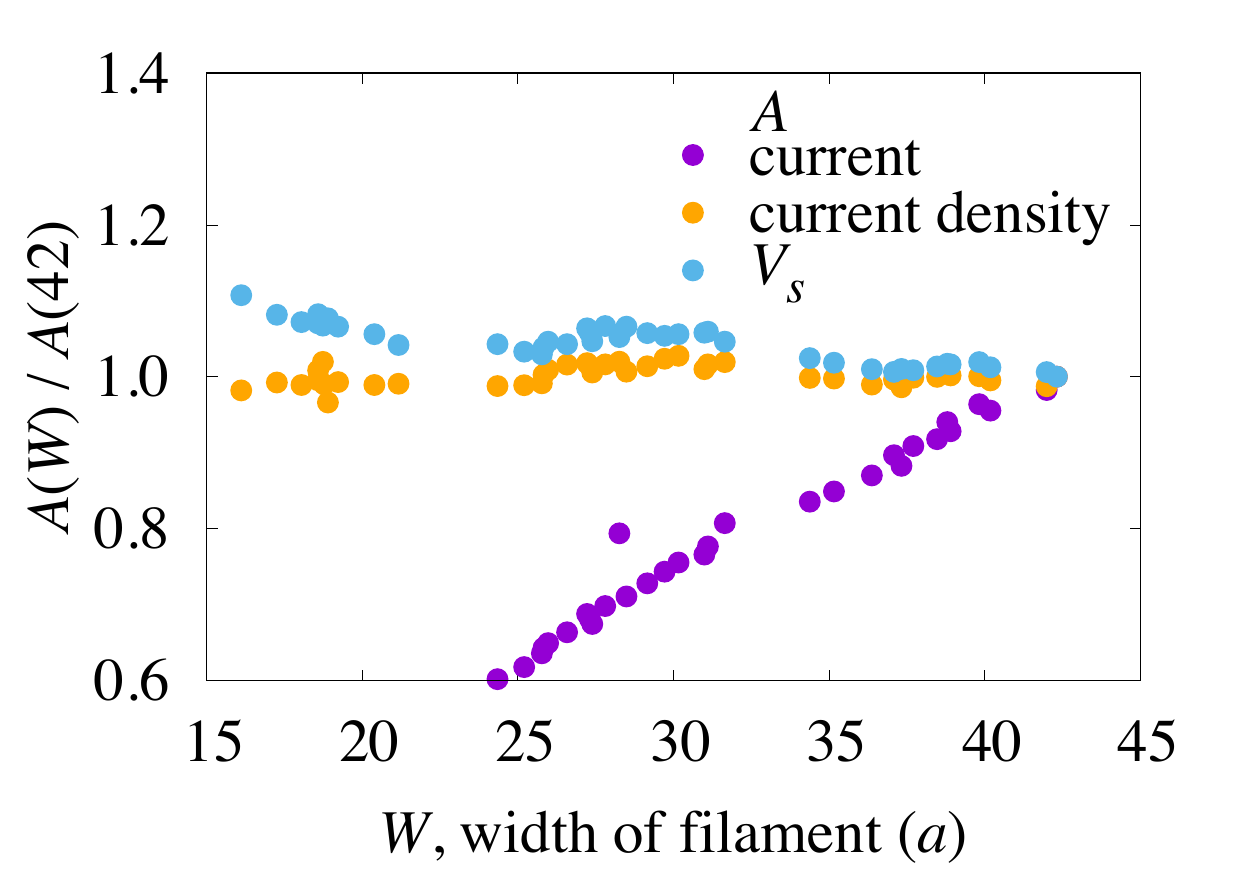}
\caption{The current-voltage relation in conductive filament during MIT.}
\label{nano-figS2}
\end{figure}

The current density, total current and sample voltage $V_s$ are plotted in Fig. \ref{nano-figS2}. The total current depends on the width of filament in a linear function. And the current density inside the filament is a property of the metallic phase, and is nearly a constant in the filamentary state through the transition. This clearly proves that the total current reduces because of the shrinking of filament. The sample voltage $V_s$, on the contrary, increases weakly during the MIT, resulting in negative $dI/dV_s$. Because the current density is constant, the increasing of $V_s$ is completely due to the fact that the \emph{resistivity} of the metallic filament increases. This may be explained with increased scattering in a narrower filament, as the scattering from domain boundaries of the metallic filament is strengthened for shorter inter-boundary distance.

In real experiments, people measure current $I$ versus total voltage $V_t$ as in Fig. \ref{nano-fig3}(a). But the intrinsic properties of the sample are more directly demonstrated with $I(V_s)$, which is the functional relation between current and the sample voltage $V_s=EL$. That is fundamentally what we plot in Fig. \ref{nano-fig3}(b). The relation of $V_s$ and $V_t$ is straightforward,
\begin{align}
V_t=V_s+IR.
\label{kirchhoff}
\end{align}
In the case that $V_s(I)$ is monotonic and single-valued, the $V_t(I)$ curve faithfully reflects all the characteristics of $V_s(I)$. But if it is non-monotonic or multi-valued, parts of $I(V_s)$ curve may be hidden from the $I-V_t$ relation. For instance, if the $I(V_s)$ is multivalued, then to fully reveal its functional relation it might be necessary to control $I$ instead of $V$. This corresponds to an essentially infinite $R$. Therefore, a convenient value of external resistance $R$ is necessary for revealing the whole picture.

Now we rewrite Eq. \eqref{kirchhoff} as
\begin{align}
\frac{V_t-V_s}{R}=I(V_s).
\end{align}
This indicates that for a given $V_t$ imposed by the dc generator, the $I$ and $V_s$ are determined by the intersection of $I-V_s$ curve with the straight line $(V_t-V_s)/R$ of slope $-1/R$.

\begin{figure}
\centering
\includegraphics[scale=0.7]{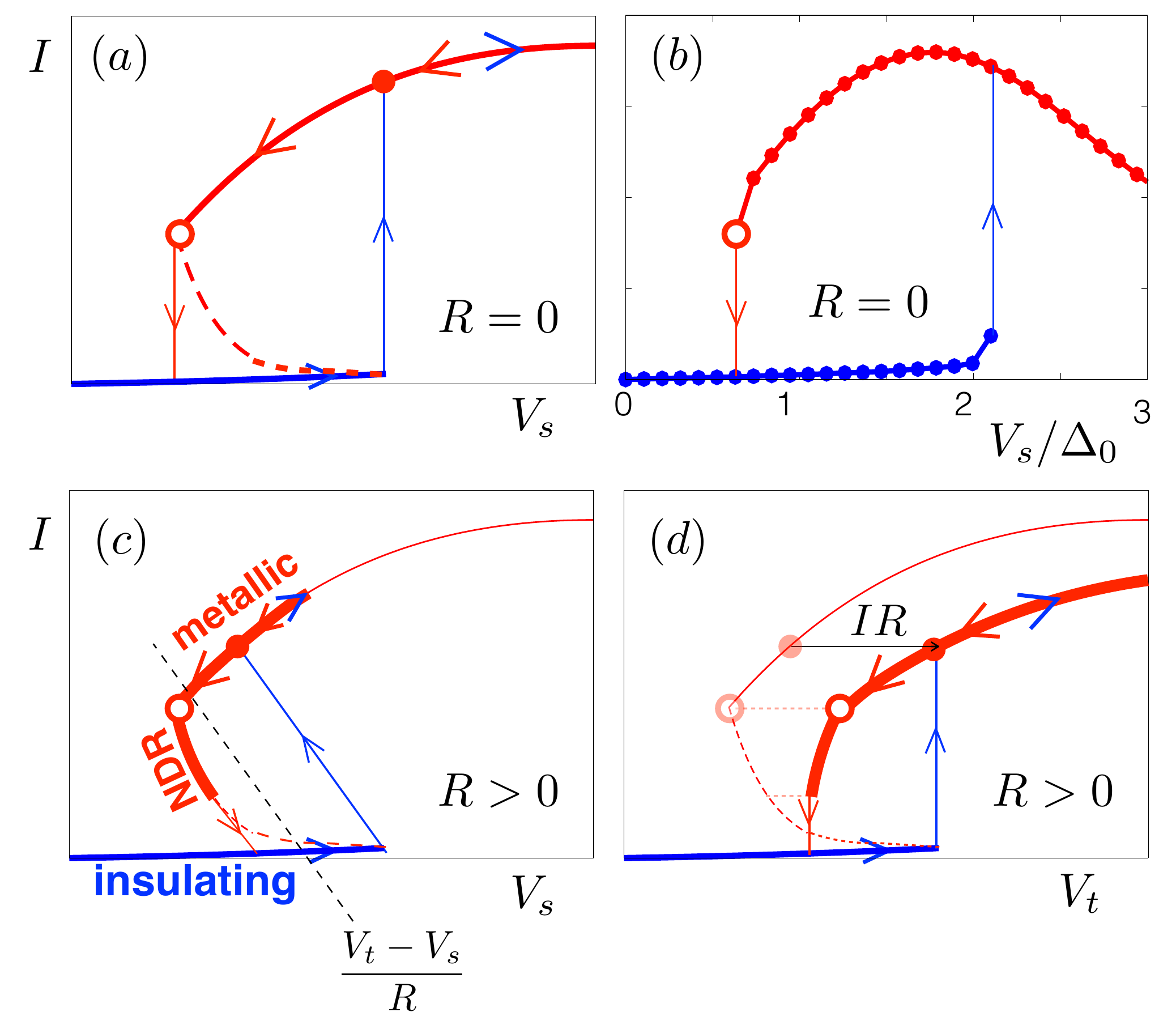}
\caption[NDR and intrinsic $I-V_s$ curve]{The NDR regime revealed due to the external resistor. (a) Schematic $I-V$ curve with external resistance $R=0$ and (b) the real numerical data for $c=0.03$. (c) Schematic $I-V_s$ curve and determination of $I-V_t$. (d) $I-V_t$ curve with external resistor $R>0$.}
\label{nano-figS1}
\end{figure}

Fig. \ref{nano-figS1} shows an S-shaped $I(V_s)$ curve and how the $I(V_t)$ curve is determined out of it. In our case, the the S-shaped $I-V_s$ relation is due to the shrinking of the filament under decreasing voltage bias\cite{ridley}. The NDR regime extends from the fully insulating phase to the fully metallic phase, and on the branch the system is a mixture of metallic and insulating regions, conceptually resembling the mixed phase of water in solid-liquid phase transition. 

For the $R=0$ case in Fig. \ref{nano-figS1}(a-b), a single trivial solution exists at $V_t=V_s=0$ with current $I=0$. As $V_t$ slowly increases, the solution continuously evolves out of equilibrium on the insulating branch (blue solid line). In the multi-valued regime, three solutions actually exist for the same $V_t$ but the system stays insulating due to continuity until $V_t=V_\text{IMT}$. When $V_t$ reaches the end of insulating branch, it cannot go back to the NDR branch (red dashed line) since $V_t$ is always increasing, therefore if jumps to the conducting branch (red solid line) as that is the only solution for $V_t>V_\text{IMT}$. This is the discontinuous insulator-to-metal transition. As $V_t$ is decreasing, the system initially remains metallic, and similarly jumps to the insulating branch at the end of metallic branch $V_t=V_\text{MIT}$. In summary, the NDR curve will not be revealed in the case $R=0$, and the hysteretic $I-V_t$ relation only features uniform metallic and insulating solutions. On the other hand, with $R>0$, the sample can enter the NDR branch as $V_t$ decreases, as shown in Fig. \ref{nano-figS1}(c-d).

It is not hard to prove the following condition, 
\begin{align}
R>R_\text{min},\text{ with }\frac{1}{R_\text{min}}=\text{min}\left\{\left|\frac{dI(V_s)}{dV_s}\right|;V_s\in[V_\text{MIT},V_\text{IMT}]\right\},
\end{align}
which guarantees that the full NDR branch is probed and no appearance of any hysteresis. When $V_s(I)$ has horizontal slope at some point on the NDR branch, the minimum of $|dI/dV_s|=0$ and $R_\text{min}\to\infty$. This current-controlled measurement has peen performed experimentally in Ref. \citenum{kim10}. In our calculation, $R=0.634$ is an intermediate value between $0$ and $R_\text{min}$, thus only part of the NDR branch is observed. 

\subsection{NDR in a large sample}
In the $I-V$ curves shown in Fig. \ref{prl-fig3}, we observed both IMT and MIT, but only after MIT the system enters NDR branch and maintains filamentary structure. In the IMT case, the filament usually extends very quickly to the whole sample, and the main feature observed is the $I-V$ relation of a completely metallic sample. Moreover, this leads to an upward kink in the inset of \ref{prl-fig3}(b). This effect is due to the fact that $E_\text{IMT}$ is significantly larger than $E_\text{MIT}$, and after IMT the system tends to have higher total current as well as larger filament width. Therefore, our $(80a)/\sqrt{2}\times(30a)/\sqrt{2}$ sample is too small to hold a conductive filament after IMT occurs. For a realistic sample used in experiments, its size is usually of several $\mu$m's, which is much larger than the size of the emerged filaments. 

\begin{figure}
\centering
\includegraphics[scale=1.1]{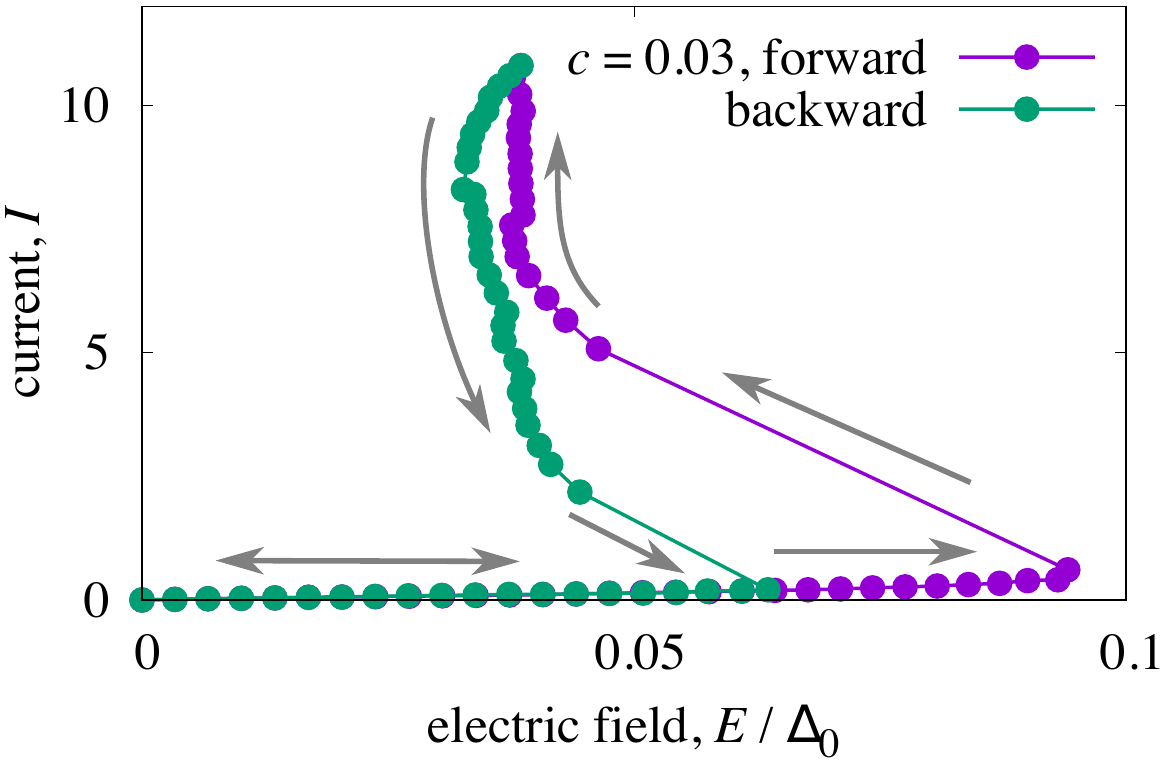}
\caption[Calculations on large sample]{$I-V$ curve for a larger sample. Its size ($(120a)/\sqrt{2}\times(30a)/\sqrt{2}$) is 50\% larger than that in the plots discussed above. Impurities are randomly distributed with concentration $c=0.03$. System undergoes IMT at $E_\text{IMT}\sim0.09\Delta_0$. In the ``forward" branch, conductive filament forms during IMT but does not extend to the whole sample. Due to the presence of filament, the system enters the NDR branch after IMT. The system is always filamentary after IMT, and never becomes completely metallic. In the backward branch, the filament narrows and finally system undergoes the MIT and returns to the high-resistance state.}
\label{lsample}
\end{figure}

To demonstrate the situation where IMT leads to a filamentary state, we perform the same calculations on a larger sample with size $(120a)/\sqrt{2}\times(30a)/\sqrt{2}$. This lattice has the same length but its width is 50\% larger than the above-mentioned one. Fig. \ref{lsample} shows the $I-V$ relation of the new sample. The total voltage $V_t$ is increased driving the system towards the IMT. At $E=E_\text{IMT}$, the system jumps to a filamentary state, and the electric field $E$ is much reduced due to the reduction of resistivity. Unlike the $I-V$ curve in Fig. \ref{prl-fig3}, the larger sample stays at the filamentary state as $V_t$ continues to increase. The current increases due to expansion of the conductive filament, and electric field decreases as we analyzed in the MIT case. When we decrease total voltage from a filamentary state, the system enters the backward $I-V$ branch, where current reduces due to narrowing of the conductive filament. Eventually the system undergoes MIT and transits to the completely insulating phase. Similar $I-V$ curve is observed in ordered insulators like VO$_2$ by a variety of experimental groups\cite{zimmers13, kim10}. It is worth noting that $I-V$ curve shows hysteresis when the conductive filament exists in the sample. This hysteresis reveals the multi-stability of filamentary solutions in non-equilibrium state. It is experimentally examined in Ref. \citenum{guenon13}, as shown in the Fig. \ref{guenon}. 

\section{Conclusion}
In this chapter, we have constructed a minimal microscopic model to study a strongly correlated lattice under high electric field. Our calculations successfully reproduce the main experimental features of resistive switching in transition metal oxides, e.g. vanadium oxides. We find that bistable phase of metallic and insulating solutions is induced by non-equilibrium effect and results in hysteretic $I-V$ curve observed in experiments. The IMT is triggered by sudden nucleation of conductive filaments, and the MIT occurs via narrowing of conductive filament, which leads to negative-differential-resistance behavior. We quantitatively verified that the RS is induced by Landau-Zener tunneling through the AF gap. And by showing $T_\text{eff}$ locally reaches $T_\text{N}$ before IMT, we reconciled the electronic mechanism with the thermal scenario in which Joule heating raises temperature and push the system to phase transition. The thermal mechanism is established if the non-equilibrium excitations created through LZ tunneling is interpreted as thermal excitations. So the electronic and thermal descriptions of the RS are essentially equivalent.

Furthermore, we have predicted strongly non-Fermi-Dirac shape for distribution function before IMT. Experiments performed with femto-second STM or photoemission may be able to resolve the dominant roles in the electronic and thermal mechanism.

\chapter{Strong-field Transport in Graphene}
\label{graphene}
In this chapter, we discuss the strong-field transport of graphene and the interplay between non-equilibrium effects and optical phonon interaction. We will concentrate on the non-equilibrium steady state of Dirac electrons, and examine whether and how current saturation occurs in the Dirac electron limit.

\section{Modeling NESS of Graphene}
\begin{figure}
\centering
\includegraphics[scale=1]{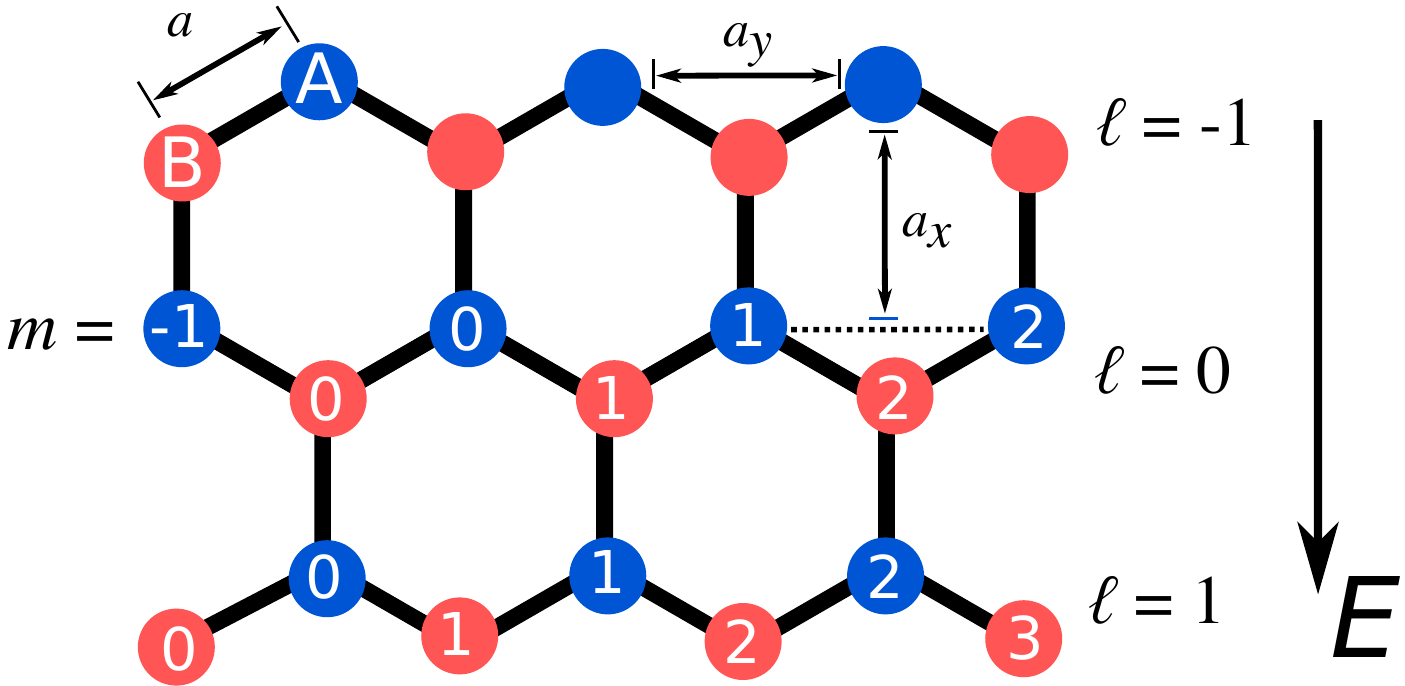}
\caption[Tight-binding model of graphene]{Tight-binding model of graphene. The A/B labels are pseudo-spins, indicating the sublattice to which the atoms belong.}
\label{grphmodel}
\end{figure}

To model the graphene under electric fields, we consider a tight-binding honeycomb lattice connected to fermion reservoirs at each site. Electrons are coupled to optical Holstein phonons . For simplicity, the phonons are treated as having uniform optical-phonon-frequency $\omega_\text{ph}$ (Einstein approximation). Suppose the electric field is along two of the six chemical bonds in a hexagon as shown in Fig. \ref{grphmodel}. The total hamiltonian reads,
\begin{align}
H=H_\text{TB}+H_\text{bath}+H_\text{ph}+H_\text{E}.
\end{align}
The hamiltonian is very similar with the one we discussed in Chap. \ref{prb} and \ref{prl}. And the difference is that $H_\text{TB}$ is now a tight-binding model on a honeycomb lattice and the interaction term $H_\text{ph}$ is the coupling of electrons with Holstein phonons,
\begin{align}
H_\text{ph}=g_0\sum_{\bm{rq}}(a_{\bm{rq}}+a^\dag_{\bm{rq}})d^\dag_{\bm{r}}d_{\bm{r}}+\sum_{\bm{rq}}\omega_\text{ph}a^\dag_{\bm{rq}}a_{\bm{rq}}.
\end{align}
with $a_{\bm{rq}}$ being the annihilation operator of optical phonons with momentum $\bm{q}$ at position $\bm{r}$. In the following discussion, we assume the electric field is along $x$-direction, and the perpendicular direction is $y$. In the following discussion, we do not consider real spin $\sigma=\uparrow,\downarrow$, and all the physical quantities, including current and carrier density, should be understood as those per spin.

To implement the non-equilibrium DMFT formulation developed in previous chapters, we organize the lattice sites with index $(\ell,m,s)$ as shown in Fig. \ref{grphmodel}. Note that the crystal structure of honeycomb is a triangular Bravais lattice with two-atom basis. The two atoms are labeled by pseudo-spin $s=A,B$. A and B atoms separately constitute two \emph{sublattice}s. We define the position vector of the basis $(\ell,m)$ being $\bm{r}_{\ell m}$ which is simply the position of the A-atom in the basis. We then divide the tight-binding hamiltonian $H_\text{TB}=H_{\text{TB},\parallel}+H_{\text{TB},\perp}$, with
\begin{align}
H_{\text{TB},\perp}&=\sum_{\ell m}\gamma\left(d^\dag_{\ell m A}d_{\ell m B}+d^\dag_{\ell m A}d_{\ell m+1, B}+H.c.\right),\nonumber\\
H_{\text{TB},\parallel}&=\sum_{\text{odd }\ell, m}\gamma\left(d^\dag_{\ell m B}d_{\ell+1, m-1, A}+H.c.\right)\nonumber\\
&+\sum_{\text{even }\ell, m}\gamma\left(d^\dag_{\ell m B}d_{\ell+1, m, A}+H.c.\right).
\end{align}
Finally, the electric field part of hamiltonian $H_E$ reads,
\begin{align}
H_E=-\sum_{\ell m s}(\bm{r}_{\ell m }+\delta_s\hat{\bm{y}})\cdot\bm{E}\left(d^\dag_{\ell m s}d_{\ell m s}+\sum_\alpha c^\dag_{\ell m s\alpha}c_{\ell m s\alpha}\right),
\end{align}
where $\delta_s=\frac{a_y}{2}\delta_{s,B}$. In equilibrium $E=0$ and $H_E=0$, the hamiltonian is diagonalized with Fourier transform $d^\dag_{\bm{k}s}=\sum_{\ell m}\text{e}^{i\bm{k}\cdot\bm{r}_{\ell m s}}d^\dag_{\ell m s}/\sqrt{N}$. In fact, under this transformation,
\begin{align}
H_{\text{TB},\perp}(\bm{k})=\gamma\left(1+\text{e}^{-ik_ya_y/2}\right)d^\dag_{\bm{k}A}d_{\bm{k}B}+H.c.,
\end{align}
where $a_y=\sqrt{3}a$ with lattice constant $a$. Now the range of $\bm{k}$ is the First Brillouin Zone $[-\pi/a_x,\pi/a_x)\times[-\pi/a_y,\pi/a_y)$. On the other hand, the parallel part becomes
\begin{align}
H_{\text{TB},\parallel}(\bm{k})=\gamma\text{e}^{-ik_ya_y/2}\text{e}^{-ik_xa_x}d^\dag_{\bm{k}A}d_{\bm{k}B}+H.c.
\end{align}
Combining the two components we can calculate the dispersion relation. Defining $\Delta_{\bm{k}}=\gamma\left(1+\text{e}^{-ik_ya_y/2}+\text{e}^{-ik_ya_y}\text{e}^{-ik_xa_x}\right)$, the energy-momentum relation reads as,
\begin{align}
\epsilon_{\bm{k}}&=\pm|\Delta_{\bm{k}}|, \quad\text{with}\nonumber\\
&=\pm\sqrt{1+4\cos\left(\frac{k_xa_x}{2}\right)\cos\left(\frac{k_ya}{2}\right)+4\cos^2\left(\frac{k_ya}{2}\right)}.
\end{align}
The $\pm$ sign corresponds to A/B sublattice. 

\subsection{Recursion relations}
To deal with the non-equilibrium situation $E\ne0$, we only diagonalize the hamiltonian in the perpendicular direction, so that we have
\begin{align}
H_{\text{TB}, \perp}(k_y)=\gamma\sum_{\ell}\left(1+\text{e}^{-ik_ya_y}\right)d^\dag_{\ell k_y A}d_{\ell,k_y,B}+H.c.,
\label{perph}
\end{align}
and the parallel part $H_{\text{TB},\parallel}$ becomes,
\begin{align}
H_{\text{TB}, \parallel}(k_y)=\sum_{\ell}\text{e}^{-i\frac{k_ya_y}{2}}d^\dag_{\ell+1,k_y, A}d_{\ell,k_y,B}+H.c.
\label{parah}
\end{align}
Now the two-dimensional problem is again reduced to one-dimensional modes with different $k_y$'s. To include the degrees of freedom due to pseudospin in a concise way, we define $2\times2$ matrix-valued Green's functions $\mathbf{G}^{r,\lessgtr}_{ss'}$ with $s,s'=A/B$. We then follow the procedures in section \ref{LZthermal} to derive the recursion relations for calculating Green's functions. We firstly define the ``one-electron" hamiltonian,
\begin{align}
&\hat{h}(k_y)=\begin{pmatrix}0&1+\exp\left(-ik_ya_y\right)\\1+\exp\left(ik_ya_y\right)&-Ea/2\end{pmatrix},\nonumber\\
\end{align}
in which the term $-Ea/2$ is due to different electrostatic potential energy for A/B electrons of the same $\ell$. Note $A$-electron only couples to $B$-electron in hamiltonian \eqref{parah} and vice versa. So hopping of electrons between $\ell$ and $\ell\pm1$ leads to matrix terms proportional to $\begin{pmatrix}0&&1\\0&&0\end{pmatrix}$ and $\begin{pmatrix}0&&0\\1&&0\end{pmatrix}$. To take into account this effect, we define
\begin{align}
\tilde{\mathbf{F}}_-&=\begin{pmatrix}0&&1\\0&&0\end{pmatrix}\mathbf{F}_{-}\begin{pmatrix}0&&0\\1&&0\end{pmatrix}=\begin{pmatrix}F_{-,BB}&&0\\0&&0\end{pmatrix}, \text{ and }\nonumber\\\tilde{\mathbf{F}}_+&=\begin{pmatrix}0&&0\\0&&F_{+,AA}\end{pmatrix},
\end{align}
Therefore, the recursion relations are written as below,
\begin{align}
\mathbf{F}^{r}_{\pm}(\omega)^{-1}&=\omega-\hat{h}(k_y)-\mathbf{\Sigma}^r(\omega)-\gamma^2\tilde{\mathbf{F}}^{r}_{\pm}(\omega\pm Ea_y),\nonumber\\
\mathbf{F}^{\lessgtr}_{\pm}(\omega)&=\mathbf{F}^{r,\pm}(\omega)\left(\mathbf{\Sigma}^{\lessgtr}(\omega)+\gamma^2\tilde{\mathbf{F}}^{\lessgtr}_{\pm}(\omega\pm Ea_y)\right)\mathbf{F}^{a}_{\pm}(\omega),
\end{align}
where $\mathbf{\Sigma}^{r,\lessgtr}=\mathbf{\Sigma}^{r,\lessgtr}_\Gamma+\mathbf{\Sigma}^{r,\lessgtr}_\text{ph}$ includes both contributions from fermion bath and from optical phonon scattering. The fermion bath part is 
\begin{align}
\mathbf{\Sigma}^r_\Gamma(\omega)&=i\Gamma\mathbb{I},\nonumber\\
\mathbf{\Sigma}^<_\Gamma(\omega)&=2i\Gamma \times\text{diag}\{f_\text{FD}(\omega),f_\text{FD}(\omega+Ea/2)\},
\end{align}
And the electron-phonon part will be discussed in the next section. After $\mathbf{F}^{r,\lessgtr}$ are computed, the local Green's functions are calculated straightforwardly using Eq. \eqref{gloc22}.

\subsection{Self energy of optical-phonon interaction}
In the second-order perturbation theory, the self energy is computed as
\begin{align}
\Sigma^>_\text{ph}(\omega)/g_0^2&=\mathcal{G}^>(\omega-\omega_\text{ph})(n_\text{ph}+1)+\mathcal{G}^>(\omega+\omega_\text{ph})n_\text{ph},\nonumber\\
\Sigma^<_\text{ph}(\omega)/g_0^2&=\mathcal{G}^<(\omega+\omega_\text{ph})(n_\text{ph}+1)+\mathcal{G}^<(\omega-\omega_\text{ph})n_\text{ph},
\end{align}
with Weiss-field Green's functions $\mathcal{G}$ and coupling constant $g_0$. The $n_\text{ph}=1/[\exp(\omega_\text{ph}/T)-1]$ is the Bose-Einstein distribution. In our calculations it is usually assumed that $\omega_\text{ph}\gg T_\text{bath}$ and $n_\text{ph}\sim 0$. As a result, the optical phonon reservoir is almost empty and absorptions of optical phonons by electrons would be very rare events. Only emissions of optical phonons are effectively relevant in this case.

After the DMFT calculation is convergent, the current per unit cell $\bar{J}$ is computed as
\begin{align}
\bar{J}&=i\gamma\langle d^\dag_{10B}d_{00A}-H.c.\rangle\nonumber\\
&=2\gamma\text{Re}G^<_{10A,00B}(t,t)\nonumber\\
&=2\gamma\sum_{k_y}\text{e}^{-ik_ya_y/2}\text{Re}G^<_{1k_yA,0k_yB}(t,t)/N_y\nonumber\\
&=\int_{\frac{-\pi}{a_y}}^{\frac{\pi}{a_y}} \frac{dk_y}{2\pi/a_y}\bar{J}_{k_y},
\end{align}
with
\begin{align}
\bar{J}_{k_y}&=2\gamma\text{e}^{-ik_ya_y/2}\text{Re}G^<_{1k_yA,0k_yB}(t,t)\nonumber\\
&=2\gamma\text{e}^{-ik_ya_y/2}\text{Re}\int \frac{d\omega}{2\pi}G^<_{1k_yA,0k_yB}(\omega).
\end{align}
Now we use Dyson's equation for non-equilibrium Green's functions and expand the final result as
\begin{align}
\bar{J}_{k_y}&=2\gamma^2\text{Re}\int \frac{d\omega}{2\pi}[G^<_{0k_yB}(\omega)F^{a}_{+,0k_yA}(\omega+Ea_y)\nonumber\\
&+G^r_{0k_yB}(\omega)F^{<}_{+,0k_yA}(\omega+Ea_y)].
\end{align}
This current contribution can be calculated with Green's functions solely from the one-dimensional problem with transverse momentum $k_y$. After it is calculated, summing over $k_y$ results in $\bar{J}$, the total current per unit cell. We should emphasize the current per unit cell is slightly different from the usually defined current density by a factor $a_y=\sqrt{3}a$. The current density is $J=\bar{J}/a_y$.

\subsection{Momentum distribution of electrons}
To compute the momentum distribution of electrons, we note that $\mathbf{n}_{\bm{k}}=-i\mathbf{G}^<_{\bm{k}\bm{0}}(t,t)
=-i\sum_{\ell m}\exp(i\bm{k}\cdot\bm{r}_{\ell m})\mathbf{G}^<_{\bm{r}_{\ell m}\bm{0}}(t,t)\nonumber$. We have used matrix-valued Green's functions and $\mathbf{n}_{\bm{k},ss'}=-iG^<_{\bm{k},ss'}(t,t)$. For simplicity of notations, we will omit the subscripts of $\bm{r}_{\ell m}$ and replace $\sum_{\ell m}$ by $\sum_{\bm{r}}$ in this section. Using the time-translational invariance of the Green's functions, time $t$ can be fixed as $0$, so the momentum distribution is calculated as
\begin{align}
\mathbf{n}_{\bm{k}}&=-i\sum_{\bm{r}}\exp(i\bm{k}\cdot\bm{r})\int \frac{d\omega}{2\pi} \mathbf{G}^<_{\bm{r}\bm{0}}(\omega)\nonumber\\
&=-i\sum_{\bm{r}}\exp(i\bm{k}\cdot\bm{r})\int \frac{d\omega}{2\pi} \sum_{\bm{r}'}\mathbf{G}^r_{\bm{r}\bm{r}'}(\omega)\mathbf{\Sigma}^<(\omega+\bm{r}'\cdot\bm{E})\mathbf{G}^a_{\bm{r}'\bm{0}}(\omega),
\end{align}
where $\mathbf{\Sigma}^<(\omega)=\mathbf{\Sigma}_\Gamma^<(\omega)+\mathbf{\Sigma}_\text{ph}^<(\omega)$ is the total lesser self energy, including components from both fermion reservoirs and optical phonon baths. Now we shift $\omega\to\omega-\bm{r}'\cdot\bm{E}$, and notice that $\mathbf{G}^r_{\bm{r+a}\bm{r'+a}}(\omega)=\mathbf{G}^r_{\bm{r}\bm{r}'}(\omega+\bm{a}\cdot\bm{E})$, with $\bm{r},\bm{r}'$ and $\bm{a}$ being lattice vectors. Therefore the formula is reduced to
\begin{align}
\mathbf{n}_{\bm{k}}&=-\frac{i}{2\pi}\int d\omega\sum_{\bm{r}\bm{r}'}\exp(i\bm{k}\cdot\bm{r})\mathbf{G}^r_{\bm{r}\bm{r}'}(\omega-\bm{r}'\cdot\bm{E})\mathbf{\Sigma}^<(\omega)\mathbf{G}^a_{\bm{r}'\bm{0}}(\omega-\bm{r}'\cdot\bm{E})\nonumber\\
&=-\frac{i}{2\pi}\int d\omega\sum_{\bm{r}\bm{r}'}\exp(i\bm{k}\cdot(\bm{r}-\bm{r}'))\mathbf{G}^r_{\bm{r}-\bm{r}',\bm{0}}(\omega)\mathbf{\Sigma}^<(\omega)[\exp(-i\bm{k}\cdot\bm{r}')\mathbf{G}^r_{-\bm{r}'\bm{0}}(\omega)]^\dag\nonumber\\
&=-\frac{i}{2\pi}\int d\omega\mathbf{G}^r_{\bm{k}}(\omega)\mathbf{\Sigma}^<(\omega)\mathbf{G}^a_{\bm{k}}(\omega),
\label{nkgrph}
\end{align}
where we have defined $\mathbf{G}^r_{\bm{k}}(\omega)=\sum_{\bm{r}}\exp(i\bm{k}\cdot\bm{r})\mathbf{G}^r_{\bm{r}\bm{0}}(\omega)$. In practical calculations, we firstly compute $\mathbf{G}^r_{\bm{r}\bm{0}}(\omega)$ and Fourier transform them to $\mathbf{G}^r_{\bm{k}}(\omega)$ in momentum space. Then $n_{\bm{k}}$ is calculated by evaluating the integral in \eqref{nkgrph}. Finally, to interpret the result $\mathbf{n}_{\bm{k}}$, we should expand it in terms of equilibrium diagonalized basis\cite{neto09}, 
\begin{align}
\psi_{\pm,\bm{k}}&=\frac{1}{\sqrt{2}}\begin{pmatrix}\text{e}^{-i\theta_{\bm{k}}/2}\\\pm\text{e}^{i\theta_{\bm{k}}/2}\end{pmatrix},\nonumber\\
&\text{with}\quad\theta_{\bm{k}}=\Delta_{\bm{k}}/|\Delta_{\bm{k}}|.
\end{align}
We define unitary transformation $U_{\bm{k}}=\begin{pmatrix}\psi_{+,\bm{k}}&\psi_{-,\bm{k}}\end{pmatrix}$, and transform the $\mathbf{n}_{\bm{k}}$,
\begin{align}
\mathbf{\tilde{n}}_{\bm{k}}=U_{\bm{k}}^\dag\mathbf{n}_{\bm{k}}U_{\bm{k}}
\end{align}
Then the particle numbers for upper/lower bands are $\mathbf{\tilde{n}}_{\bm{k},++}$ and $\mathbf{\tilde{n}}_{\bm{k},--}$. These equations will be useful to calculate the non-equilibrium momentum distribution in Section \ref{evol}.

\section{Signature of Landau-Zener tunneling}
\begin{figure}
\centering
\includegraphics[scale=1]{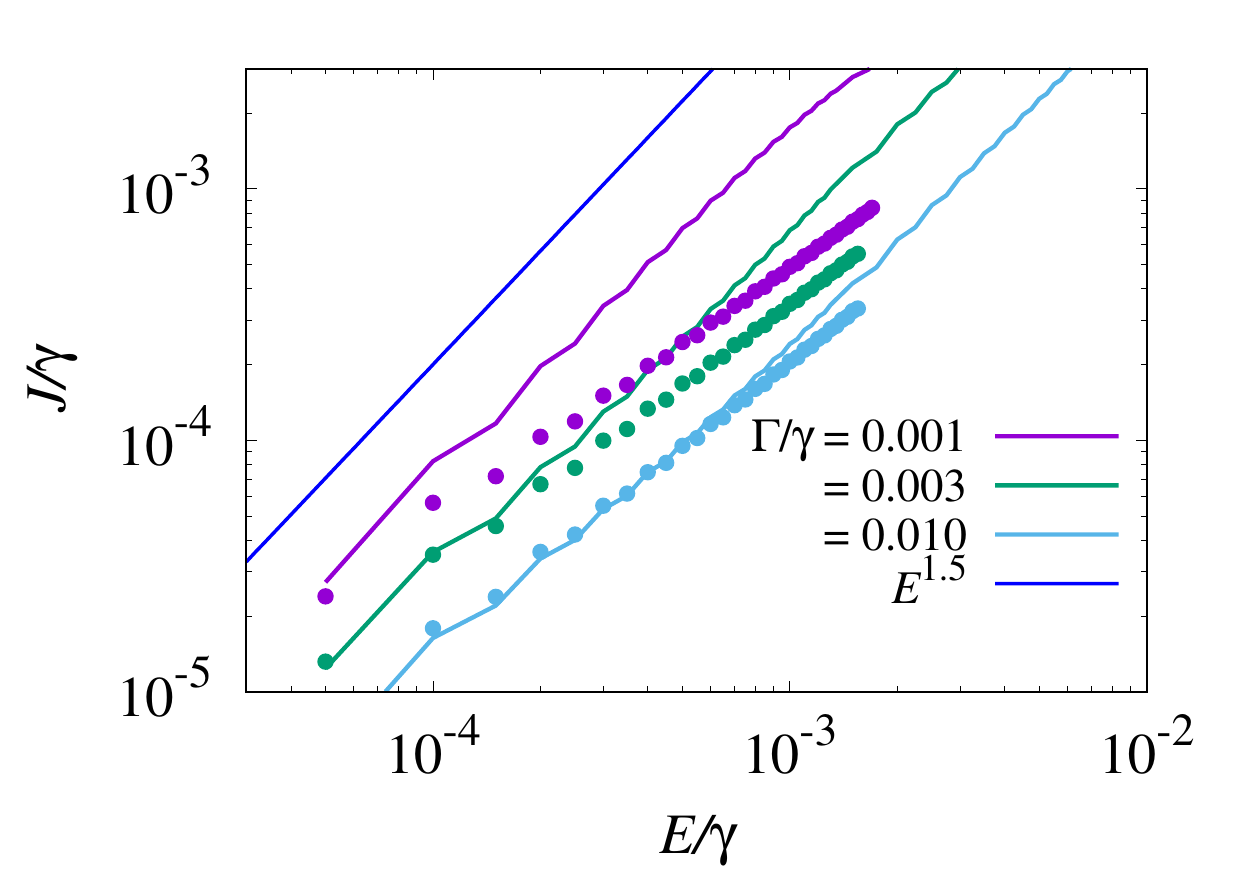}
\caption[$J-E$ relation of graphene under strong field]{$J-E$ relation of graphene under strong field. Current in the system with (dots) and without (solid lines) the optical-phonon interactions at different damping parameter $\Gamma$'s. As damping increases, the range of electric field in which $J\propto E^{1.5}$ holds expands. In the data, $\Gamma=0.001\gamma$ and $g_0^2=0.2\gamma$. Optical phonon frequency $\omega_\text{ph}$ is $0.05\gamma$.}
\label{jegrph}
\end{figure}

As discussed in Eq. \eqref{condgrph}, the weak-field conductivity of graphene per spin is given by Kubo formula
\begin{align}
\sigma_0=\frac{1}{2\pi^2}+\frac{\mu}{4\pi\Gamma},
\end{align}
where $\mu$ is the chemical potential and $\Gamma=g^2A(0)$ is the damping parameter. Note $g$ in this formula is the coupling constant to fermion reservoirs. When the chemical potential is high enough so that $\frac{\mu}{4\pi\Gamma}\gg \frac{1}{2\pi^2}$, the graphene conducts current like a metal with conductivity $\sigma_0\propto \mu$. A more interesting situation is when $\mu=0$ and the graphene sample is right at the Dirac point. The non-zero conductivity $\sigma_0=1/2\pi^2$ is a constant, and as we see in Appendix \ref{apkubo}, the effective temperature diverges in the weak-field limit. This anomaly suggests that strong non-equilibrium effect might occur even when the electric field is relatively small. To analyze the non-equilibrium steady state of graphene under electric fields, we consider the effective theory of graphene electrons at the Dirac point, which is
\begin{align}
h_\text{eff}=v_F[\sigma_x (p_x+Et)+\tau\sigma_y p_y],
\label{grpheff}
\end{align}
where the relative momentum $\bm{p}$ is measured from the center of a Dirac cone. The $v_F=3a\gamma/2$ is the Fermi velocity and $\tau=\pm1$ corresponds to the valley dof\cite{neto09}.  Defining unitary transformation
\begin{align}
U=\frac{1}{\sqrt{2}}\begin{pmatrix}1+i&&1-i\\1+i&&-1-i\end{pmatrix},
\end{align}
it is straightforward to verify that under this unitary transformation, the hamiltonian \eqref{grpheff} is equivalent to
\begin{align}
U^\dag h_\text{eff}U=v_F\begin{pmatrix}p_x+Et&&p_y\\p_y&&-(p_x+Et)\end{pmatrix},
\end{align}
which is nothing but the typical Landau-Zener hamiltonian with gap $\Delta=v_Fp_y$. Every transverse mode with fixed $p_y$ is thus a dissipative Landau-Zener tunneling problem. It has been discussed in Chap. \ref{prl} that if particles are initially in the lower band, then the tunneling rate to the upper band is
\begin{align}
\gamma_\text{LZ}=\exp(-\pi v_Fp_y^2/E).
\end{align} 
This suggests that only when $p_y\sim E$, significant particles will be excited to the upper band. On the other hand, in the semi-classical picture, electrons are accelerated by the electric field so that canonical $\bar{p}_x(t)=p_x+Et$ and relax in a rate $\tau^{-1}_\Gamma=2\Gamma$. In addition, since LZ tunneling typically occurs when the canonical momentum $\bar{p}_x(t)$ reaches the direct band gap at $\bar{p}_x=0$, those excited electrons can only reach a range $p_x\lesssim E\tau_\Gamma$. Combining these facts, an ansatz can be proposed for momentum distribution
\begin{align}
n_{\bm{p}}=\theta[\delta(E)-p_y]\theta[-\delta(E)+p_y]\theta(E\tau_\Gamma+p_x)\theta(p_x),
\label{nkansatz}
\end{align}
with $\delta(E)\propto\sqrt{E/v_F}$. This distribution is only non-zero in a rectangular box with length $L=E\tau_\Gamma$ and width $W=\sqrt{E/v_F}$. When $L\gg W$, this jet-like distribution is almost completely aligned with the electric field and the averaged velocity should be close to $v_F$, resulting in $J\propto v_FLW\propto E^{1.5}$. This argument is verified in Fig. \ref{jegrph}, where the current density $J$ is plotted versus electric field in log scale. We compare the case with optical phonon coupling $g_0=0$ to that with $g_0\ne0$. In the former case, the only dissipative mechanism is fermion reservoirs, and a $J\propto E^{1.5}$ scaling law is shown (solid lines). This is a signature of Landau-Zener mechanism. Under the electric field, electrons are excited from lower band ($\epsilon_{\bm{k}}<0$) to the upper band ($\epsilon_{\bm{k}}>0$). This is an example of the Schwinger effect\cite{schwinger51,Vandecasteele10,Rosenstein10,Kao10} of pair production in vacuum state. 

When optical phonon interaction is considered, the $J-E$ curve deviates from the non-interacting $1.5$-power law. Interestingly, the damping parameter $\Gamma$ changes the range where $1.5$ scaling holds. The stronger the damping $\Gamma$ is, the more robust the $J\propto E^{1.5}$ relation is. This is explained by the fact that optical phonon emission only takes effect when the energies of excited electrons approach $\omega_\text{ph}$. So the electric field at which deviation from $1.5$-law occurs should satisfy $E_\text{dev}\tau_\Gamma\sim \omega_\text{ph}/v_F$. As a result, we have $E_\text{dev}$ proportional to damping $\Gamma$. In experiments, this superlinear 1.5-power behavior is observed in low-mobility devices\cite{Vandecasteele10}. 

\section{Evolution of momentum distribution under external field}
\label{evol}
To further understand the current saturation due to optical phonon scattering, we look at the evolution of momentum distribution $n_{\bm{k}}$ under electric fields. 

\begin{figure}
\centering
\includegraphics[scale=0.4]{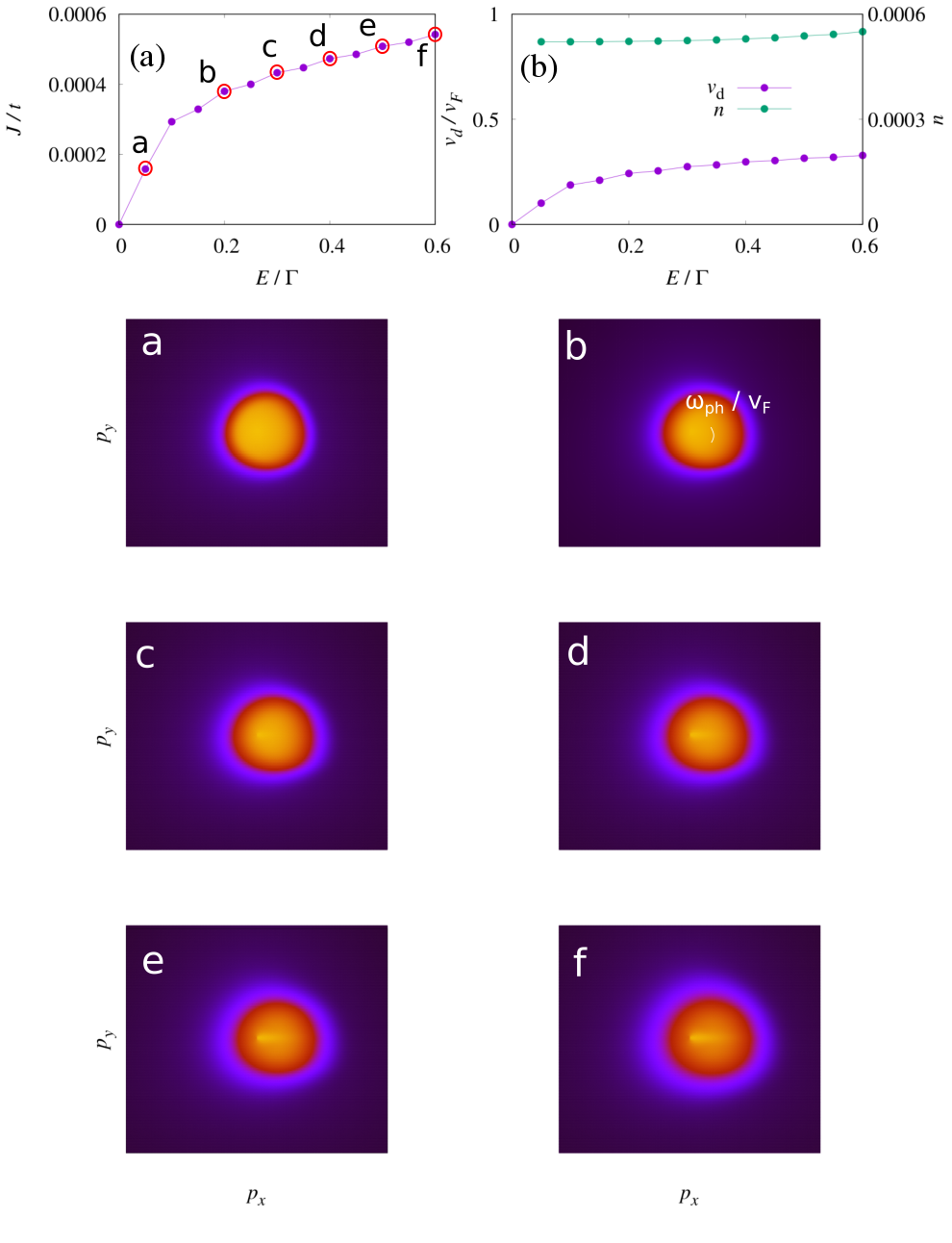}
\caption[Current and momentum distribution away from Dirac point]{Current and momentum distribution away from Dirac point. (a) Saturation of current under electric fields and (b) drift velocity $v_d$ and total current carriers number $n$ under electric fields. The $v_d$ saturates like current, and $n$ is almost unchanged. The color maps show momentum distributions of electrons at corresponding electric fields. $\mu=0.1\gamma$ in this case. Fermi sea is shifted at small electric fields. Its displacement is nearly unchanged at high fields. }
\label{nk10}
\end{figure}

The first situation is when $\mu>0$, and a finite Fermi sea exists around the center of the Dirac cone. Fig. \ref{nk10} shows the current and momentum distributions at a variety of electric fields. The Fermi sea is shifted along the field-direction when electric field is applied. However, at high electric fields, the Fermi sea is reluctant to shift, resulting in saturation of the current. This is due to strong coupling with optical phonons, so that electrons quickly lose energy to phonon baths as soon as the energy it gains from electric power reaches $\omega_\text{ph}$. The total density of current carriers $n$ is basically unchanged. The drift velocity $v_d$ is defined as 
\begin{align}
v_d=J/n,
\end{align}
which increases and saturates following the trend of $J$.

\begin{figure}
\centering
\includegraphics[scale=0.4]{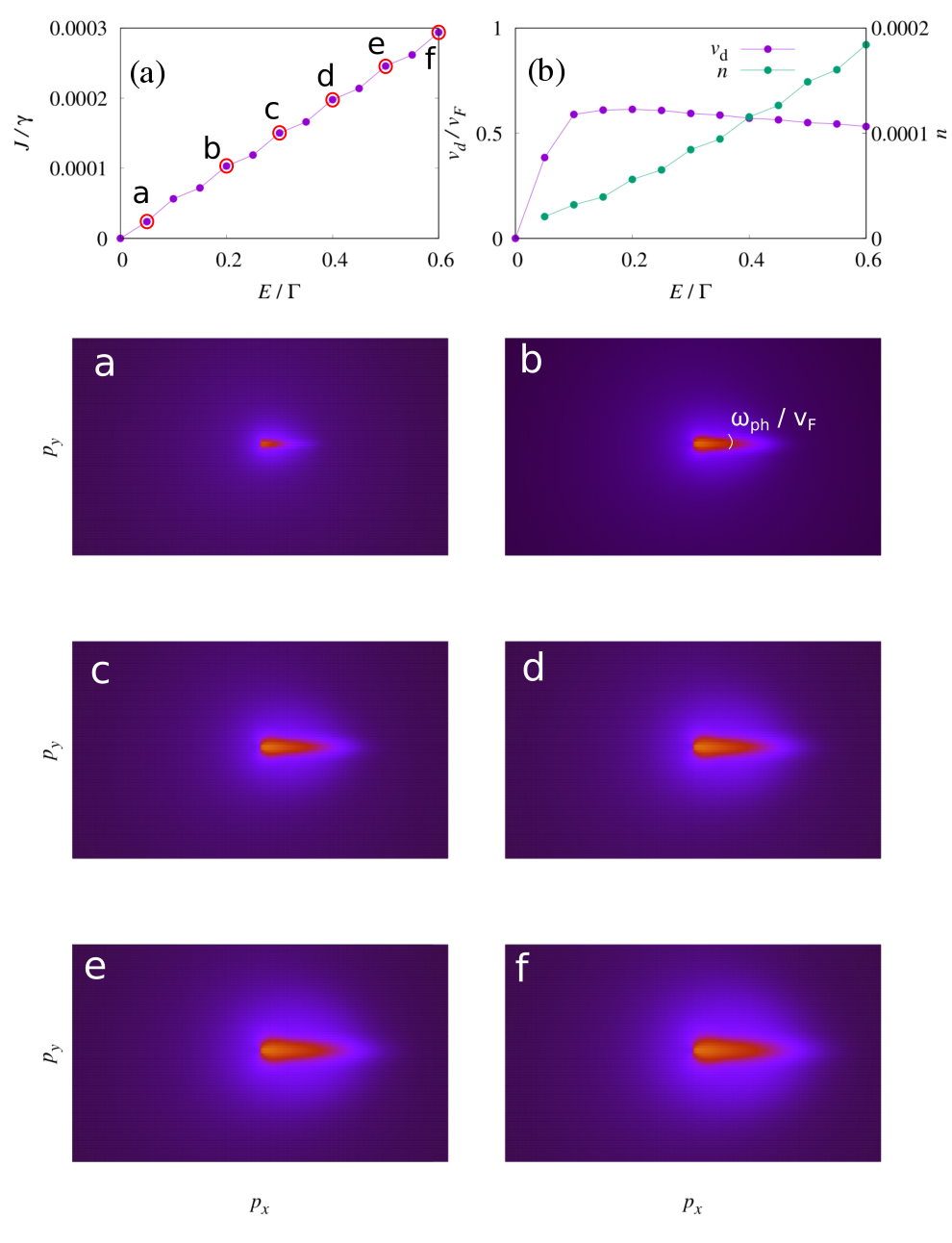}
\caption[Current and momentum distribution at Dirac point]{Current and momentum distribution at Dirac point. (a) current which scales linearly with electric field. (b) drift velocity $v_d$ and total number of current carriers $n$. the drift velociy overshoots and then decreases due to optical phonon emission. And the total number of excitations $n$ increases monotonically with $E$. The net result is the linearly increasing current $J=nv_d$. Colormaps show distribution of electrons (upper band) at a variety of electric fields. The distribution of holes is essentially identical. }
\label{nk00}
\end{figure}

Surprisingly, although current saturates when $\mu\ne0$ in Fig. \ref{nk10}(a), it increases almost linearly without saturation for the $\mu=0$ case as shown in Fig. \ref{jegrph} and Fig. \ref{nk00}(a). Moreover, number of excitations in this case, shown in Fig. \ref{nk00}(b), increases with the electric field, whereas the drift velocity $v_d=J/n$ saturates and slowly decreases. The linearly increasing current is the net result of a saturated velocity $v_d$ and an increasing number of current carriers $n$. To explain this phenomenology, we look at the momentum distributions. 

In the case of $\mu=0$, the Fermi surface is reduced to points at the center of the Dirac cone. When electric field is applied, the electrons are excited to the upper band due to Landau-Zener tunneling, leading to a jet-like charge distribution along the field-direction ($x$-direction). Holes in the lower band have the same momentum distribution. The shape of the distribution verifies the ansatz we proposed in Eq. \eqref{nkansatz}, with jet length $\sim 2\omega_\text{ph}/v_F$ upon saturation of $v_d$. At higher electric fields, the length of the ``jet" is fixed due to optical phonon interaction, whereas the jet width thickens, leading to increasing $n$ and reducing $v_d$. 

\section{Conclusion}
In this chapter, we implement the dissipative lattice model in a honeycomb crystal structure to discuss the non-equilibrium steady state in graphene. We discussed the critical role that Landau-Zener effect plays in the electronic transport in graphene. The calculations indicate that under strong electric field a jet-like distribution of electrons/holes forms when the system is at Dirac point, while the semiclassical picture of a shifted Fermi sea holds for systems away from Dirac point ($\mu\ne0$). For a system at Dirac point, we identify a parameter regime in which the $J\propto E^{1.5}$ relation is predicted for smaller electric fields. This can be explained with a simplified model in which excitations only exist in a narrow ``jet" in the momentum space where $p_y\lesssim\sqrt{E/v_F}$ and $p_x\lesssim E\tau$. In the presence of optical phonon interaction, the current deviates from the $1.5$-power law and transits to a $J\propto E$ relation. With optical phonons, the length of the jet-like distribution in field-direction is controlled by $\omega_\text{ph}/v_F$, with its width thickening due to continued formation of electron-hole excitations. This effect leads to increasing number of current carriers $n$ as well as saturated $v_d$. In contrary with high-carrier-density samples, graphene with its chemical potential at Dirac point shows no saturation of electric current at strong electric fields. The results discussed in this chapter are directly comparable to experimental data. For further details, we refer the reader to the publication \cite{li2018}.

\begin{ubbackmatter}
\appendix

\chapter{Kubo formula}
\label{apkubo}
The linear conductivity at small field of a non-interacting dissipative lattice model can be calculated with Kubo formula. For simplicity, we will calculate the current-current correlation function in imaginary time and then analytically continue it to real time. With Matsubara frequency $i\nu$ and momentum $q=0$ (uniform response), the optical conductivity is 
\begin{align}
\sigma(i\nu)=\frac{i}{i\nu}\frac{1}{L\beta}\sum_{k,n}v_k^2G_k(i\omega_n)G_k(i\omega_n+i\nu),
\end{align}
with group velocity $v_k=2\gamma\sin(k)$. The Matsubara Green's function is
\begin{align}
G_k(i\omega_n)=\frac{1}{i\omega_n-\epsilon_k+i\Gamma(\omega_n/|\omega_n|)}=\int d\epsilon\frac{\rho_0(\epsilon-\epsilon_k)}{i\omega_n-\epsilon},
\end{align}
where $\rho_0(\epsilon)=\Gamma/\pi(\epsilon^2+\Gamma^2)^{-1}$. We then perform the Matsubara summation and subsequently the analytic continuation $i\nu\to\omega+i\eta$ for finite $\omega$, 
\begin{align}
\sigma(\omega)=\frac{i}{\omega}\sum_kv^2_k\int d\epsilon_1\int d\epsilon_2\rho_0(\epsilon_1-\epsilon_k)\rho_0(\epsilon_2-\epsilon_k)\frac{f(\epsilon_1)-f(\epsilon_2)}{\omega+\epsilon_1-\epsilon_2+i\eta}.
\end{align}
To obtain the dc-conductivity, we need to take the real part and the static limit $\omega\to0$. Assuming zero temperature, we have the following result,
\begin{align}
\sigma_0=\frac{4\gamma^2\Gamma}{\pi}\int_0^{2\pi}\frac{d\boldsymbol{k}}{(2\pi)^d}\frac{|\boldsymbol{v_k}|^2}{\left(\Gamma^2+\epsilon_{\boldsymbol{k}}^2\right)^2},
\end{align}
in the $d$-dimensional system.

In 1D tight-binding chain, we have $\epsilon_k=-2\gamma \cos(k)$ and $v_k=2\gamma\sin(k)$, which gives
\begin{align}
\sigma_0&=\frac{4\gamma^2\Gamma^2}{\pi}\int_0^{2\pi}\frac{dk}{2\pi}\frac{\sin^2k}{(\Gamma^2+4\gamma^2\cos^2k)^2}\nonumber\\
&=\frac{2\gamma^2}{\pi\Gamma\sqrt{\Gamma^2+4\gamma^2}}.
\end{align}
The zero-energy spectral weight can be carried out in a similar manner,
\begin{align}
A_\text{loc}(0)=\int_0^{2\pi} \frac{dk}{2\pi} \frac{\Gamma/\pi}{4\gamma^2\cos^2k+\Gamma^2}=\frac{1}{\pi\sqrt{\Gamma^2+4\gamma^2}}.
\end{align}
Hence the effective temperature can be carried out analytically for one-dimensional TB chain. In the case of 2D TB lattice, the same equations hold, but the procedure is more complicated. It is difficult to get a closed formula of conductivity. And in the limit of $\Gamma\to0$, an approximate result is obtained,
\begin{align}
\sigma_0&=\frac{4\Gamma^2}{\pi}\int \frac{dk_xdk_y}{(2\pi)^2}\frac{\sin^2k_x+\sin^2k_y}{\left[\Gamma^2+4\gamma^2(\cos^2k_x+\cos^2k_y)^2\right]^2}\nonumber\\
&\approx 4\Gamma\int \frac{dk_xdk_y}{(2\pi)^2}\frac{\sin^2k_x+\sin^2k_y}{\Gamma^2+4\gamma^2(\cos^2k_x+\cos^2k_y)^2}\delta\left[4\gamma^2(\cos^2k_x+\cos^2k_y)^2\right]\nonumber\\
&\propto \frac{1}{\Gamma},
\end{align}
where we have used the relation,
\begin{align}
\lim_{\Gamma\to0}\frac{\Gamma/\pi}{x^2+\Gamma^2}=\delta(x).
\end{align}
The spectral function (DoS) needs more attentions, since it has a singularity at $\omega=0$. The singular behavior invalidates the Sommerfeld expansion of the RHS of Eq. \ref{eflux}. To get an estimate of $A_\text{loc}(\omega)$ at $\omega\ll\gamma$, we note
\begin{align}
A_\text{loc}(\omega)&=\frac{1}{2\pi}\int_{\epsilon_k=\omega}\frac{dS/(2\pi)}
{|\nabla_k\epsilon_k|}\nonumber\\
&=\frac{1}{(2\pi)^2}\int_{\epsilon_k=\omega}\frac{dS}{2\gamma\sqrt{\sin^2 k_x+\sin^2 k_y}}.
\end{align}
This is a complicated integration in general, but when $\omega \ll \gamma$, we can simplify it and obtain an approximate formula. First we notice the symmetry of the surface $\epsilon_k=\omega$, thus the integration can be done in the first quadrant $k_x,k_y>0$. Then we make variable substitution,
\begin{align}
K_1=\frac{1}{2}(k_x+k_y)\nonumber\\
K_2=\frac{1}{2}(k_x-k_y),
\end{align}
with Jacobi determinant $J=\left|\partial(k_x,k_y)/\partial(K_1,K_2)\right|=2$, and $\epsilon_K=-4\gamma \cos K_1\cos K_2$. Then the integration becomes
\begin{align}
A_\text{loc}(\omega)&=4\times\frac{1}{(2\pi)^2}\int_{-\frac{\pi}{2}+\delta_\omega}^{\frac{\pi}{2}-\delta_\omega}\frac{2dK_2}{4\gamma\sqrt{\cos^2 K_1\sin^2 K_2+\sin^2 K_1\cos^2 K_2}}\nonumber\\
&=4\times\frac{1}{(2\pi)^2}\int_{-\frac{\pi}{2}+\frac{\omega}{4\gamma}}^{\frac{\pi}{2}-\frac{\omega}{4\gamma}}F(K_2,\omega)dK_2,
\end{align}
with $K_1=K_1(\omega, K_2)$ is determined by $\omega=-4\gamma\cos K_1\cos K_2$, with the condition $K_1>0$ satisfied in the first quadrant of ($k_x,k_y$) plane. $\delta_\omega$ is the value of $K_2$ when the curve $\epsilon_K=\omega$ reaches the boundary of FBZ and $K_1=\pi\pm K_2$. It is determined with $\omega=4\gamma \cos\delta_\omega$. The factor $4$ comes from four quadrants. When $\omega\to0$, we have $K_1\to\frac{\pi}{2}$ and the integrand $F(K_2,\omega\to0)$ is finite and smooth except for $K_2=\pm\frac{\pi}{2}$. Therefore, for small $\omega$ the integration is approximately
\begin{align}
A_\text{loc}(\omega)&\approx4\times\frac{1}{(2\pi)^2}\int_{-\frac{\pi}{2}+\delta_\omega}^{\frac{\pi}{2}-\delta_\omega}dK_2[F(K_2,0)+\mathcal{O}(\omega)]\nonumber\\
&=4\times\frac{1}{(2\pi)^2}\int_{-\frac{\pi}{2}+\delta_\omega}^{\frac{\pi}{2}-\delta_\omega}\frac{dK_2}{2\gamma|\cos K_2|}+\mathcal{O}(\omega)\nonumber\\
&\approx 4\times\frac{1}{(2\pi)^2\gamma}\log\left[\cot\left(\frac{1}{4}\sqrt{\frac{\omega}{\gamma}}\right)\right]+\mathcal{O}(\omega)\nonumber\\
&\approx -\frac{2}{(2\pi)^2\gamma}\log\omega+\mathcal{O}(\omega),
\end{align}
and only the leading term $\sim \log\omega$ is critical for getting an estimate of effective temperature.

Last but not the least, we consider the situation of Dirac electrons in graphene, i.e., two-dimensional linearized dispersion relation $\epsilon_{\bm{k}}=c|\boldsymbol{k}|$. Assuming chemical potential is $\mu$, the evaluation of spectral function $A_\text{loc}(\mu)$ is straightforward, and the conductivity can be computed as
\begin{align}
\sigma_0&=\frac{\Gamma^2}{\pi}\int_0^\infty\frac{2\pi kdk}{(2\pi)^2}\frac{c^2}{[\Gamma^2+(ck-\mu)^2]^2}\nonumber\\
&=\frac{1}{2\pi^2}\int_0^\infty\frac{kdk}{\Gamma^2}\frac{c^2}{\left[1+\left(\frac{ck-\mu}{\Gamma}\right)^2\right]}\nonumber\\
&=\frac{1}{2\pi^2}\left\{\int_0^\infty \frac{x}{(1+x^2)^2}dx+\frac{\mu}{\Gamma}\int_0^\infty\frac{1}{(1+x^2)^2}dx\right\}\nonumber\\
&=\frac{1}{4\pi^2}+\frac{1}{8\pi}\frac{\mu}{\Gamma}.
\end{align}
In this result, the second term corresponds to the regular conductance due to finite chemical potential $\mu$. The first term is non-zero even for $\mu=0$, resulting in the universal minimum conductivity in graphene\cite{dassarma11}. In fact, this term is $\frac{1}{2\pi}\frac{e^2}{h}$ when constants are restored. It should be multiplied by degeneracy of spin and valley, and the ``lower half" of the Dirac cone (of holes) should be counted. Then the universal conductivity is $2g_vg_s\times \frac{1}{2\pi}\frac{e^2}{h}=\frac{4e^2}{\pi h}$, as verified in Ref. \citenum{Miao1530, danneau08}.

\chapter{Anisotropic effective temperature}
\label{apateff}
Now we discuss the direction-resolved effective temperature in 2D lattice model, to support the anisotropy observed in electric-field-driven IMT and filament formation. We will derive the following formula
\begin{align}
T_\text{eff}\sim\frac{|\boldsymbol{E}\cdot\boldsymbol{v}_F|}{\Gamma},
\end{align}
which is applied to the regions inside metallic domains, to account for the strong dependence of Joule heating on the crystallographic direction with respect to the field direction. This expression also elucidates how Joule heating sets the temperature as the result of a balance of external field $\boldsymbol{E}$ and energy dissipation $\Gamma$.

In the metallic regime, we can neglect the on-site Coulomb interaction. When the spatial and temporal inhomogeneties of the system are at the mesoscopic scale, we consider the the gradient expansion up to first order to obtain the Quantum Boltzmann Equation, 
\begin{align}
[\partial_T+\boldsymbol{v(k)}\cdot\nabla_{\boldsymbol{X}}+\boldsymbol{E}\cdot\nabla_{\boldsymbol{k}}]f(X;k)=2\Gamma[f_0(\omega-\mu_{\boldsymbol{X}})-f(X;k)],
\end{align}
where $X=(T,\boldsymbol{X})$ are spacetime coordinates and $k=(\omega,\boldsymbol{k})$ are their corresponding Fourier components. The velocity $\boldsymbol{v}(\boldsymbol{k})=\nabla_{\boldsymbol{k}}\epsilon(\boldsymbol{k})$ with the dispersion relation $\epsilon(\boldsymbol{k})$. Note that the Quantum Boltzmann Equation takes into account the non-trivial band structure $\epsilon(\boldsymbol{k})$, whereas the conventional BTE (Eq. \eqref{bte}) only considers $\bm{v}(\bm{k})=\bm{k}/m$. With Coulomb interaction ignored in our case, the RHS of the equation is only due to scattering with the degrees of freedom of the local fermion baths, which are equilibrium at temperature $T_\text{bath}$ and chemical potential $\mu_{\bm{X}}=-\bm{E}\cdot\bm{X}$. $f_0(\epsilon)=[1+\exp(\epsilon/T_\text{bath})]^{-1}$ is the Fermi-Dirac distribution function. 

In the steady state, the time-dependence of $f(X;k)$ drops out, and the Quantum Boltzmann Equation becomes,
\begin{align}
[\boldsymbol{v(k)}\cdot\nabla_{\boldsymbol{X}}+\boldsymbol{E}\cdot\nabla_{\boldsymbol{k}}]f(X;k)=2\Gamma[f_0(\omega+\bm{E}\cdot\bm{X})-f(X;k)].
\end{align}
In a first-order approximation, one can check that the term $\bm{E}\cdot\nabla$ can be neglected by expanding $f(X;k)$ in a power series of $\bm{E}$. The resulting equation reduces to,
\begin{align}
\boldsymbol{v(k)}\cdot\nabla_{\boldsymbol{X}}f(X;k)=2\Gamma[f_0(\omega+\bm{E}\cdot\bm{X})-f(X;k)].
\end{align}
Take the limit of $T_\text{bath}\to0$, the equation can be solved analytically,
\begin{align}
f(\bm{X};\omega,\bm{k})&=\Theta(-\omega-\bm{E}\cdot\bm{X})+\nonumber\\
&+\frac{1}{2}[\text{sign}(\omega+\bm{E}\cdot\bm{X})+\text{sign}(\bm{E}\cdot\bm{v}(\bm{k}))]\exp\left(-2\Gamma\left|\frac{\omega+\bm{E}\cdot\bm{X}}{\bm{E}\cdot\bm{v}(\bm{k})}\right|\right),
\end{align}
where $\Theta(x)$ is the Heaviside step function. For wave-vectors $\bm{k}^*$ such that $\bm{E}\perp\bm{k}^*$, this equation becomes
\begin{align}
f(\bm{X};\omega,\bm{k}^*)=\Theta(-\omega-\bm{E}\cdot\bm{X}),
\end{align}
which is simply the zero-temperature Fermi-Dirac distribution. And for wave vectors such that $\bm{v}(\bm{k})\parallel \bm{E}$, the distribution function is far from the zero-temperature F-D function. In general, the effective temperature depending on $\bm{k}$ reads,
\begin{align}
T_\text{eff}\sim\frac{|\bm{E}\cdot\bm{v}(\bm{k})|}{\Gamma},
\end{align}
therefore in the weak-field limit in a non-interacting model, electrons traveling in the field direction have higher effective temperature than those traveling perpendicularly. In the weak-field limit, as the current is mostly contributed by electrons at Fermi surface, which have velocity $\bm{v_F}$, we reach the conclusion,
\begin{align}
T_\text{eff}\sim\frac{|\bm{E}\cdot\bm{v_F}|}{\Gamma}.
\end{align}
By redefining the electric potential slope with Hartree-Fock mean-field, $\bm{E}\to\tilde{\bm{E}}_\sigma(\bm{X})\equiv \bm{E}-\nabla_{\bm{X}}[U\langle n_{-\sigma}(\bm{X}\rangle]$, the expression is approximately generalized to the interacting model, showing a dependence on the inhomogeneous non-equilibrium distribution of charge and order parameter.
\clearpage
\references[Bibliography]{thesisbib}
\end{ubbackmatter}

\begin{thebibliography}{10}

\bibitem{chudnovskii98}
F.~Chudnovskii, A.~Pergament, G.~Stefanovich, P.~Metcalf, and J.~Honig,
  ``Switching phenomena in chromium-doped vanadium sesquioxide,'' {\em Journal
  of applied physics}, vol.~84, no.~5, pp.~2643--2646, 1998.

\bibitem{sujay15}
S.~Singh, G.~Horrocks, P.~M. Marley, Z.~Shi, S.~Banerjee, and
  G.~Sambandamurthy, ``Proliferation of metallic domains caused by
  inhomogeneous heating near the electrically driven transition in
  ${\mathrm{vo}}_{2}$ nanobeams,'' {\em Phys. Rev. B}, vol.~92, p.~155121, Oct
  2015.

\bibitem{pan14}
F.~Pan, S.~Gao, C.~Chen, C.~Song, and F.~Zeng, ``Recent progress in resistive
  random access memories: materials, switching mechanisms, and performance,''
  {\em Materials Science and Engineering: R: Reports}, vol.~83, pp.~1--59,
  2014.

\bibitem{waser07}
R.~Waser and M.~Aono, ``Nanoionics-based resistive switching memories,'' {\em
  Nature materials}, vol.~6, no.~11, pp.~833--840, 2007.

\bibitem{waser09}
R.~Waser, R.~Dittmann, G.~Staikov, and K.~Szot, ``Redox-based resistive
  switching memories--nanoionic mechanisms, prospects, and challenges,'' {\em
  Advanced materials}, vol.~21, no.~25-26, pp.~2632--2663, 2009.

\bibitem{kumai99}
R.~Kumai, Y.~Okimoto, and Y.~Tokura, ``Current-induced insulator-metal
  transition and pattern formation in an organic charge-transfer complex,''
  {\em Science}, vol.~284, no.~5420, pp.~1645--1647, 1999.

\bibitem{jeong13}
J.~Jeong, N.~Aetukuri, T.~Graf, T.~D. Schladt, M.~G. Samant, and S.~S. Parkin,
  ``Suppression of metal-insulator transition in vo2 by electric field--induced
  oxygen vacancy formation,'' {\em Science}, vol.~339, no.~6126,
  pp.~1402--1405, 2013.

\bibitem{oka03}
T.~Oka, R.~Arita, and H.~Aoki, ``Breakdown of a mott insulator: a nonadiabatic
  tunneling mechanism,'' {\em Physical review letters}, vol.~91, no.~6,
  p.~066406, 2003.

\bibitem{oka10}
T.~Oka and H.~Aoki, ``Dielectric breakdown in a mott insulator: Many-body
  schwinger-landau-zener mechanism studied with a generalized bethe ansatz,''
  {\em Phys. Rev. B}, vol.~81, p.~033103, Jan 2010.

\bibitem{oka12}
T.~Oka, ``Nonlinear doublon production in a mott insulator: Landau-dykhne
  method applied to an integrable model,'' {\em Phys. Rev. B}, vol.~86,
  p.~075148, Aug 2012.

\bibitem{sugimoto08}
N.~Sugimoto, S.~Onoda, and N.~Nagaosa, ``Field-induced metal-insulator
  transition and switching phenomenon in correlated insulators,'' {\em Phys.
  Rev. B}, vol.~78, p.~155104, Oct 2008.

\bibitem{eckstein10}
M.~Eckstein, T.~Oka, and P.~Werner, ``Dielectric breakdown of mott insulators
  in dynamical mean-field theory,'' {\em Phys. Rev. Lett.}, vol.~105,
  p.~146404, Sep 2010.

\bibitem{janod15}
E.~Janod, J.~Tranchant, B.~Corraze, M.~Querr{\'e}, P.~Stoliar, M.~Rozenberg,
  T.~Cren, D.~Roditchev, V.~T. Phuoc, M.-P. Besland, {\em et~al.}, ``Resistive
  switching in mott insulators and correlated systems,'' {\em Advanced
  Functional Materials}, vol.~25, no.~40, pp.~6287--6305, 2015.

\bibitem{guiot13}
V.~Guiot, L.~Cario, E.~Janod, B.~Corraze, V.~T. Phuoc, M.~Rozenberg,
  P.~Stoliar, T.~Cren, and D.~Roditchev, ``Avalanche breakdown in gata4se8-
  xtex narrow-gap mott insulators,'' {\em Nature communications}, vol.~4,
  p.~1722, 2013.

\bibitem{stoliar13}
P.~Stoliar, L.~Cario, E.~Janod, B.~Corraze, C.~Guillot-Deudon,
  S.~Salmon-Bourmand, V.~Guiot, J.~Tranchant, and M.~Rozenberg, ``Universal
  electric-field-driven resistive transition in narrow-gap mott insulators,''
  {\em Advanced materials}, vol.~25, no.~23, pp.~3222--3226, 2013.

\bibitem{sblee}
S.~Lee, S.~Chae, S.~Chang, J.~Lee, S.~Park, Y.~Jo, S.~Seo, B.~Kahng, and
  T.~Noh, ``Strong resistance nonlinearity and third harmonic generation in the
  unipolar resistance switching of nio thin films,'' {\em Applied Physics
  Letters}, vol.~93, no.~25, p.~252102, 2008.

\bibitem{driscoll09}
T.~Driscoll, H.-T. Kim, B.-G. Chae, M.~Di~Ventra, and D.~Basov,
  ``Phase-transition driven memristive system,'' {\em Applied physics letters},
  vol.~95, no.~4, p.~043503, 2009.

\bibitem{duchene71}
J.~Duchene, M.~Terraillon, P.~Pailly, and G.~Adam, ``Filamentary conduction in
  vo2 coplanar thin-film devices,'' {\em Applied Physics Letters}, vol.~19,
  no.~4, pp.~115--117, 1971.

\bibitem{zimmers13}
A.~Zimmers, L.~Aigouy, M.~Mortier, A.~Sharoni, S.~Wang, K.~G. West, J.~G.
  Ramirez, and I.~K. Schuller, ``Role of thermal heating on the voltage induced
  insulator-metal transition in ${\mathrm{vo}}_{2}$,'' {\em Phys. Rev. Lett.},
  vol.~110, p.~056601, Jan 2013.

\bibitem{mcwhan73}
D.~McWhan, A.~Menth, J.~Remeika, W.~Brinkman, and T.~Rice, ``Metal-insulator
  transitions in pure and doped v 2 o 3,'' {\em Physical Review B}, vol.~7,
  no.~5, p.~1920, 1973.

\bibitem{guenon13}
S.~Gu{\'e}non, S.~Scharinger, S.~Wang, J.~Ram{\'\i}rez, D.~Koelle, R.~Kleiner,
  and I.~K. Schuller, ``Electrical breakdown in a v2o3 device at the
  insulator-to-metal transition,'' {\em EPL (Europhysics Letters)}, vol.~101,
  no.~5, p.~57003, 2013.

\bibitem{meric08}
I.~Meric, M.~Y. Han, A.~F. Young, B.~Ozyilmaz, P.~Kim, and K.~L. Shepard,
  ``Current saturation in zero-bandgap, top-gated graphene field-effect
  transistors,'' {\em Nature nanotechnology}, vol.~3, no.~11, pp.~654--659,
  2008.

\bibitem{barreiro09}
A.~Barreiro, M.~Lazzeri, J.~Moser, F.~Mauri, and A.~Bachtold, ``Transport
  properties of graphene in the high-current limit,'' {\em Physical review
  letters}, vol.~103, no.~7, p.~076601, 2009.

\bibitem{shishir09}
R.~Shishir, D.~Ferry, and S.~Goodnick, ``Room temperature velocity saturation
  in intrinsic graphene,'' in {\em Journal of Physics: Conference Series},
  vol.~193, p.~012118, IOP Publishing, 2009.

\bibitem{fang11}
T.~Fang, A.~Konar, H.~Xing, and D.~Jena, ``High-field transport in
  two-dimensional graphene,'' {\em Physical Review B}, vol.~84, no.~12,
  p.~125450, 2011.

\bibitem{ramamoorthy15}
H.~Ramamoorthy, R.~Somphonsane, J.~Radice, G.~He, C.-P. Kwan, and J.~Bird,
  ````freeing'' graphene from its substrate: Observing intrinsic velocity
  saturation with rapid electrical pulsing,'' {\em Nano letters}, vol.~16,
  no.~1, pp.~399--403, 2015.

\bibitem{perebeinos10}
V.~Perebeinos and P.~Avouris, ``Inelastic scattering and current saturation in
  graphene,'' {\em Physical Review B}, vol.~81, no.~19, p.~195442, 2010.

\bibitem{QDS}
U.~Weiss, {\em Quantum Dissipative Systems}.
\newblock World Scientific, 2008.

\bibitem{Tsuji09}
N.~Tsuji, T.~Oka, and H.~Aoki, ``Nonequilibrium steady state of photoexcited
  correlated electrons in the presence of dissipation,'' {\em Phys. Rev.
  Lett.}, vol.~103, p.~047403, 2009.

\bibitem{RMP-NEQDMFT}
H.~Aoki, N.~Tsuji, M.~Eckstein, M.~Kollar, T.~Oka, and P.~Werner,
  ``Nonequilibrium dynamical mean-field theory and its applications,'' {\em
  Rev. Mod. Phys.}, vol.~86, pp.~779--837, Jun 2014.

\bibitem{Caldeira-Leggett83}
A.~O. Caldeira and A.~J. Leggett, ``Path integral approach to quantum brownian
  motion☆ path integral approach to quantum brownian motion☆,'' {\em Path
  integral approach to quantum Brownian motion}, vol.~121, no.~3, p.~587, 1983.

\bibitem{jauho94}
A.-P. Jauho, N.~S. Wingreen, and Y.~Meir, ``Time-dependent transport in
  interacting and noninteracting resonant-tunneling systems,'' {\em Phys. Rev.
  B}, vol.~50, pp.~5528--5544, Aug 1994.

\bibitem{Turkowski-Freericks}
V.~Turkowski and J.~K. Freericks, ``Nonlinear response of bloch electrons in
  infinite dimensions,'' {\em Phys. Rev. B}, vol.~71, p.~085104, 2005.

\bibitem{Jauho-Wilkins}
A.~P. Jauho and J.~W. Wilkins, ``Theory of high-electric-field quantum
  transport for electron-resonant impurity systems,'' {\em Phys. Rev. B},
  vol.~29, p.~1919, 1984.

\bibitem{graf-vogl}
M.~Graf and P.~Vogl, ``Electromagnetic fields and dielectric response in
  empirical tight-binding theory,'' {\em Phys. Rev. B}, vol.~51, p.~4940, 1995.

\bibitem{jong-prb}
J.~Han, ``Solution of electric-field-driven tight-binding lattice coupled to
  fermion reservoirs,'' {\em Phys. Rev. B}, vol.~87, p.~058119, 2013.

\bibitem{Gellmann-Goldberger}
M.~Gell-Mann and M.~L. Goldberger, ``The formal theory of scattering,'' {\em
  Phys. Rev.}, vol.~91, pp.~398--408, Jul 1953.

\bibitem{jong-prb07}
J.~E. Han, ``Mapping of strongly correlated steady-state nonequilibrium system
  to an effective equilibrium,'' {\em Phys. Rev. B}, vol.~75, p.~125122, Mar
  2007.

\bibitem{jong-prb06}
J.~E. Han, ``Quantum simulation of many-body effects in steady-state
  nonequilibrium: Electron-phonon coupling in quantum dots,'' {\em Phys. Rev.
  B}, vol.~73, p.~125319, Mar 2006.

\bibitem{ashcroft78}
N.~W. Ashcroft, N.~D. Mermin, and S.~Rodriguez, {\em Solid state physics}.
\newblock AAPT, 1978.

\bibitem{lebwohl70}
P.~A. Lebwohl and R.~Tsu, ``Electrical transport properties in a
  superlattice,'' {\em Journal of applied physics}, vol.~41, no.~6,
  pp.~2664--2667, 1970.

\bibitem{dassarma11}
S.~Das~Sarma, S.~Adam, E.~H. Hwang, and E.~Rossi, ``Electronic transport in
  two-dimensional graphene,'' {\em Rev. Mod. Phys.}, vol.~83, pp.~407--470, May
  2011.

\bibitem{Miao1530}
F.~Miao, S.~Wijeratne, Y.~Zhang, U.~C. Coskun, W.~Bao, and C.~N. Lau,
  ``Phase-coherent transport in graphene quantum billiards,'' {\em Science},
  vol.~317, no.~5844, pp.~1530--1533, 2007.

\bibitem{meir-wingreen}
Y.~Meir and N.~S. Wingreen, ``Landauer formula for the current through an
  interacting electron region,'' {\em Phys. Rev. Lett.}, vol.~68,
  pp.~2512--2515, Apr 1992.

\bibitem{kotliar-rmp}
A.~Georges, G.~Kotliar, W.~Krauth, and M.~J. Rozenberg, ``Dynamical mean-field
  theory of strongly correlated fermion systems and the limit of infinite
  dimensions,'' {\em Rev. Mod. Phys.}, vol.~68, pp.~13--125, Jan 1996.

\bibitem{georges96}
A.~Georges, G.~Kotliar, W.~Krauth, and M.~J. Rozenberg, ``Dynamical mean-field
  theory of strongly correlated fermion systems and the limit of infinite
  dimensions,'' {\em Rev. Mod. Phys.}, vol.~68, pp.~13--125, Jan 1996.

\bibitem{prange-kadanoff}
R.~E. Prange and L.~P. Kadanoff, ``Transport theory for electron-phonon
  interactions in metals,'' {\em Phys. Rev.}, vol.~134, pp.~A566--A580, May
  1964.

\bibitem{Tsuji08}
N.~Tsuji, T.~Oka, and H.~Aoki, ``Correlated electron systems periodically
  driven out of equilibrium: $\text{Floquet}+\text{DMFT}$ formalism,'' {\em
  Phys. Rev. B}, vol.~78, p.~235124, Dec 2008.

\bibitem{Aron-prl12}
C.~Aron, G.~Kotliar, and C.~Weber, ``Dimensional crossover driven by an
  electric field,'' {\em Phys. Rev. Lett.}, vol.~108, p.~086401, Feb 2012.

\bibitem{Amaricci12}
A.~Amaricci, C.~Weber, M.~Capone, and G.~Kotliar, ``Approach to a stationary
  state in a driven hubbard model coupled to a thermostat,'' {\em Phys. Rev.
  B}, vol.~86, p.~085110, Aug 2012.

\bibitem{Aron-prb12}
C.~Aron, ``Dielectric breakdown of a mott insulator,'' {\em Phys. Rev. B},
  vol.~86, p.~085127, Aug 2012.

\bibitem{joura08}
A.~V. Joura, J.~K. Freericks, and T.~Pruschke, ``Steady-state nonequilibrium
  density of states of driven strongly correlated lattice models in infinite
  dimensions,'' {\em Phys. Rev. Lett.}, vol.~101, p.~196401, Nov 2008.

\bibitem{yamada75}
K.~Yamada, ``Perturbation expansion for the anderson hamiltonian. iv,'' {\em
  Progress of Theoretical Physics}, vol.~54, no.~2, p.~316, 1975.

\bibitem{goldhaber-gordon98}
D.~Goldhaber-Gordon, H.~Shtrikman, D.~Mahalu, D.~Abusch-Magder, U.~Meirav, and
  M.~A. Kastner, ``Kondo effect in a single-electron transistor,'' {\em
  Nature}, vol.~391, pp.~156--159, 01 1998.

\bibitem{cronenwett98}
S.~M. Cronenwett, T.~H. Oosterkamp, and L.~P. Kouwenhoven, ``A tunable kondo
  effect in quantum dots,'' {\em Science}, vol.~281, no.~5376, pp.~540--544,
  1998.

\bibitem{altshuler09}
B.~L. Altshuler, V.~E. Kravtsov, I.~V. Lerner, and I.~L. Aleiner, ``Jumps in
  current-voltage characteristics in disordered films,'' {\em Physical review
  letters}, vol.~102, no.~17, p.~176803, 2009.

\bibitem{hansmann13}
P.~Hansmann, A.~Toschi, G.~Sangiovanni, T.~Saha-Dasgupta, S.~Lupi, M.~Marsi,
  and K.~Held, ``Mott--hubbard transition in v2o3 revisited,'' {\em physica
  status solidi (b)}, vol.~250, no.~7, pp.~1251--1264, 2013.

\bibitem{berglund69}
C.~Berglund, ``Thermal filaments in vanadium dioxide,'' {\em IEEE Transactions
  on Electron Devices}, vol.~16, no.~5, pp.~432--437, 1969.

\bibitem{ridley}
B.~Ridley, ``Specific negative resistance in solids,'' {\em Proceedings of the
  Physical Society}, vol.~82, no.~6, p.~954, 1963.

\bibitem{kim10}
H.-T. Kim, B.-J. Kim, S.~Choi, B.-G. Chae, Y.~W. Lee, T.~Driscoll, M.~M.
  Qazilbash, and D.~Basov, ``Electrical oscillations induced by the
  metal-insulator transition in vo 2,'' {\em Journal of applied physics},
  vol.~107, no.~2, p.~023702, 2010.

\bibitem{neto09}
A.~C. Neto, F.~Guinea, N.~M. Peres, K.~S. Novoselov, and A.~K. Geim, ``The
  electronic properties of graphene,'' {\em Reviews of modern physics},
  vol.~81, no.~1, p.~109, 2009.

\bibitem{schwinger51}
J.~Schwinger, ``On gauge invariance and vacuum polarization,'' {\em Physical
  Review}, vol.~82, no.~5, p.~664, 1951.

\bibitem{Vandecasteele10}
N.~Vandecasteele, A.~Barreiro, M.~Lazzeri, A.~Bachtold, and F.~Mauri,
  ``Current-voltage characteristics of graphene devices: Interplay between
  zener-klein tunneling and defects,'' {\em Phys. Rev. B}, vol.~82, p.~045416,
  Jul 2010.

\bibitem{Rosenstein10}
B.~Rosenstein, M.~Lewkowicz, H.~C. Kao, and Y.~Korniyenko, ``Ballistic
  transport in graphene beyond linear response,'' {\em Phys. Rev. B}, vol.~81,
  p.~041416, Jan 2010.

\bibitem{Kao10}
H.~C. Kao, M.~Lewkowicz, and B.~Rosenstein, ``Ballistic transport, chiral
  anomaly, and emergence of the neutral electron-hole plasma in graphene,''
  {\em Phys. Rev. B}, vol.~82, p.~035406, Jul 2010.

\bibitem{li2018}
J.~Li and J.~E. Han, ``Nonequilibrium excitations and transport of dirac
  electrons in electric-field-driven graphene,'' {\em Phys. Rev. B}, vol.~97,
  p.~205412, May 2018.

\bibitem{danneau08}
R.~Danneau, F.~Wu, M.~F. Craciun, S.~Russo, M.~Y. Tomi, J.~Salmilehto, A.~F.
  Morpurgo, and P.~J. Hakonen, ``Shot noise in ballistic graphene,'' {\em Phys.
  Rev. Lett.}, vol.~100, p.~196802, May 2008.

\end{thebibliography}
\end{document}